\documentclass[aps,prx,reprint,superscriptaddress,nofootinbib]{revtex4-1}

\usepackage{graphicx}
\usepackage{xcolor}
\usepackage{amsmath}
\usepackage{amssymb}

\begin{document}


\title{Ensemble nonequivalence and Bose-Einstein condensation in weighted networks}

\author{Qi Zhang}
\email[]{zhang@lorentz.leidenuniv.nl}
\affiliation{Lorentz Institute for Theoretical Physics, Leiden University, Niels Bohrweg 2, 2333 CA Leiden, The Netherlands}
\author{Diego Garlaschelli}
\email[]{garlaschelli@lorentz.leidenuniv.nl, diego.garlaschelli@imtlucca.it}
\affiliation{Lorentz Institute for Theoretical Physics, Leiden University, Niels Bohrweg 2, 2333 CA Leiden, The Netherlands}
\affiliation{IMT School of Advanced Studies, Piazza San Francesco 19, 55100 Lucca, Italy}

\date{\today}

\begin{abstract}
The asymptotic (non)equivalence of canonical and microcanonical ensembles, describing systems with soft and hard constraints respectively, is a central concept in statistical physics. Traditionally, the breakdown of ensemble equivalence (EE) has been associated with nonvanishing relative canonical fluctuations of the constraints in the thermodynamic limit. Recently, it has been reformulated in terms of a nonvanishing relative entropy density between microcanonical and canonical probabilities. The earliest observations of EE violation required phase transitions or long-range interactions. More recent research on binary networks found that an extensive number of local constraints can also break EE, even in absence of phase transitions. 
Here we study for the first time ensemble nonequivalence in weighted networks with local constraints.
Unlike their binary counterparts, these networks can undergo a form of Bose-Einstein condensation (BEC) producing a core-periphery structure where a finite fraction of the link weights concentrates in the core. This phenomenon creates a unique setting where local constraints coexist with a phase transition.
We find surviving relative fluctuations only in the condensed phase, as in more traditional BEC settings. 
However, we also find a non-vanishing relative entropy density for all temperatures, signalling a breakdown of EE due to the presence of an extensive number of constraints, irrespective of BEC.
Therefore, in presence of extensively many local constraints, vanishing relative fluctuations no longer guarantee EE.
\end{abstract}


\maketitle

\section{Introduction}
Statistical ensembles were introduced by Gibbs~\cite{gibbs1902elementary} to mathematically describe systems at thermodynamic equilibrium, i.e. where certain conserved macroscopic properties (such as the total energy) are constant, while the microscopic state (i.e. the state of all the microscopic constituents) is subject to fluctuations. 
For a system with $n$ units and discrete degrees of freedom, a statistical ensemble is a probability distribution $P(\mathbf{W})$ over the collection $\mathcal{W}_n=\{\mathbf{W}\}$ of all the possible (unobserved) microscopic states of the system, given a set of measurable macroscopic properties. 
Clearly, $P(\mathbf{W})\ge 0$ for all $\mathbf{W}\in\mathcal{W}_n$ and 
$\sum_{\mathbf{W}\in\mathcal{W}_n}P(\mathbf{W})=1$.
This distribution conceptualizes the fact that, ideally, repeated observations of the microscopic state would retrieve different (and independent) outcomes.
It can be viewed as the probability distribution that maximizes the Gibbs-Shannon entropy functional
\begin{equation}\label{eq_shannon}
  S[P]\equiv-\sum_{\mathbf{W}\in\mathcal{W}_n}P(\mathbf{W})\ln P(\mathbf{W}),
\end{equation}
under a set of (macroscopic) constraints, therefore being maximally noncommittal with respect to missing (microscopic) information~\cite{jaynes1957information}.

Depending on the choice of the macroscopic properties being constrained, different statistical ensembles can be constructed. 
The \emph{microcanonical ensemble} $P_\textrm{mic}(\mathbf{W}|E)$ is used to describe systems with fixed total energy $E$ (energetic isolation), while the \emph{canonical ensemble} $P_\textrm{can}(\mathbf{W}|\beta)$ is used to describe systems with fixed (inverse) temperature $\beta$ (thermal equilibrium)~\cite{gibbs1902elementary}. 
For a physical system, the inverse temperature $\beta$ equals $1/kT$ where $T$ is the absolute temperature and $k$ is Boltzmann's constant. 
In both ensembles, the microscopic state $\mathbf{W}$ is random, but the randomness is governed differently by the two distributions $P_{\textrm{can}}(\mathbf{W}|{\beta})$ and $P_\textrm{mic}(\mathbf{W}|E)$.
In particular, while the microcanonical ensemble assigns each realized configuration $\mathbf{W}$ a constant (deterministic) value $E(\mathbf{W})=E$ of the total energy (which can therefore be regarded as a `hard' constraint corresponding to energetic isolation), the canonical ensemble assigns configurations a fluctuating (random) energy with a certain expected value $\langle E\rangle_{\beta}=\sum_{\mathbf{W}\in\mathcal{W}_n}P_\textrm{can}(\mathbf{W}|\beta)E(\mathbf{W})$ and a positive standard deviation $\sigma_{\beta}(E)>0$ (i.e. the energy plays the role of a `soft' constraint resulting from the contact with a heat bath at fixed inverse temperature $\beta$).

\subsection{Conjugate ensembles}
The two ensembles can be made conjugate to each other by choosing a specific value $E^*$ and simultaneously setting the total energy $E(\mathbf{W})$ of each realized configuration $\mathbf{W}$ in the microcanonical ensemble equal to  $E(\mathbf{W})\equiv E^*$ and the inverse temperature $\beta$ in the canonical ensemble to the corresponding value $\beta\equiv\beta^*$ such that the resulting average value $\langle E\rangle_{\beta^*}$ of the fluctuating total energy under the canonical probability $P_\textrm{can}(\mathbf{W}|\beta^*)$ equals $E^*$, i.e. 
\begin{equation}
\langle E\rangle_{\beta^*}\equiv E^*.
\label{eq_conjugate}
\end{equation}

For systems with finite size, the two conjugate ensembles are unavoidably different, because in the microcanonical ensemble the hardness of the constraint implies extra dependencies among the state of the microscopic constituents with respect to the canonical case. 
However, in the \emph{thermodynamic limit} (i.e. when the number of units in the system goes to infinite) and under certain `natural' circumstances, the two associated probabilistic descriptions are expected to become effectively equivalent (i.e. the canonical fluctuations and the microcanonical dependencies are both expected to play an asymptotically vanishing role) as a result of some form of the law of large numbers. 
This idea, which dates back to Gibbs himself~\cite{gibbs1902elementary} and has continued to attract a lot of interest until presently~\cite{campa2009statistical,touchette2015equivalence,squartini2015breaking,squartini2017reconnecting}, goes under the name of \emph{ensemble equivalence} (EE).
When EE holds, one can treat the two ensembles as asymptotically interchangeable, and hence use any of them based on mathematical or computational convenience. For instance, analytical calculations are signficantly easier in the canonical ensemble, while numerical randomizations of an initial configuration can be carried out more naturally in the microcanonical ensemble.

Most statistical physics textbooks still convey the message that EE is expected to hold in general as a sort of principle at the basis of ensemble theory. 
The possible breakdown of EE, also known as \emph{ensemble nonequivalence} (EN), is still not discussed systematically in the literature.
However, several observations of EN have been documented over the past decades~\cite{ellis2000large,blume1971ising,barre2001inequivalence,ellis2004thermodynamic,lynden1999negative,chavanis2003gravitational,d2000negative,barre2007ensemble,radin2013phase,ellis2002nonequivalent,kastner2010nonequivalence,campa2009statistical,squartini2015breaking}. 
These observations motivated various efforts aimed at elucidating both the possible physical mechanisms at the origin of EN and, in parallel, its proper mathematical definition(s).

\subsection{Physical mechanisms for ensemble (non)equivalence}
Traditionally, the `natural' circumstances generally invoked to ensure EE mainly concern the presence of (loosely speaking) `at most weak' interactions between the constituents of the system. 
This condition is automatically realized when the system consists of independent units or units with short-range interactions and sufficiently high temperature (to stay away from possible low-temperature phases with broken symmetries, for which the canonical average value of the energy is no longer the typical value). 
Indeed, violations of EE have been documented in presence of long-range interactions (e.g. in gravitational systems) or phase transitions (e.g. in interacting spin systems)~\cite{ellis2000large,blume1971ising,barre2001inequivalence,ellis2004thermodynamic,lynden1999negative,chavanis2003gravitational,d2000negative,barre2007ensemble,radin2013phase,campa2009statistical,touchette2015equivalence}.

However, recent research on complex systems encountered beyond the usual realm of physics has found an additional mechanism that can break EE, even in presence of weak (or no) interactions: namely, the presence of \emph{an extensive number of local constraints}~\cite{squartini2015breaking,squartini2017reconnecting}. 
This situation is frequently found in networks with constraints on the number of links (\emph{degree}) of each node.
More specifically, the \emph{binary configuration model}~\cite{squartini2017maximum} is a widely used null model of graphs with a given \emph{degree sequence}, i.e. a given vector of node degrees.
The model captures many properties found in real-world networks, because the local character of the degree constraint can accomodate the strong structural heterogeneity typically observed across nodes in real networks. 
Unlike the traditional thermodynamic example where the total energy (and possibly a small, finite number of additional macroscopic properties) is a global and unique constraint for the system, networks with given node degrees are systems with as many constraints as the number of fundamental units, i.e. where constraints are \emph{extensive in number and local in nature.}
This situation has been found to break the equivalence of the corresponding canonical and microcanonical ensembles, even without long-range interactions or phase transitions~\cite{squartini2015breaking,garlaschelli2016ensemble,roccaverde2019breaking}.
Notably, since systems with local constraints are generic models for virtually any heterogenous system, the new mechanism significantly widens the range of real-world cases for which EE may break down. 
This novel result deserves further research.

\subsection{Mathematical definitions of ensemble (non)equivalence\label{sec:math}}
Besides the aforementioned advances in the study of the physical mechanisms at the origin of EN, significant progress has been made in the mathematical characterization of EN as well.
Traditionally, the informal criterion~\cite{gibbs1902elementary} used to test whether two conjugate ensembles are equivalent is checking whether the \emph{relative fluctuations} of the constraint in the canonical ensemble, e.g. the ratio $\sigma_{\beta^*}(E)/E^*$ of the canonical standard deviation to the average value of the energy, vanish in the thermodynamic limit.
If this happens, then the canonical fluctuations of the total energy are negligible with respect to the total energy itself and, intuitively, the energy in the canonical ensemble can be though of as an effectively deterministic quantity, very much like in the conjugated microcanonical ensemble. 
Similarly, the extra dependencies among the microscopic constituents in the microcanonical ensemble are expected to play a smaller and smaller role, thus coming closer to the canonical case.

More recent approaches have considered different rigorous definitions of EE, which can be beautifully summarized~\cite{touchette2015equivalence} as the following three notions: \emph{thermodynamic} equivalence (convexity of the microcanonical entropy density), \emph{macrostate} equivalence (equality of the expected values of macroscopic quantities under the two ensembles), and \emph{measure} equivalence (vanishing of the relative entropy density between the microcanonical and canonical probability distributions). Under mild conditions, these notions have been shown to be equivalent~\cite{touchette2015equivalence}. In this paper, we use measure equivalence as it is more transparently related to the ensemble probabilities.

As a useful result, research on the relationship between statistical physics and combinatorics has revealed that the relative entropy between the microcanonical and the canonical probability distributions is, under certain conditions, asymptotically proportional to the logarithm of the determinant of the covariance matrix of the \emph{effective} constraints in the canonical ensemble~\cite{squartini2017reconnecting}. 
The effective constraints are those that are neither redundant, i.e. trivially replicating other constraints, nor degenerate, i.e. deterministically restricting the canonical and microcanonical configurations in exactly the same way. For instance, formally imposing `two' constraints where one is the total energy $E$ and the other one is twice the total energy $2E$ is clearly a redundant choice: the effective number of constraints is just one in this case. As another example, if in addition to the energy $E$ we impose its square value $E^2$, then for those values of the Lagrange multipliers such that $\langle E^2\rangle=\langle E\rangle^2$ the variance of the energy will be zero also in the canonical ensemble: $E$ will become degenerate and deterministically equal to its imposed value in both ensembles, so not contributing any difference between the two (by contrast, for parameter values such that $\langle E^2\rangle>\langle E\rangle^2$ there are no allowed configurations in the microcanonical ensemble because the hard values of $E^2$ and $E$ become conflicting, thereby breaking the equivalence with the canonical one). In general, if the problem is not ill-posed from the beginning, the number of effective constraints coincides with the number of enforced constraints. However, it may happen that some constraints become ineffective for certain degenerate values of the parameters. In any case, for a given parameter value the number of effective constraints coincides with the rank of the covariance matrix of all imposed constraints~\cite{squartini2017reconnecting}.

Since nonequivalence in the measure sense corresponds to the (super)extensivity of the relative entropy, studying the asymptotic behaviour of the determinant of the (effective) covariance matrix is enough in order to assess ensemble nonequivalence.
It is worth noticing that, if there is a single constraint (say, the total energy $E$), then the determinant of the covariance matrix coincides with the corresponding variance $\sigma^2_{\beta^*}(E)$ and the relative entropy grows asymptotically (under the necessary hypotheses) as $\ln \sigma^2_{\beta^*}(E)=2\ln \sigma_{\beta^*}(E)$. On the other hand, since $E$ is a global constraint, it is generally extensive in the number $n$ of units of the system. Therefore the vanishing of the relative fluctuations, i.e. $\sigma_{\beta^*}(E)=o(E)$, implies that the relative entropy is subextensive, i.e. the relative entropy density vanishes in the thermodynamic limit. This suggests that, \emph{in presence of a global constraint, the vanishing of the relative fluctuations implies ensemble equivalence}.\footnote{Note that the converse is not necessarily true. However, observing ensemble equivalence and non-vanishing relative fluctuations simultaneously requires some rather uncommon circumstances: for instance, if $\sigma_{\beta^*}(E)$ grows like $E^\alpha$ with $\alpha\ge 1$ while the entropy still grows like $E$, then the relative entropy is still subextensive, while the relative fluctuations do not vanish.}
How this picture changes in presence of an extensive number of local constraints has not been investigated yet. In particular, whether the vanishing of relative fluctuations still implies ensemble equivalence remains an open question.

\subsection{The contribution of this paper}
This paper connects to both lines of research described above (physical mechanisms and mathematical definitions for EN) and its aim is therefore twofold.
On the one hand, we aim at investigating for the first (to the best of our knowledge) time the phenomenon of EN in a model system that combines the presence of an extensive number of local constraints with a phase transition. 
On the other hand, we aim at understanding whether the intuitive criterion of vanishing relative fluctuations of the constraints still ensures EE in this more general setting.

Concretely, we consider the \emph{weighted configuration model}~\cite{squartini2017maximum}, namely a model of \emph{weighted} (as opposed to binary) networks with given \emph{strength sequence}, i.e. with given values of the \emph{strength} (sum of the weights of incident links) of each node.
The weighted character of the model allows for the emergence of a phase transition that is impossible to observe in the corresponding binary configuration model, namely Bose-Einstein Condensation (BEC)~\cite{park2004statistical,garlaschelli2009generalized}. 
For the sake of clarity, it is worth mentioning here that, although a form of BEC in networks was identified for the first time in growing binary graphs~\cite{bianconi2001bose}, the notion we refer to here refers to static networks and as such can only occur in weighted networks~\cite{park2004statistical}. Indeed, while the configuration model for weighted networks obeys Bose-Einstein statistics, the configuration model for binary networks obeys Fermi-Dirac statistics~\cite{park2004statistical,garlaschelli2009generalized,garlaschelli2013low}.
BEC can arise in our model by appropriately tuning the strength sequence.
In particular, we are going to show that we can make the strength sequence temperature-dependent and generate BEC by picking a sufficiently low temperature, below a certain critical value. The simplest such setting is one where the network has a `core-periphery' structure, with BEC appearing in the core.

We find that, for all temperatures and irrespective of whether BEC emerges, the canonical and microcanonical ensembles are always nonequivalent as signalled by a nonvanishing relative entropy density.
On the other hand, the relative fluctuations of all the constraints vanish when BEC is absent, while some of them do not vanish when BEC is present.
This shows that the relative fluctuations cannot distinguish between equivalence and nonequivalence of the ensembles in this more general case where multiple constraints are present. In fact, what they do is detecting the presence of BEC.
Therefore the traditional criterion for EE based on the vanishing of the relative fluctuations is no longer valid in presence of an extensive number of constraints, even when applied simultaneously to all constraints.
These results enrich our understanding of the phenomenology of EN and shed more light on its relationship with both the extensivity of the constraints and the presence of phase transitions.

The remainder of this paper is organized as follows. In Sec.~\ref{sec_ensembles} we rigorously define  the canonical and microcanonical ensembles of weighted networks with given strength sequence.
In Sec.~\ref{sec_EN} we introduce two criteria for the (non)equivalence of the ensembles, one based on the relative entropy between the corresponding probability distributions (measure equivalence) and one based on the relative fluctuations of the constraints. 
In Sec.~\ref{sec_BEC} we study in detail a model defined by the simplest family of strength sequences, driven by a temperature parameter, such that we can observe both a BEC and a non-BEC phase.
In Sec.~\ref{sec_conclusions} we offer our conclusions. Finally, the Appendix contains useful calculations needed to establish the scaling of the relative entropy in all the regimes considered.

\section{Canonical and microcanonical ensembles of weighted networks\label{sec_ensembles}}
In this section, we introduce the definition of weighted networks and of canonical and microcanonical ensembles of weighted networks with given strength sequence.

\begin{figure*}[t]
  \includegraphics[width=\textwidth]{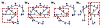}
\caption{Example of four weighted networks $\mathbf{W}_1, \mathbf{W}_2, \mathbf{W}_3, \mathbf{W}_4$ having the same number $n=8$ of nodes (labelled from $a$ to $h$) and the same strength sequence $\vec{s}(\mathbf{W}_i)=\vec{s}^*=(11,6,5,13,8,5,2,4)$ ($i=1,4$), but different structure ($\mathbf{W}_i\ne\mathbf{W}_j$ for all $i\ne j$).
The dashed blocks highlight the links from node $a$ to its neighbours: in different networks, node $a$ can have different neighbours and, importantly, different distributions of link weights (more or less concentrated on specific neighbours). 
More homogeneous choices of $\vec{s}^*$ would result in less concentrated link weights, while more heterogeneous choices of $\vec{s}^*$ would result in more concentrated link weights.
}
\label{weighted-strength-network}
\end{figure*}

\subsection{Weighted network ensembles}
Weighted networks are widely used to describe systems with a large number of components and heterogeneous patterns of interaction~\cite{newman2006structure}. 
We represent a possible configuration of a weighted network with $n$ nodes as an $n\times n$ weighted adjacency matrix $\mathbf{W}$. 
Each entry of the matrix $w_{ij}$ ($1\leq i\leq n, 1\le j\leq n)$ denotes the weight of the link between node $i$ and node $j$, which is taken from the set $\mathbb{N}$ of natural numbers (including zero, which corresponds to the absence of a link between $i$ and $j$). 
In this work, we only consider undirected networks without self-loops, thus the weighted matrix $\mathbf{W}$ is a symmetric matrix ($w_{ij}=w_{ji}$ for all $i,j$) and its diagonal elements are zero ($w_{ii}=0$ for all $i$). The number of independent entries of each such matrix is therefore ${n\choose 2}=n(n-1)/2$.

An ensemble of weighted networks on $n$ nodes is the discrete (infinite) set $\mathcal{W}_n=\mathbb{N}^{n(n-1)/2}$ of all available configurations for the matrix $\mathbf{W}$ and a probability distribution $P(\mathbf{W})$ over $\mathcal{W}_n$ that is specified by a given vector $\vec{C}(\mathbf{W})$ of constraints, which can be enforced either as a soft constraint (canonical ensemble) or as a hard constraint (microcanonical ensemble)~\cite{squartini2017maximum}. 
So the matrix $\mathbf{W}$ is a possible outcome of a random variable.
We will consider the weighted configuration model, for which the constraints are the strengths of all nodes, i.e. the strength sequence $\vec{C}(\mathbf{W})=\vec{s}(\mathbf{W})$, where the strength $s_i(\mathbf{W})$ of node $i$ is a local sum of all the link weights that connect $i$ to its neighbours in the particular network $\mathbf{W}$:
\begin{equation}
s_i(\mathbf{W})=\sum_{j=1}^{n}w_{ij},\qquad i=1,n.
\label{strength_node}
\end{equation}
Clearly, the number of scalar constraints is $n$, which coincides with the number of nodes, so this model is a perfect example of a system subject to an extensive number of local constraints.
In fact, it is the weighted counterpart of the binary configuration model, where EN driven by local constraints was observed for the first time~\cite{squartini2015breaking}.

A crucial consequence of the presence of local constraints in weighted networks is illustrated in Fig.\ref{weighted-strength-network}, where we show an example of different networks with the same strength sequence. 
For a given choice of $\vec{s}^*$, in different realizations of the network each node $i$ can have different neighbours and different distributions of weights on the links that connect it to those neighbours.
In particular, the strength $s_i^*$ can be more or less concentrated on specific neighbours (a property that is usually quantified by the so-called \emph{disparity}~\cite{serrano2006correlations}). 
However, more homogeneous choices of $\vec{s}^*$ unavoidably result in less concentrated link weights, while more heterogeneous choices of $\vec{s}^*$ impose more concentrated link weights.
This fact will allow us to consider (in Sec.~\ref{sec_BEC}) different structural regimes ranging between two extreme limits: a constant (infinite-temperature) strength sequence implying that on average each node is connected to its neighbours in an equally strong way, and a `step-like' (zero-temperature) strength sequence implying an extreme concentration of link weights among a small subset of the $n$ nodes. In between these two limits, a certain critical temperature separates a `non-condensed' (high-temperature) phase from a `condensed' (low-temperature) phase featuring the properties of BEC.

\subsection{Canonical weighted network ensemble\label{sec_cano}}
We first discuss how to implement the strength sequence constraint mathematically in the canonical ensemble (soft constraint)~\cite{squartini2017maximum}.
Recall that in the traditional canonical ensemble the inverse temperature $\beta^*$ is the only (scalar) parameter of the canonical probability distribution $P_{\textrm{can}}(\mathbf{W}|{\beta}^*)$, conjugate to a certain (scalar) total energy $E^*$ in the corresponding microcanonical ensemble.
Explicitly, $P_{\textrm{can}}(\mathbf{W}|{\beta}^*)$ is the Boltzmann distribution $P_{\textrm{can}}(\mathbf{W}|{\beta}^*)=e^{-\beta^* E(\mathbf{W})}/Z(\beta^*)$ with inverse temperature $\beta^*=1/kT^*$.
By contrast, in our setting the canonical distribution $P_{\textrm{can}}(\mathbf{W}|\vec{\beta}^*)$ has to depend on an $n$-dimensional vector $\vec{\beta}^*$ of parameters, conjugate to the $n$-dimensional constraint $\vec{s}^*$ which, in turn, defines the microcanonical ensemble.
This distribution is found by maximizing the Gibbs-Shannon entropy functional defined in Eq.~\eqref{eq_shannon} under the soft constraint
\begin{equation}
\langle \vec{s}\rangle_{\vec{\beta}^*}=\sum_{\mathbf{W}\in\mathcal{W}_n}P_{\textrm{can}}(\mathbf{W}|\vec{\beta}^*)\vec{s}(\mathbf{W})\equiv\vec{s}^*
\label{eq_soft}
\end{equation}
which generalizes the conjugacy condition in Eq.~\eqref{eq_conjugate}.
The solution to the maximization problem sees $\vec{\beta}^*$ play the role of a vector of Lagrange multipliers coupled to the strength sequence $\vec{s}^*$, and is given by~\cite{squartini2017maximum}
\begin{equation}\label{Can_probability}
P_{\textrm{can}}(\mathbf{W}|\vec{\beta}^*)=\frac{e^{-H(\mathbf{W},\vec{\beta}^*)}}{Z(\vec{\beta}^*)},
\end{equation}
where the (network) \emph{Hamiltonian} $H(\mathbf{W},\vec{\beta})=\vec{\beta}\cdot\vec{s}(\mathbf{W})$ is the linear combination of the node strengths, the \emph{partition function} $Z(\vec{\beta})=\sum_{\mathbf{W}\in\mathcal{W}_n}e^{-H(\mathbf{W},\vec{\beta})}$ is a normalization constant, and $\vec{\beta}^*$ is the unique parameter value realizing Eq.~\eqref{eq_soft}.
Note that Eq.~\eqref{Can_probability} has still the form of the Boltzmann distribution, with the important caution that \emph{the inverse temperature has been reabsorbed into the Hamiltonian.} 
Therefore, to keep the parallel with the traditional physical situation, in our setting the Hamiltonian should be thought of as the inverse temperature times the energy, and $\vec{\beta}^*$ as the inverse temperature times a vector of `fields', each coupled to a different constraint.
Clearly, since the probability in Eq.~\eqref{Can_probability} must be dimensionless, the product $\vec{\beta}\cdot\vec{s}(\mathbf{W})$ must be dimensionless as well.
In Sec.~\ref{sec_BEC}, we will notice that the Hamiltonian can be further reinterpreted as also incorporating a `chemical potential' governing the expected weight of the links in the network~\cite{garlaschelli2013low}.

Notably, $P_{\textrm{can}}(\mathbf{W}|\vec{\beta}^*)$ depends on $\mathbf{W}$ only through $\vec{s}(\mathbf{W})$. In particular, it gives the same value $P_{\textrm{can}}(\mathbf{W}^*|\vec{\beta}^*)$ to any network $\mathbf{W}^*$ such that $\vec{s}(\mathbf{W}^*)=\vec{s}^*$.
Explicitly, given the definition of node strength in Eq.~\eqref{strength_node}, the network Hamiltonian can be written for a generic value of $\vec{\beta}$ as
\begin{equation}\label{Hamiltonian}
H(\mathbf{W},\vec{\beta})=\sum_{i=1}^{n}\sum_{i<j}(\beta_i+\beta_j)w_{ij}.
\end{equation}
The partition function can be easily shown~\cite{squartini2017maximum} to be
\begin{equation}\label{partition-function dewd}
Z(\vec{\beta})=\prod_{i=1}^{n}\prod_{i<j}\frac{1}{1-e^{-(\beta_i+\beta_j)}}
\end{equation}
provided that $\beta_i+\beta_j>0$ for all $i,j$ (otherwise, the model admits no solution).
The canonical probability distribution therefore factorizes over pairs of nodes as
\begin{equation}\label{can-dewd-1}
P_{\textrm{can}}(\mathbf{W}|\vec{\beta})=
\prod_{i=1}^{n}\prod_{i<j}q_{ij}(w_{ij}|\vec{\beta}),
\end{equation}
where
\begin{equation}\label{probability0wij}
  q_{ij}(w|\vec{\beta})=\frac{e^{-(\beta_i+\beta_j)w}}{[1-e^{-(\beta_i+\beta_j)}]^{-1}}
\end{equation} 
is the probability that the weight of the link between nodes $i$ and $j$ takes the particular value $w$. Therefore different pairs of nodes are statistically independent in the canonical ensemble (while they are not in the microcanonical one).

Note that $q_{ij}(w|\vec{\beta})$ is a geometric distribution~\cite{squartini2015unbiased,squartini2017maximum} with expected value
\begin{eqnarray}
\langle w_{ij} \rangle_{\vec{\beta}}&=& 
\sum_{w\in\mathbb{N}}w\,q_{ij}(w|\vec{\beta})\nonumber\\
&=&\sum_{w\in\mathbb{N}}w\,\frac{e^{-(\beta_i+\beta_j)w}}{[1-e^{-(\beta_i+\beta_j)}]^{-1}}\nonumber\\
&=&\frac{e^{-(\beta_i+\beta_j)}}{1-e^{-(\beta_i+\beta_j)}}
\label{average-edges-1-1}
\end{eqnarray}
(representing the expected weight of the link connecting nodes $i$ and $j$) and variance
\begin{eqnarray}
\textrm{Var}_{\vec{\beta}}(w_{ij})&=&\frac{e^{-(\beta_i+\beta_j)}}{[1-e^{-(\beta_i+\beta_j)}]^2}\nonumber\\
&=&\langle w_{ij}\rangle_{\vec{\beta}}(1+\langle w_{ij}\rangle_{\vec{\beta}}).
\end{eqnarray}
As we will discuss in detail is Sec.~\ref{sec_BEC}, Eq.~\eqref{average-edges-1-1} has the form of Bose-Einstein statistics, where $\langle w_{ij} \rangle_{\vec{\beta}}$ plays the role of an expected occupation number for the state labeled by nodes $i$ and $j$. 
In an appropriate `low-temperature' regime, BEC can emerge into the model through the divergence of the occupation number $\langle w_{ij} \rangle_{\vec{\beta}}$ for one (possibly degenerate) `ground state' (corresponding to $\beta_i+\beta_j\to 0^+$), while the occupation number for the other states remains finite~\cite{park2004statistical,squartini2017maximum,garlaschelli2009generalized}. 
This discussion requires a series of considerations that we leave for later.
For the moment, we notice that Eq.~\eqref{average-edges-1-1} allows us to determine the special value ${\vec{\beta}}^*$ corresponding to the given strength sequence ${\vec{s}}^*$.
Summing over all nodes $j\ne i$, the average value of the strength of node $i$ is 
\begin{equation}\label{average-strength-1}
  \langle s_{i} \rangle_{\vec{\beta}} =\sum_{j\ne i}\langle w_{ij} \rangle_{\vec{\beta}}=\sum_{j\neq i}\frac{e^{-(\beta_i+\beta_j)}}{1-e^{-(\beta_i+\beta_j)}},
\end{equation}
whence we can reformulate Eq.~\eqref{eq_soft} as
\begin{equation}\label{dewd-strength-1}
\sum_{j\neq i}\langle w_{ij}\rangle_{\vec{\beta}^*}=\sum_{j\neq i}\frac{e^{-(\beta^*_i+\beta^*_j)}}{1-e^{-(\beta^*_i+\beta^*_j)}}\equiv s^*_i\quad i=1,n
\end{equation}
which fixes the unique parameter value $\vec{\beta}^*$. Notably, this value is also the one that maximizes the \mbox{(log-)likelihood}~\cite{garlaschelli2008maximum,squartini2017maximum}, i.e.
\begin{equation}
\vec{\beta}^*=\textrm{argmax}_{\vec{\beta}}\ln P_{\textrm{can}}(\mathbf{W}^*|\vec{\beta}),
\label{eq_argmax}
\end{equation}
where, again, $\mathbf{W}^*$ is any configuration such that $\vec{s}(\mathbf{W}^*)=\vec{s}^*$.
In general, it is not possible to write $\vec{\beta}^*$ explicitly as a function of $\vec{s}^*$.
However, Eq.~\eqref{dewd-strength-1} or equivalently Eq.~\eqref{eq_argmax} can be efficiently solved numerically~\cite{squartini2015unbiased,squartini2017maximum} using various algorithms that have been coded for this purpose~\cite{maxsam,meh}.
In any case, a general property (that we will use later) is that for any two nodes $i$ and $j$ with the same expected strength ($s^*_i=s^*_j$), the corresponding parameters $\beta^*_i$ and $\beta^*_j$ obey the same equation in~\eqref{dewd-strength-1} and are therefore equal. In other words, $s^*_i=s^*_j$ implies $\beta^*_i=\beta^*_j$. 

Once $\vec{\beta}^*$ is calculated, we can plug it back into Eq.~\eqref{can-dewd-1} to finally obtain the probability distribution
\begin{equation}
  P_{\textrm{can}}(\mathbf{W}|\vec{\beta}^*)=\prod_{i=1}^{n}\prod_{i<j}\frac{e^{-(\beta^*_i+\beta^*_j)w_{ij}}}{[1-e^{-(\beta^*_i+\beta^*_j)}]^{-1}}
\label{eq_Pcan}
\end{equation}
that characterizes the canonical ensemble entirely.
For practical purposes, $\vec{\beta}^*$ can be inserted into Eq.~\eqref{probability0wij} to obtain the link weight probability $q_{ij}(w|\vec{\beta}^*)$, from which several expected network properties can be calculated very directly.
For instance, besides the expected link weight $\langle w_{ij}\rangle_{\vec{\beta}^*}$, we can calculate the probability that nodes $i$ and $j$ are connected by a link, irrepective of the weight of the latter, as follows:
\begin{eqnarray}
p^*_{ij}&\equiv& 
\langle \Theta(w_{ij}) \rangle_{\vec{\beta}^*}\nonumber\\
&=&\sum_{w=1}^{\infty}w\,q_{ij}(w|\vec{\beta}^*)\nonumber\\
&=&1-q_{ij}(0|\vec{\beta}^*)\nonumber\\
&=&e^{-(\beta^*_i+\beta^*_j)}\nonumber\\
&=&\frac{\langle w_{ij}\rangle_{\vec{\beta}^*}}{1+\langle w_{ij}\rangle_{\vec{\beta}^*}},
\label{pij}
\end{eqnarray}
where $\Theta(x)$ denotes the Heaviside step function, defined as $\Theta(x)=1$ if $x>0$ and  $\Theta(x)=0$ if $x\le 0$.
Note that, if $i$ and $j$ belong to the condensed state where the expected link weight $\langle w_{ij}\rangle_{\vec{\beta}^*}$ diverges ($\beta^*_i+\beta^*_j\to 0^+$), then they become deterministically connected, i.e. $p^*_{ij}\to 1^-$. By contrast, non-condensed states have $\langle w_{ij}\rangle_{\vec{\beta}^*}<\infty$, $\beta^*_i+\beta^*_j>0$, and $p^*_{ij}<1$. The analogy with BEC will be discussed in much more detail in Sec.~\ref{sec_BEC}.

Besides the structural properties, one of the key quantities that we will need in order to determine EE (or the lack thereof) is the resulting canonical entropy $S^*_{\textrm{can}}$, obtained by inserting Eq.~\eqref{eq_Pcan} into Eq.~\eqref{eq_shannon}: 
\begin{eqnarray}
 S^*_{\textrm{can}}&\equiv &S[P_{\textrm{can}}(\mathbf{W}|\vec{\beta}^*)]\nonumber\\
&=&\langle H(\mathbf{W},\vec{\beta}^*) \rangle+\ln Z(\vec{\beta}^*)\nonumber\\
 &=&\vec{\beta}^*\cdot\langle \vec{s}(\mathbf{W})\rangle+\ln Z(\vec{\beta}^*)\nonumber\\
 &=&\vec{\beta}^*\cdot\vec{s}^*+\ln Z(\vec{\beta}^*)\nonumber\\
&=&-\ln P_{\textrm{can}}(\mathbf{W}^*|\vec{\beta}^*).\label{can-entropy-G}
\end{eqnarray}
Note that the calculation of the canonical entropy $S^*_{\textrm{can}}$ of the entire weighted network ensemble only requires the knowledge of the probability of one generic network $\mathbf{W}^*$ with strength sequence $\vec{s}^*$, which is in turn directly calculated through Eq.~\eqref{eq_Pcan}.

\subsection{Microcanonical weighted network ensemble\label{sec:micro}}
We now come to the microcanonical ensemble.
Its governing probability distribution $P_{\textrm{mic}}(\mathbf{W}|\vec{s}^*)$ can be obtained by maximizing the Gibbs-Shannon entropy functional in Eq.~\eqref{eq_shannon} under the hard constraint
\begin{equation}
\vec{s}(\mathbf{W})=\vec{s}^*
\label{eq_hard}
\end{equation}
that applies to each network $\mathbf{W}$ realized (with positive probability) in the set $\mathcal{W}_n$.
The solution is obviously the uniform probability distribution
\begin{equation}\label{Mic_probability_distribution}
P_{\textrm{mic}}(\mathbf{W}|\vec{s}^*)=
\left\{\begin{array}{cc}
\Omega_{\vec{s}^*}^{-1} & \vec{s}(\mathbf{W})=\vec{s}^* \\
0 & \vec{s}(\mathbf{W})\neq\vec{s}^*
\end{array}
\right.
\end{equation}
where $\Omega_{\vec{s}^*}$ is the number of networks for which the hard constraint in Eq.~\eqref{eq_hard} is realized. An implicit assumption throughout this paper is that the particular strength sequence $\vec{s}^*$ is \emph{graphic}, i.e. it can be realized by at least one network, so that $\Omega_{\vec{s}^*}>0$.
In this case as well, the (microcanonical) entropy is obtained by inserting Eq.~\eqref{Mic_probability_distribution} into Eq.~\eqref{eq_shannon}:
\begin{eqnarray}
S^*_{\textrm{mic}}&\equiv&S[P_{\textrm{mic}}(\mathbf{W}|\vec{s}^*)]\nonumber\\
&=&\ln\Omega_{\vec{s}^*}\nonumber\\
&=&-\ln P_{\textrm{mic}}(\mathbf{W}^*|\vec{s}^*),\label{Microcanonica-entropy}
\end{eqnarray}
which is also known as Boltzmann entropy.
Note that $\mathbf{W}^*$ has the same meaning as in Eq.~\eqref{can-entropy-G}, therefore both the canonical and microcanonical entropies are equal to minus the log of the corresponding probability, evaluated in any state $\mathbf{W}^*$ realizing the hard constraint in Eq.~\eqref{eq_hard}.

Note that, although the derivation of $P_{\textrm{mic}}(\mathbf{W}|\vec{s}^*)$ is formally much more direct than that of $P_{\textrm{can}}(\mathbf{W}|\vec{\beta}^*)$ in the conjugate canonical ensemble, its explicit calculation is more challenging, as it requires the combinatorial enumeration of all the $\Omega_{\vec{s}^*}$ weighted networks with strength sequence $\vec{s}^*$ (as a side remark, it is precisely the local nature of $\vec{s}^*$ that makes the calculation of $\Omega_{\vec{s}^*}$ daunting).
Here, we will employ a recently proposed saddle-point asymptotic formula, for a generic discrete system under a $K$-dimensional vector $\vec{C}^*$ of \emph{effective} (see Sec.~\ref{sec:math}) constraints, for the number $\Omega_{\vec{C}^* }$ of microcanonical configurations~\cite{squartini2017reconnecting}. 
The formula uses only conjugate canonical quantities, namely the canonical entropy $S^*_{\textrm{can}}$ and the $K\times K$ covariance matrix $\mathbf{\Sigma}^*$ among the $K$ constraints in the  canonical ensemble, and reads~\cite{squartini2017reconnecting}
\begin{equation}\label{Prob-micr-st-1}
\Omega_{\vec{C}^* }={\frac{e^{S^*_{\textrm{can}}}}{\sqrt{\det(2\pi\mathbf{\Sigma^*})}}\prod_{k=1}^{K}[1+O(1/\lambda^*_k)]},
\end{equation}
where $\{\lambda^*_k\}_{k=1}^K$ are the eigenvalues of $\mathbf{\Sigma}^*$.
The symbol $O(x)$ indicates a quantity with a finite limit when divided by $x$ as $n\to\infty$, i.e. $O(x)$ is asymptotically of the same order as $x$. 
Note that, since covariance matrices are positive semidefinite, $\lambda^*_k\ge 0$ for all $k$. Moreover, since the constraints are assumed to be non-redundant, then $\lambda^*_k> 0$ for all $k$~\cite{squartini2017reconnecting} (if some of the constraints were redundant, there would be certain zero eigenvalues rendering the above equation inapplicable; that is why the formula should be applied to a maximal set of $K$ non-redundant constraints).
Finally, if these eigenvalues grow sufficiently fast as $n\to\infty$, then the product on the right hand side becomes irrelevant, in which case the knowledge of $S^*_{\textrm{can}}$ and $\det(2\pi\mathbf{\Sigma^*})$ is enough in order to characterize the asymptotics of $\Omega_{\vec{C}^*}$.

In our setting where $\vec{C}^*=\vec{s}^*$ and $K=n$ (node strengths are all mutually independent as it is not possible to guess any individual node strength from the knowledge of the other $n-1$ ones), we calculate the entries of $\mathbf{\Sigma^*}$ as
\begin{eqnarray}
{\Sigma}^*_{ij}&\equiv&\textrm{Cov}_{\vec{\beta}^*}(s_i,s_j)\nonumber\\
&=&\langle s_i s_j\rangle_{\vec{\beta}^*}-\langle s_i\rangle_{\vec{\beta}^*}\langle s_j\rangle_{\vec{\beta}^*}\nonumber\\
&=&\left.\frac{\partial^2\ln Z(\vec{\beta})}{\partial{\beta_i}\partial{\beta_j}}\right|_{\vec{\beta}=\vec{\beta}^*}\label{Sigma_unit}
\end{eqnarray}
where $Z(\vec{\beta})$ is given by Eq.~\eqref{partition-function dewd}.
An explicit calculation gives
\begin{eqnarray}
{\Sigma}^*_{ii}&=&\textrm{Var}_{\vec{\beta}^*}(s_i)\nonumber\\
&=&\sum_{j\neq i}\frac{e^{-(\beta^*_i+\beta ^*_j)}}{[1-e^{-(\beta^*_i+\beta ^*_j)}]^2}\nonumber\\
&=&\sum_{j\neq i}\langle w_{ij}\rangle_{\vec{\beta}^*}(1+\langle w_{ij}\rangle_{\vec{\beta}^*})\label{covariance-matrix-dewd-diagonal}
\end{eqnarray}
for the diagonal entries (i.e. the variances of the constraints) and
\begin{eqnarray}
 {\Sigma}^*_{ij}&=&\textrm{Cov}_{\vec{\beta}^*}(s_i,s_j)\nonumber\\
&=&\textrm{Var}_{\vec{\beta}^*}(w_{ij})\nonumber\\
&=&\frac{e^{-(\beta^*_i+\beta^*_j)}}{[1-e^{-(\beta^*_i+\beta^*_j)}]^2}\nonumber\\
&=&\langle w_{ij}\rangle_{\vec{\beta}^*}(1+\langle w_{ij}\rangle_{\vec{\beta}^*})\quad(i\neq j)\label{non-diagonal-matrix-dewd}
\end{eqnarray}
for the off-diagonal entries (i.e. the covariances between distinct constraints).

We finally obtain
\begin{eqnarray}
S^*_{\textrm{mic}}&=&\ln\Omega_{\vec{s}^*}\label{Mic-shannon-entropy-1}\\
&=&S^*_{\textrm{can}}-\ln \sqrt{\det(2\pi \mathbf{\Sigma}^*)}+\sum_{k=1}^{n}\ln[1+O(1/\lambda^*_k)]\nonumber
\end{eqnarray}
where we have used $\sqrt{\det(2\pi \mathbf{\Sigma}^*)}=\prod_{k=1}^n\sqrt{2\pi\lambda^*_k}$.
Note that the eigenvalues $\{\lambda^*_k\}_{k=1}^n$ are positive, the $n$ node strengths being linearly independent constraints~\cite{squartini2017reconnecting}.
In principle, in order to compute (the leading term of) Eq.~\eqref{Mic-shannon-entropy-1} explicitly, we need to specify a value for $\vec{s}^*$, calculate the resulting matrix  $\mathbf{\Sigma^*}$, and the eigenvalues of the latter.
However, in Sec.~\ref{sec_EN} we show that the knowledge of the diagonal elements of the covariance matrix is enough for our purposes. 
This result is then used in Sec.~\ref{sec_BEC} when we consider specific choices of $\vec{s}^*$.
\begin{figure*}[tbp]
  \includegraphics[width=\textwidth]{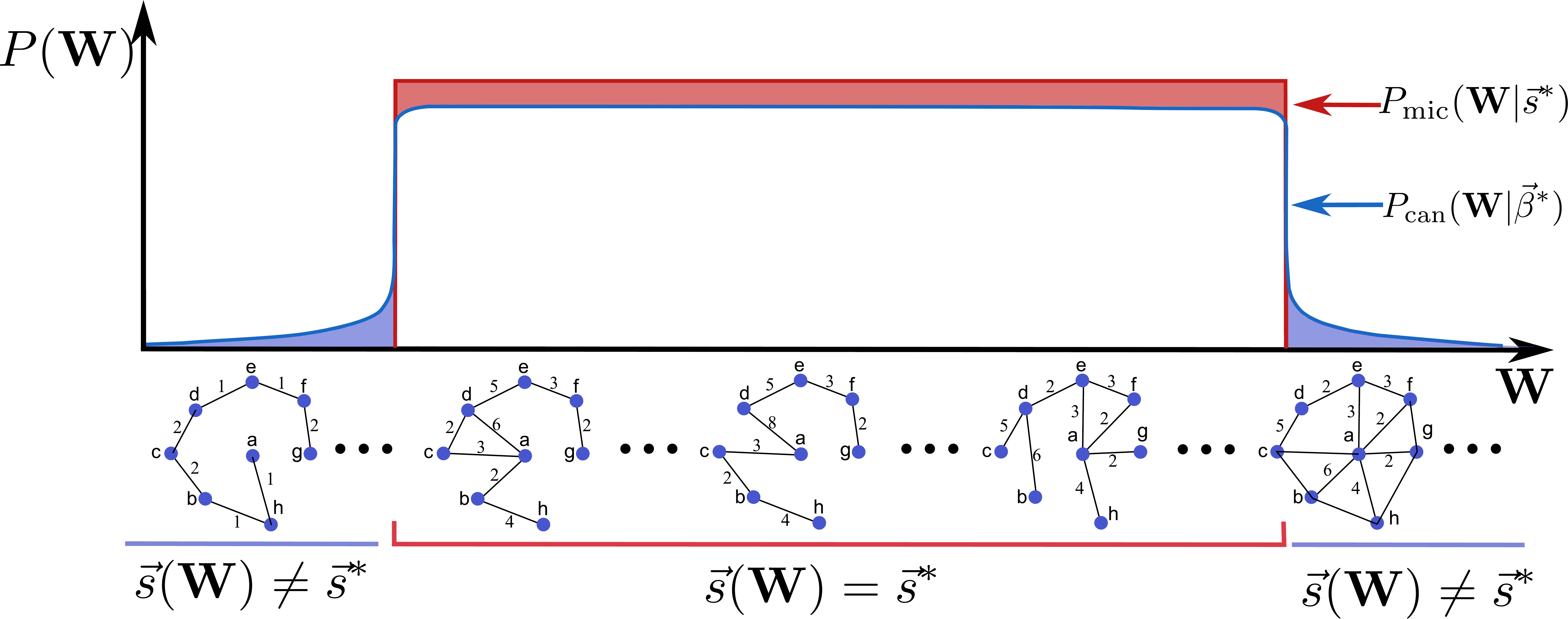}
  \caption{Illustration of the nonequivalence of ensembles of weighted networks with given strength sequence $\vec{s}^*$ in the measure sense. 
Schematically, the $x$ axis represents all weighted networks $\mathbf{W}\in\mathcal{W}_n$. Here, $n=8$ and $\vec{s}^*=(11,6,5,13,8,5,2,4)$. 
The $\Omega_{\vec{s}^*}$ networks matching the particular strength sequence $\vec{s}^*$, i.e. those for which $\vec{s}(\mathbf{W})=\vec{s}^*$, are represented in the middle. 
The $y$ axis represents the canonical and microcanonical probabilities for each network. 
The microcanonical distribution $P_{\textrm{mic}}(\mathbf{W}|\vec{s}^*)$ assigns zero probability to the networks for which $\vec{s}(\mathbf{W})\neq\vec{s}^*$, and uniform probability $P_{\textrm{mic}}(\mathbf{W}^*|\vec{s}^*)=\Omega_{\vec{s}^*}^{-1}$ to each network $\mathbf{W}^*$ for which $\vec{s}(\mathbf{W}^*)=\vec{s}^*$.
By contrast, the conjugate canonical distribution $P_{\textrm{can}}(\mathbf{W}|\vec{\beta}^*)$ assigns positive probability to all the networks in $\mathcal{W}_n$ and therefore has `tails' that extend all over the $x$ axis. 
Normalization implies that, while also the canonical probability gives a constant value $P_{\textrm{can}}(\mathbf{W}^*|\vec{\beta}^*)$ to each network $\mathbf{W}^*$ with the strength sequence $\vec{s}^*$, this value is (for $n$ finite) smaller (lower plateau) than the corresponding microcanonical one (upper plateau): $P_{\textrm{can}}(\mathbf{W}^*|\vec{\beta}^*)<P_{\textrm{mic}}(\mathbf{W}^*|\vec{s}^*)$.
Indeed, the blue and red areas should be equal because of normalization.
Intuitively, the two ensembles become equivalent in the thermodynamic limit if, sufficiently fast as $n\to\infty$, the canonical tails vanish (the blue areas disappear) and the canonical plateau `catches up' with the microcanonical one (the red area disappears). 
Measure equivalence formalizes this `sufficiently fast' rigorously, finding that EE corresponds to the condition $\lim_{n\to\infty}[\ln P_{\textrm{mic}}(\mathbf{W}^*|\vec{s}^*)-\ln P_{\textrm{can}}(\mathbf{W}^*|\vec{\beta}^*)]/n=0$.
It turns out that this is not the case for the model discussed here: canonical and microcanonical ensembles of weighted networks with given strength sequence are not equivalent.
}
  \label{EN_distribution}
\end{figure*}

\section{Equivalence and nonequivalence of weighted network ensembles\label{sec_EN}}
In this section we use the knowledge of the canonical and microcanonical probability distributions derived in the previous section in order to establish two criteria for the equivalence of ensembles of weighted networks, namely the one based on the vanishing of the relative entropy density between the two distributions~\cite{touchette2015equivalence} and the traditional one based on the vanishing of the canonical relative fluctuations of the constraints~\cite{gibbs1902elementary}. 

\subsection{Relative entropy density}
As we have anticipated, EE can be stated mathematically using three different notions, namely \emph{thermodynamic}, \emph{macrostate} and \emph{measure} equivalence~\cite{touchette2015equivalence}. 
These definitions turn out to be, under mild assumptions, essentially equivalent~\cite{touchette2015equivalence}. 
Here, we use the definition in the measure sense, which has been recently formulated explicitly for binary network ensembles~\cite{squartini2015breaking,garlaschelli2016ensemble,roccaverde2019breaking,squartini2017reconnecting} and is based on the vanishing of a suitable \emph{relative entropy density} between the microcanonical and canonical probability distributions.
Our calculations generalize those results to the case of weighted networks, for which measure equivalence has not been studied yet.

In general, the relative entropy (or Kullback-Leibler divergence) between two distributions $P$ and $Q$, both having support over a discrete set $\mathcal{W}_n=\{\mathbf{W}\}$ of configurations in analogy with Eq.~\eqref{eq_shannon}, is defined as
\begin{equation}\label{eq_Kull}
D[P||Q]\equiv\sum_{\mathbf{W}\in\mathcal{W}_n}P(\mathbf{W})\ln \frac{P(\mathbf{W})}{Q(\mathbf{W})}
\end{equation}
and quantifies `how far' the distribution $P$ is from the reference distribution $Q$~\cite{kullback1951information}.
When $P$ and $Q$ represent the microcanonical and canonical distributions respectively, it can be shown~\cite{squartini2015breaking,squartini2017reconnecting} that $D[P||Q]$ reduces to the difference between the canonical and microcanonical entropy, which can be both estimated on a single configuration realizing the hard constraint (as we have indeed shown in the previous section for our weighted network model). Moreover, its calculation asymptotically requires only the canonical covariance matrix $\mathbf{\Sigma^*}$ between the constraints.

We are now going to see how these general results apply to our specific case.
A visual illustration of the idea behind measure equivalence for our ensembles of weighted networks with given strength sequence is provided in Fig.~\ref{EN_distribution}.
Following Eq.~\eqref{eq_Kull}, the relative entropy between $P_{\textrm{mic}}(\mathbf{W}|\vec{s}^*)$ and $P_{\textrm{can}}(\mathbf{W}|\vec{\beta}^*)$ is defined as
\begin{eqnarray}D^*&\equiv& D[P_{\textrm{mic}}(\mathbf{W}|\vec{s}^*)||P_{\textrm{can}}(\mathbf{W}|\vec{\beta}^*)]\nonumber\\
&=&
\sum_{\mathbf{W}\in\mathcal{W}_n}P_{\textrm{mic}}(\mathbf{W}|\vec{s}^*)\ln\frac{P_{\textrm{mic}}(\mathbf{W}|\vec{s}^*)}{P_{\textrm{can}}(\mathbf{W}|\vec{\beta}^*)}.\label{Relative-entropy}
\end{eqnarray}
By inserting Eq.~\eqref{Mic_probability_distribution} into Eq.~\eqref{Relative-entropy} and using the fact that $P_{\textrm{can}}(\mathbf{W}|\vec{\beta}^*)$ has the same value for any network $\mathbf{W}^*$ matching the hard constraint $\vec{s}(\mathbf{W}^*)=\vec{s}^*$, we confirm that $D^*$ can be estimated pointwise on $\mathbf{W}^*$ as
\begin{equation}\label{Relative-entropy-simplified}
D^*=
\ln P_{\textrm{mic}}(\mathbf{W}^*|\vec{s}^*)-\ln P_{\textrm{can}}(\mathbf{W}^*|\vec{\beta}^*).
\end{equation}
Moreover, using Eqs.~\eqref{can-entropy-G} and~\eqref{Microcanonica-entropy}, we also confirm that it reduces to the entropy difference
\begin{equation}
D^*=S^*_{\textrm{can}}-S^*_{\textrm{mic}}.
\label{stocazzo}
\end{equation}
Now, Eq.~\eqref{Mic-shannon-entropy-1} immediately allows us to obtain
\begin{equation}
D^*=\ln \sqrt{\det(2\pi \mathbf{\Sigma}^*)}+\sum_{k=1}^{n}\ln[1+O(1/\lambda^*_k)]
\label{eq_KL}
\end{equation}
which depends only on the eigenvalues of the canonical covariance matrix $\mathbf{\Sigma^*}$, whose diagonal and off-diagonal entries are given in Eqs.~\eqref{covariance-matrix-dewd-diagonal} and~\eqref{non-diagonal-matrix-dewd} respectively.

The definition of measure equivalence is the vanishing of the relative entropy \emph{density}, i.e. of the ratio $D^*/n$, in the thermodynamic limit~\cite{touchette2015equivalence}.
Explicitly, EE in the measure sense corresponds to 
\begin{equation}\label{specific-relative-entropy}
d^*\equiv\lim_{n\rightarrow\infty}\frac{D^*}{n}=0
\end{equation}
or equivalently~\cite{squartini2017reconnecting}
\begin{equation}\label{equaltiv-1}
D^*=o(n),
\end{equation}
where $o(n)$ denotes a quantity that, if divided by $n$, vanishes as $n\to\infty$.
Equation~\eqref{stocazzo} allows us to understand the above definition of macrostate EE as follows. The microcanonical entropy $S^*_{\textrm{mic}}$ is the logarithm of the number of accessible configurations under hard constraints, while the canonical entropy $S^*_{\textrm{can}}$ is the logarithm of the corresponding `effective' number of configurations under soft constraints. EE requires that, as $n$ increases, the typical configurations of the system under the two ensembles become the same, i.e. that the (effective) numbers of configurations in the two ensembles become closer to each other. This cannot happen if, as we keep adding one more unit to the system, the difference between the two entropies (i.e. the relative entropy $D^*$) keeps increasing by an arbitrary amount. The criterion in Eq.~\eqref{specific-relative-entropy} establishes that if the entropy difference per node, i.e. $D^*/n$, does not vanish as $n$ diverges, then the two ensembles cannot be equivalent.

Equation~\eqref{equaltiv-1} implies that, in order to assess whether the system is under EE, we do not need the exact value of $D^*$, but only its leading order with respect to $n$.
Then from Eq.~\eqref{eq_KL} we see that, since the term $1+O(1/\lambda^*_k)$ is at most of the same order as $\sqrt{2\pi\lambda^*_k}$, the presence of $O(1/\lambda^*_k)$ does not affect ensemble (non)equivalence:
\begin{equation}\label{eq:Sorder}
D^*=O\left(\sum_{k=1}^{n}\ln\lambda^*_k\right)=O(\ln\det \mathbf{\Sigma}^*).
\end{equation}
So, in order to check whether Eq.~\eqref{equaltiv-1} holds, it is ultimately enough to check whether
\begin{equation}
\ln\det \mathbf{\Sigma}^*=o(n).
\label{eq:ultimate}
\end{equation}
On the other hand, since our hypothesis of non-redundant constraints implies $\lambda^*_k> 0$ for all $k$ (as discussed in Sec.~\ref{sec:micro}), we see that the contribution of the term $\sum_{k=1}^{n}\ln[1+O(1/\lambda^*_k)]$ to the relative entropy in Eq.~\eqref{eq_KL} is at most $O(n)$. So if $\ln \sqrt{\det(2\pi \mathbf{\Sigma}^*)}$ grows faster than $O(n)$ then Eq.~\eqref{eq_KL} reduces to the stronger result
\begin{equation}
D^*\approx\ln \sqrt{\det(2\pi \mathbf{\Sigma}^*)}\approx\frac{1}{2}\ln \det\mathbf{\Sigma}^*,
\label{eq_KLleading}
\end{equation}
i.e. the leading term of the relative entropy  (not only its leading order) can be calculated exactly from $\ln \det\mathbf{\Sigma}^*$ (throughout this paper, the symbol ``$x\approx y$'' indicates that the leading term of $x$ and $y$ is asymptotically the same, i.e. the two quantities differ by a quantity that vanishes if divided by either $x$ or $y$ as $n\to\infty$).
Depending on the regimes considered later on in the paper, different techniques for calculating (the leading order of) the determinant of the covariance matrix $\mathbf{\Sigma}^*$ can be used. We will discuss these techniques when needed, and refer to the Appendix for explicit calculations. We will show that, except in a certain zero-temperature limit, the conditions ensuring Eq.~\eqref{eq_KLleading} are met and the leading order of the relative entropy can be calculated exactly.

\subsection{Relative fluctuations of the constraints}
We now consider the relative fluctuations of the constraints, whose behaviour in the thermodynamic limit is, historically, the traditional criterion used to check whether statistical ensembles are equivalent~\cite{gibbs1902elementary}. 
In the standard situation, where the canonical and microcanonical ensembles are defined through a single scalar constraint on the total energy $E$, the relative fluctuations are captured by a single scalar quantity $r^*\equiv \sigma_{\beta^*}(E)/{E^*}$ representing the ratio of the canonical standard deviation of the energy to the mean energy itself. 
EE is then associated with the vanishing of $r^*$ as $n\to\infty$.
In statistics, $r^*$ is called the \emph{coefficient of variation} of the random variable $E$ with expected value $E^*$ and variance $\sigma^2_{\beta^*}(E)$
 
In the case of networks with local constraints, there are $n$ coefficients of variation to consider. 
They have been calculated for both the binary and the weighted versions of the configuration model~\cite{squartini2015unbiased}.
In extreme summary, those results show that, in the binary case (where there is a constraint on the degree $k^*_i$ for each node $i=1,n$), an \emph{upper bound} $1/\sqrt{k^*_i}$ for the relative fluctuation $r_i^*=\sigma_{\beta^*}(k_i)/{k^*_i}$ can been established. 
By contrast, in the weighted case (corresponding to the model considered here with a constraint on the strength $s^*_i$ for each node $i=1,n$), the value $1/\sqrt{s^*_i}$ becomes a \emph{lower bound} for the relative fluctuation $r_i^*=\sigma_{\beta^*}(s_i)/{s^*_i}$~\cite{squartini2015unbiased}.
Those results have two implications.

First, in the binary case the only regime for which general conclusions can be drawn about the vanishing of the relative fluctuations is the so-called \emph{dense} regime where the expected degree of all nodes diverges, hence $r_i^*\to 0$. In the opposite \emph{sparse} regime where the average degree of all nodes is finite, we only have a finite upper bound for the relative fluctuations, but their actual value depends on the specific network. In general, however, the decreasing behaviour of the upper bound for the relative fluctuations in binary networks with increasing degrees is opposite to that of the relative entropy density, which increases as the expected degree increases~\cite{squartini2015breaking,garlaschelli2016ensemble,roccaverde2019breaking}.

Second, in the weighted case we have a somewhat opposite situation where we can only conclude that, in the sparse regime where the expected strengths are finite (apart from possible hubs), the relative fluctuations do not vanish. 
By contrast, in the dense case where the expected strengths diverge, the relative fluctuations can in principle pick any value. 
Confusingly, even if in the weighted case the lower bound for the relative fluctuations of the strengths goes to zero for nodes with diverging strength $s_i^*$, previous results seem to indicate that the realized value of $r_i^*$ in networks with heterogeneous strength sequence actually \emph{increases} for higher $s_i^*$~\cite{squartini2015unbiased}. This behaviour suggests that weighted networks, due to the many possible ways in which weight can accumulate on links, can behave very differently from binary networks. This observation requires further research and strengthens our motivation for studying the relative fluctuations in weighted networks in different scenarios (ranging from homogeneous to heterogenous concentrations of link weights), in conjunction with the relative entropy and, in general, ensemble (non)equivalence.

Using Eq.~\eqref{covariance-matrix-dewd-diagonal}, we can immediately calculate the standard deviation of each constraint $s_i$ around its expected value $s^*_i$ as
\begin{eqnarray}\label{variance-si-1}
 \sigma_{\vec{\beta}^*}(s_{i})&=&\sqrt{\textrm{Var}_{\vec{\beta}^*}(s_i)}\nonumber\\
&=&\sqrt{\sum_{i\neq j}  \frac{e^{-(\beta^*_i+\beta^*_j)}}{[1-e^{-(\beta^*_i+\beta^*_j)}]^2}}\nonumber\\
&=&\sqrt{\sum_{i\neq j}\langle w_{ij}\rangle_{\vec{\beta}^*}(1+\langle w_{ij}\rangle_{\vec{\beta}^*})}\nonumber\\
&=&\sqrt{s^*_i+\sum_{i\neq j}\langle w_{ij}\rangle^2_{\vec{\beta}^*}}.
\end{eqnarray}
The relative fluctuation of the strength is therefore
\begin{equation}\label{ratio-variance-streng1}
  r_i^*=\frac{\sigma_{\vec{\beta}^*}(s_i)}{s^*_i}=\sqrt{\frac{1}{s^*_i}+\frac{\sum_{i\neq j}\langle w_{ij}\rangle_{\vec{\beta}^*}^2}{(\sum_{i\neq j}\langle w_{ij}\rangle_{\vec{\beta}^*})^2}},
\end{equation}
Since $\langle w_{ij}\rangle_{\vec{\beta}^*}\ge0$ for all $i,j$, we have  $(\sum_{i\neq j}\langle w_{ij}\rangle_{\vec{\beta}^*})^2\ge\sum_{i\neq j}(\langle w_{ij}\rangle_{\vec{\beta}^*})^2$, showing that $1/\sqrt{s_i^*}$ is indeed a lower bound for $r_i^*$. 
When studying the asymptotic behaviour of the relative fluctuations in the thermodynamic limit, we will be interested in whether the limit
\begin{eqnarray}
\rho_i^*\equiv\lim_{n\to\infty}r_i^*
\label{eq:rho}
\end{eqnarray}
is zero (vanishing relative fluctuations) or positive (nonvanishing relative fluctuations) for each node $i=1,n$.

\section{BEC in weighted networks\label{sec_BEC}}
In physical systems composed of bosons, i.e. particles obeying Bose-Einstein statistics, BEC is a phase transition whereby, below a certain critical temperature, a finite fraction of the total number of particles condenses in the ground state, i.e. the state with lowest energy (or more generally in a finite number of states with lowest energy).
BEC was theoretically predicted by Satyendra Nath Bose and Albert Einstein in 1924~\cite{BEC}, and it has since then been observed in various physical systems.
Models of BEC have been studied in different statistical ensembles in the standard case with only global constraints (total energy and/or total number of particles)~\cite{navez1997fourth,holthaus1998condensate,mullin2003bose,chatterjee2014fluctuations,tarasov2015grand,crisanti2019condensation}.
Although the detailed phenomenology exhibited by these models depends on the choice of the energy and the structure of the interactions, it is generally found that EE breaks down in the condensed (low-temperature) phase, as signalled by nonvanishing relative fluctuations of the constraints.

In this Section, we are going to show that a form of BEC, even if quite different from that found in more traditional physical settings, can also appear in our ensembles of weighted networks. 
The possible onset of BEC in our system creates an ideal situation where an EE-breaking phase transition can be studied in combination with an additional and unrelated mechanism for the breakdown of EE, i.e. the presence of local constraints, which is always active in both the condensed and the non-condensed phases.
To illustrate our results, we first make some important clarifications in order to establish a rigorous link from weighted network ensembles to Bose-Einstein statistics and then study the different phases of the model.

\subsection{Bose-Einstein statistics in weighted networks}
As we have already recalled, weighted networks with a constraint on the strength sequence obey Bose-Einstein statistics, as opposed to binary networks that obey Fermi-Dirac statistics~\cite{park2004statistical,garlaschelli2009generalized,garlaschelli2013low}. 
Indeed, inserting Eq.~\eqref{Hamiltonian} into Eq.~\eqref{Can_probability} we get the probability of a configuration for a gas of free particles in the \emph{grandcanonical ensemble}\footnote{In the grandcanonical ensemble, both the total energy and the total number of particles are treated as soft constraints. With respect to the canonical ensemble, the appearance of the number of particles as an additional soft constraint requires the introduction of an extra Lagrange multiplier, the chemical potential. Interestingly, in the context of BEC a fourth (so-called `Maxwell's Demon') ensemble has also been introduced where the total number of particles is soft while the total energy is hard~\cite{navez1997fourth}.}, where the pair $i,j$ labels an energy state, the weight $w_{ij}$ is the number of particles in that state (occupation number), and the sum $\beta_i+\beta_j$ can be interpreted as 
\begin{equation}
\beta_i+\beta_j=\frac{\epsilon_{ij}-\mu(T)}{kT}.
\label{eq_epsilon}
\end{equation}
In the latter expression, $\epsilon_{ij}$ represents the energy of the state, $1/kT$ is the inverse temperature, and $\mu(T)$ is the \emph{chemical potential} (required to fix the same expected overall number of particles for all values of $T$)~\cite{garlaschelli2013low}.
Indeed, as we discussed in Sec.~\ref{sec_cano}, in our setting the energy and temperature (and in this case, the chemical potential as well) are all reabsorbed into $\vec{\beta}$.
Therefore we can interpret the link weight $w_{ij}$ as the number of `elementary particles' of weight, i.e. the number of quanta of unit weight, populating the link between nodes $i$ and $j$~\cite{park2004statistical,garlaschelli2009generalized}. 
The total number of such particles in the system is the total weight $W$ of all links in the network:
\begin{equation}
W(\mathbf{W})=\sum_{i=1}^n\sum_{j<i}w_{ij}=\frac{1}{2}\sum_{i=1}^n s_i(\mathbf{W}).
\label{eq_W}
\end{equation}
The `weighted' property $w_{ij}\in\mathbb{N}$, which leads to Eqs.~\eqref{partition-function dewd},~\eqref{can-dewd-1} and~\eqref{probability0wij}, corresponds to the possibility that the same state (pair of nodes) is occupied by indefinitely many particles (subject to the average number dictated by the chemical protential), which is a property of bosons.
By contrast, in binary networks one has to impose $w_{ij}\in(0,1)$, which is a property of fermions~\cite{park2004statistical}. An extensive treatment of the role of chemical potential and temperature in binary networks can be found in~\cite{garlaschelli2013low}.
Here, to properly interpret what the weighted model is doing, we should give a series of clarifications. 

First, we should make a clear distrinction beween the $n$ `units' of our system (i.e. the nodes of the network) and the $W$ `particles' of weight that, as a formal analogy, can be interpreted as populating the links of the network.
The former are the real constituents of our physical system, while the latter are a mathematical abstraction used to represent the nature of the interactions (links) between such constituents. 
If we imagine doubling the size of our network, we should imagine doubling the number $n$ of nodes: indeed, we can imagine the network `growing in size' by adding one single node at a time, but we cannot imagine adding one single \emph{pair} of nodes at a time, without actually adding $n$ new pairs. One should also not be tempted to regard node pairs as the fundamental units by the fact that, mathematically, the $n(n-1)/2$ variables $\{w_{ij}\}$ involving different pairs of nodes are independent random variables: actually, this only occurs in the canonical ensemble and would in any case not be true for more general choices of the constraints.
Moreover, even the $n(n-1)/2$ independent node pairs in the canonical ensemble cannot be assigned independent values of the parameters, since there are only $n$ parameters corresponding to the Lagrange multipliers attached to each node. Explicit (and strong) consequences of this fact will be illustrated precisely in the context of BEC.
Therefore the physical size of our system is $n$, and this is why in Eq.~\eqref{specific-relative-entropy} we defined the relative entropy density as the relative entropy divided by $n$ in the first place.
How the total weight $W$ varies with the system size $n$ depends on a specific property, i.e. on how we make the entries of $\vec{s}^*$ (and the resulting value of $W^*=\sum_{i=1}^n s^*_i/2$) scale with $n$. 
For instance, we may choose to be in the sparse regime where $\vec{s}^*$ remains finite as $n\to\infty$, or in the dense regime where $\vec{s}^*$ diverges as $n\to\infty$.
As we show below, the latter is the relevant case for BEC to emerge.

Second, we stress that, irrespective of the above, we always consider a hard number $n$ of nodes, and this is why we compare (only) the canonical (soft value of $\vec{s}$) and microcanonical (hard value of $\vec{s}$) ensembles of networks, both for fixed $n$ (which sets the dimension of $\vec{s}$).
We do \emph{not} consider the grandcanonical ensemble \emph{of network configurations} where $n$ is soft.
The grandcanonical ensemble introduced in the aforementioned analogy with systems of bosons is a different one; it may be denoted as an ensemble of \emph{weight quanta} in a network with fixed $n$ and originates from the fact that the Hamiltonian in Eq.~\eqref{Hamiltonian}, and consequently the total link weight (not $n$) in Eq.~\eqref{eq_W}, is a fluctuating quantity in the canonical ensemble of network configurations.
The fluctuations in the (rescaled) energy $H$ (canonical ensemble of network configurations) are seen as fluctuations in the particle number $W$ (grandcanonical ensemble of weight quanta) in the Bose-Einstein analogy. Fluctuations in the particle number have been the subject of many studies in the literature on BEC~\cite{navez1997fourth,holthaus1998condensate,mullin2003bose,chatterjee2014fluctuations,tarasov2015grand,crisanti2019condensation}.
Note that, in both canonical and microcanonical ensembles, the individual link weights $\{w_{ij}\}$ are fluctuating quantities, despite the fact that the total link weight $W^*$ is constant in the microcanonical ensemble.
Therefore the numbers of `weight particles' of individual links fluctuate in both ensembles.

Third, while we necessarily discuss the (non)equivalence of the canonical and microcanonical ensembles in the thermodynamic limit $n\to\infty$, the total weight $W^*$ can (and, across the canonical ensemble, will in any case) be arbitrarily large even for finite $n$. 
Indeed, the phase transition that we are about to discuss (namely, BEC) does not \emph{per se} require the limit $n\to\infty$, while it definitely requires the limit $W^*\to\infty$.
Abstractly, these two limits (and the associated phenomena of EN and BEC respectively) may appear as mathematically unrelated. 
However, in practice they are physically related once the scaling of $\vec{s}^*$ with $n$ is specified.
In particular, we are going to show that, in order to observe BEC, we need be in a dense regime where $W^*=O(n^2)$. 
This ensures that, when taking the thermodynamic limit $n\to\infty$ in order to study ensemble (non)equivalence, we are automatically implying $W^*\to\infty$ so that we can check for BEC at the same time.

Last, we recall that $\vec{\beta}\cdot\vec{s}(\mathbf{W})$ has to be dimensionless in order to ensure that the probability is a number. Therefore, since $w_{ij}$ is dimensionless, so are $s_i^*$ and $\beta_i^*$. In turn, this implies that both sides of Eq.~\eqref{eq_epsilon} must be dimensionless. On the other hand, when modelling a real system, the `energy' $\epsilon_{ij}$ may represent any physical `cost' associated to the link between nodes $i$ and $j$ (more precisely, it represents the cost of reinforcing $w_{ij}$ by a unit of weight) and may therefore carry its own unit of measure (e.g. it may depend on some distance between nodes $i$ and $j$). Necessarily, the chemical potential $\mu(T)$ carries the same units as the energy. As for the `temperature' $T$, it may be chosen to be dimensionless as it merely represents a control parameter (this is the choice that we will make later); alternatively, it may carry the same units of the energy if it is useful that temperature and energy live on the same scale. Irrespective of this choice, in our setting the `Boltzmann constant' $k$ is simply a constant that takes care of all dimensional units of measure and makes the ratio on the right hand side of Eq.~\eqref{eq_epsilon} dimensionless.

\subsection{Core-periphery networks\label{sec:corper}}
With the above clarifications, we can finally go back to our model.
In the traditional physical situation, in the canonical ensemble the energy $\epsilon_{ij}$ of each state $i,j$ is a constant and the temperature $T$ can be varied. 
Clearly, $\epsilon_{ij}$ is independent of $T$, while the chemical potential $\mu(T)$ is chosen, as a function of temperature, in order to realize the correct ($T$-independent) expected total number $\langle W\rangle^*\equiv W^*$ of particles for all values of $T$.
In this `direct problem', every state will therefore have an expected occupation number governed by $\epsilon_{ij}$, $T$ and $\mu(T)$.
In our `inverse' setting, $T^*$ and $\mu^*$ are instead reabsorbed into $\vec{\beta}^*$, which in turn is induced by the chosen value of the strength sequence  (rather than the other way around).
We should therefore regard $\vec{s}^*=\vec{s}^*(T)$ and $\vec{\beta}^*=\vec{\beta}^*(T)$ as $T$-dependent, while $W^*$ remains $T$-independent.
This means that the chemical potential $\mu^*(T)$ should be such that
\begin{equation}
\sum_{i=1}^n {s}_i^*(T)=2W^*\quad \forall T\ge 0.
\label{eq:totalW}
\end{equation}

BEC emerges when, below a certain critical temperature $T_c$, the occupation number of the state with minimum energy $\epsilon_\textrm{min}=\textrm{min}_{i,j}\{\epsilon_{ij}\}$ (ground state), or of a finite number of states with lowest energy, becomes so large that it reaches a finite fraction of the total number $W^*$ of particles.
Clearly, this requires the existence of at least two different energy levels (the ground state and at least one `excited' state).
Therefore the simplest way to obtain BEC in our model is by considering a strength sequence of the form 
\begin{equation}
s^*_i(T)=\left\{\begin{array}{ll}s^*_+(T)&i=1,n_+\\s^*_-(T)&i=n_++1,n\end{array}\right.\quad s^*_+(T)\ge s^*_-(T),
\label{eq_bis}
\end{equation}
i.e. by partitioning the $n$ nodes into two classes, which we call \emph{core} and \emph{periphery}: the core has a finite number 
\begin{equation}
n_+=O(1)
\label{n+}
\end{equation}
of nodes, each having a `large' strength $s^*_+(T)$, while the periphery has an extensive number 
\begin{equation}
n_-=n-n_+=O(n)
\label{n-}
\end{equation}
of nodes, each having a `small' strength $s^*_-(T)$.
What we mean precisely by `small' and `large' will be clarified below.
For the moment, we notice that the BEC phase ($T\!<\!T_c$) corresponds to picking a `condensed' value of $\vec{s}^*(T\!<\!T_c)$ such that, in the thermodynamic limit, the core takes up a finite fraction of the total weight $W^*$ of all links in the network, despite having a finite size.  
In particular, in the zero-temperature limit \emph{all} the total weight $W^*$ is in the core.
By contrast, the non-condensed phase $T\!>\!T_c$ is one where $\vec{s}^*(T\!>\! T_c)$ is such that no individual link receives a finite fraction of $W^*$.
In particular, the infinite-temperature limit should be such that the energy difference between ground and excited states becomes ineffective, i.e. $s^*_+(T\!\to\!\infty)=s^*_-(T\!\to\!\infty)$.
The different phases can be efficiently monitored by introducing a temperature-dependent order parameter $Q^*(T)$, as we show below.

We stress that, since we are ultimately interested in the relative fluctuations of the canonical constraints and in the relative entropy that can be asymptotically calculated purely from canonical quantities according to Eq.~\eqref{eq:Sorder}, practically we only need to study the canonical ensemble. The only check we need to make is that, whenever we speak of the system being in a certain `phase', this statement does not depend on the particular ensemble. In other words, we need to show that the order parameter has always the same value in the canonical and microcanonical ensembles.

Before studying the individual phases, let us make some general considerations, valid for all values of $T$.
We first find the value of $\vec{\beta}^*(T)$ corresponding to the value of $\vec{s}^*(T)$ in Eq.~\eqref{eq_bis}.
As we mentioned, $s^*_i(T)=s^*_j(T)$ implies $\beta^*_i(T)=\beta^*_j(T)$, therefore the entries of $\vec{\beta}^*(T)$ take only two values  $\beta^*_+(T)$ and $\beta^*_-(T)$ such that
\begin{equation}
\beta^*_i(T)=\left\{\begin{array}{ll}\beta^*_+(T)&i=1,n_+\\\beta^*_-(T)&i=n_++1,n\end{array}\right.\quad \beta^*_+(T)\le\beta^*_-(T).
\label{eq_bibeta}
\end{equation}
These values solve the $n$ equations in~\eqref{dewd-strength-1}, which here reduce to the two independent equations
\begin{eqnarray}
(n_+-1)w^*_+(T)+n_-w^*_0(T)&\equiv&s^*_+(T)\label{s_hub}\\
(n_--1)w^*_-(T)+n_+w^*_0(T)&\equiv&s^*_-(T)\label{s--}
\end{eqnarray}
where we have defined
\begin{equation}\label{w+}
  w^*_+(T)=\frac{e^{-2\beta^*_+(T)}}{1-e^{-2\beta^*_+(T)}}
\end{equation}
as the expected link weight $\langle w_{ij}\rangle_{\vec{\beta}^*(T)}$ between any two nodes in the core ($i,j=1,n_+$),
\begin{equation}\label{w-}
  w^*_-(T)=\frac{e^{-2\beta^*_{-}(T)}}{1-e^{-2\beta^*_{-}(T)}}
\end{equation}
as the expected link weight $\langle w_{ij}\rangle_{\vec{\beta}^*(T)}$ between any two nodes in the periphery ($i,j=n_++1,n$), and
\begin{eqnarray}
  w^*_0(T)&=&\frac{e^{-\beta^*_{-}(T)-\beta^*_+(T)}}{1-e^{-\beta^*_{-}(T)-\beta^*_+(T)}}
\nonumber\\
&=&\frac{\sqrt{w^*_+(T)w^*_-(T)}}{\sqrt{1+w^*_+(T)}\sqrt{1+w^*_-(T)}-\sqrt{w^*_+(T)w^*_-(T)}}\nonumber\\
&=&\frac{1}{\sqrt{1+1/w^*_+(T)}\sqrt{1+1/w^*_-(T)}-1}\label{w0}
\end{eqnarray}
as the expected link weight $\langle w_{ij}\rangle_{\vec{\beta}^*(T)}$ between any node in the core and any node in the periphery ($i=1,n_+$ and $j=n_++1,n$ or $j=1,n_+$ and $i=n_++1,n$).
Note that $\beta^*_+(T)\le\beta^*_-(T)$ implies $w^*_-(T)\le w^*_0(T)\le w^*_+(T)$.

Now, solving Eqs.~\eqref{s_hub} and~\eqref{s--}, we obtain the explicit values of 
$\beta^*_+(T)$ and $\beta^*_-(T)$ appearing in Eq.~\eqref{eq_bibeta}:
\begin{equation}
\beta^*_\pm(T)=\frac{1}{2}\ln\frac{1+w^*_\pm(T)}{w^*_\pm(T)}=\frac{1}{2}\ln\left(1+\frac{1}{w^*_\pm(T)}\right).
\label{eq_betas}
\end{equation}
Also note from Eq.~\eqref{w0} that 
\begin{equation}
\beta^*_+(T)+\beta^*_-(T)=\ln\left(1+\frac{1}{w^*_0(T)}\right).
\label{eq_beta0}
\end{equation}
Also, using Eq.~\eqref{pij} we can define 
\begin{equation}
p^*_\pm(T)\equiv e^{-2\beta_\pm^*(T)}=\frac{w^*_\pm(T)}{1+w^*_\pm(T)}
\label{pijpm}
\end{equation}
as the probability that a link exists (irrespective of its weight) between any two core-core ($+$) or any two periphery-periphery ($-$) nodes, and
\begin{equation}
p^*_0(T)\equiv e^{-\beta_+^*(T)-\beta_-^*(T)}=\frac{w^*_0(T)}{1+w^*_0(T)}
\label{pij0}
\end{equation}
as the probability that a link exists (irrespective of its weight) between a core node and a periphery node.

From Eq.~\eqref{eq_epsilon}, we notice that the existence of the two values above for the entries of $\vec{\beta}^*(T)$ implies that there are three energy levels, associated with the energies
\begin{eqnarray}
\epsilon_+^*&=&\mu^*(T)+2kT\beta^*_+(T),\label{eq_epsi+}\\
\epsilon_-^*&=&\mu^*(T)+2kT\beta^*_-(T),\label{eq_epsi-}\\
\epsilon_0^*&=&\mu^*(T)+kT[\beta^*_+(T)+\beta^*_-(T)]=\frac{\epsilon_+^*+\epsilon_-^*}{2},\label{eq_epsi0}
\end{eqnarray}
where $\epsilon_+^*\le\epsilon_0^*\le\epsilon_-^*$ (we recall that all energy values are finite and independent of both $T$ and $n$).
The appearance of \emph{three} distinct energy levels out of just \emph{two} values of the fundamental Lagrange multipliers confirms the interpretation that the true units of the system are the nodes and not the node pairs: it would indeed be impossible for our system to exhibit exactly two energy states, or in general to engineer an arbitrary number of energy states for the node pairs, since the only arbitrary values are those that can be attached to nodes, not to node pairs.
Also note that all the three levels above are degenerate: the $n_+(n_+-1)/2$ pairs of nodes in the core have the same expected link weight $w_+^*(T)$ and energy $\epsilon_+^*$, the $n_-(n_--1)/2$ pairs of nodes in the periphery have the same expected link weight $w_-^*(T)$ and energy $\epsilon_-^*$, and the $n_+n_-$ pairs of nodes across core and periphery have the same expected link weight $w_0^*(T)$ and energy $\epsilon_0^*$.
Therefore the ground state has energy $\epsilon^*_\textrm{min}=\epsilon_+^*$ and degeneracy $n_+(n_+-1)/2$. These degeneracies are dictated by the numbers of nodes in the two sets and cannot be assigned arbitrarily.

The occupation number of the ground state (with energy $\epsilon_+^*$) coincides with the expected weight of all links between core nodes (total `core-core' weight):
\begin{equation}\label{whub-1}
W_+^*(T)=\frac{n_+(n_+-1)}{2}w^*_+(T).
\end{equation}
Similarly, the occupation number of the first excited state (with energy $\epsilon_0$) coincides with the expected weight of all links between nodes across core and periphery (total `core-periphery' weight):
\begin{equation}\label{W0}
W_0^*(T)=n_+n_-w^*_0(T).
\end{equation}
Finally, the occupation number of the second excited state (with energy $\epsilon_-$) coincides with the expected weight of all links between periphery nodes (total `periphery-periphery' weight):
\begin{equation}\label{W-}
W_-^*(T)=\frac{n_-(n_--1)}{2}w^*_-(T).
\end{equation}
By writing $W^*$ as the sum of its core-core, core-periphery and periphery-periphery components, we get
\begin{eqnarray}
W^*&=&W^*_+(T)+W^*_0(T)+W^*_-(T)\nonumber\\
&=&\frac{n_+(n_+-1)}{2}w^*_+(T)+{n_+n_-}w^*_0(T)\nonumber\\
&&+\frac{n_-(n_--1)}{2}w^*_-(T).\label{eq:components}
\end{eqnarray}
Using Eq.~\eqref{eq:totalW}, the total weight can also be expressed as
\begin{equation}
W^*=\frac{n_+s^*_+(T)+n_-s^*_-(T)}{2}
\label{eq:totalW2}
\end{equation}
which, through Eqs.~\eqref{s_hub} and~\eqref{s--}, indeed reduces to Eq.~\eqref{eq:components}.

We stress again that the chemical potential $\mu^*(T)$ appearing in Eqs.~\eqref{eq_epsi+},~\eqref{eq_epsi-} and~\eqref{eq_epsi0} plays the role of a global Lagrange multiplier ensuring that, for all values of $T$, the total expected weight is $W^*$.
Note that the $T$-independence of $W^*$ allows us to conclude immediately that its value should be of order 
\begin{equation}
W^*=O(n^2)
\label{W*O}
\end{equation}
because, in particular, in the non-condensed phase all the $n(n-1)/2$ individual link weights $w^*_\pm$, $w^*_0$ must be finite by definition.
As we have anticipated, this result ensures that in the thermodynamic limit ($n\to\infty$) we automatically have $W^*\to\infty$, so that we can study ensemble (non)equivalence and BEC simultaneously, thereby `physically' connecting two otherwise mathematically unrelated limits.
We also note that, irrespective of temperature, the network is always in the dense regime.
We can therefore introduce the average expected link weight
\begin{equation}
w^*\equiv\frac{2W^*}{n(n-1)}=O(1),
\label{w*}
\end{equation}
which is a $T$-independent, finite parameter of our model, controlling the overall link weight in the network.
Clearly,
\begin{equation}
w^*_-(T)\le w^*\le w^*_+(T)\quad\forall T.
\label{eq:relation}
\end{equation}

It is good to remark again that, in our `inverse' problem (construction of the conjugate canonical and microcanonical ensembles), the parameters of the model are the values of the constraints, which here reduce to the two (diverging when $n\to\infty$) numbers $s_\pm^*(T)$. However, to allow consistent comparisons for different temperatures, not all strength sequences are allowed, but only those that can be obtained from one another by varying $T$. In particular the values $s_\pm^*(T)$ have to be specified for each value of $T$ and be such that the total weight is always $W^*$. Indeed Eq.~\eqref{eq:totalW2} shows that only two of the three quantities $s_\pm^*(T)$, $W^*$ are independent. 
By contrast, the traditional `direct' problem in physics sees the three energies $\epsilon^*_\pm$ and $\epsilon^*_0$ (which do not depend on $T$) as the parameters of the model, plus either $w^*$ or the chemical potential $\mu^*(T)$.
However, Eq.~\eqref{eq_epsi0} shows that $\epsilon_0^*=(\epsilon_+^*+\epsilon_-^*)/2$, indicating only two independent values of the energy (say $\epsilon_\pm^*$), as a result of the fact that there are only two types of nodes.
Moreover, we may set the minimum energy $\epsilon^*_+\equiv0$ without loss of generality, because any overall energy shift can be reabsorbed into the chemical potential. 
We can therefore rename the only remaining independent value of the energy as $\epsilon^*_-\equiv\epsilon^*>0$, and similarly $\epsilon^*_0=\epsilon^*/2>0$.
Using these replacements into Eqs.~\eqref{eq_epsi+} and~\eqref{eq_epsi-}, and combining the two equations, we get
\begin{equation}
\mu^*(T)=-2kT\beta^*_+(T)=\epsilon^*-2kT\beta^*_-(T)
\label{eq:mimmo0}
\end{equation}
which is a convenient expression for solving the `direct' problem.
Rearranging, we obtain
\begin{equation}
\epsilon^*=2kT\left[\beta^*_-(T)-\beta^*_+(T)\right].
\label{eq:mimmo1}
\end{equation}
and, using Eqs.~\eqref{eq_betas} and~\eqref{pijpm},
\begin{eqnarray}
e^{-\epsilon^*/kT}&=&e^{2[\beta^*_+(T)-\beta^*_-(T)]}\nonumber\\
&=&\frac{p^*_-(T)}{p^*_+(T)}\nonumber\\
&=&\frac{w^*_-(T)[1+w^*_+(T)]}{w^*_+(T)[1+w^*_-(T)]},
\label{eq:mimmo2}
\end{eqnarray}
which shows how the energy difference $\epsilon^*$ between periphery-periphery ($+$) and core-core ($-$) states is related to the corresponding connection probabilities $p^*_\pm(T)$ and expected link weights $w^*_\pm(T)$.
Therefore the most compact way of parametrizing the direct problem is by specifying only the two finite, positive and $T$-independent numbers $\epsilon^*$ and $w^*$, and explore the resulting network properties by finding $\mu(T)$ (as a function of $\epsilon^*$, $w^*$ and $T$) and varying $T$ as a control parameter. This will indeed allow us to easily explore the different (high- and low-temperature) phases consistently. 

\begin{figure*}[t]
  \includegraphics[width=\textwidth]{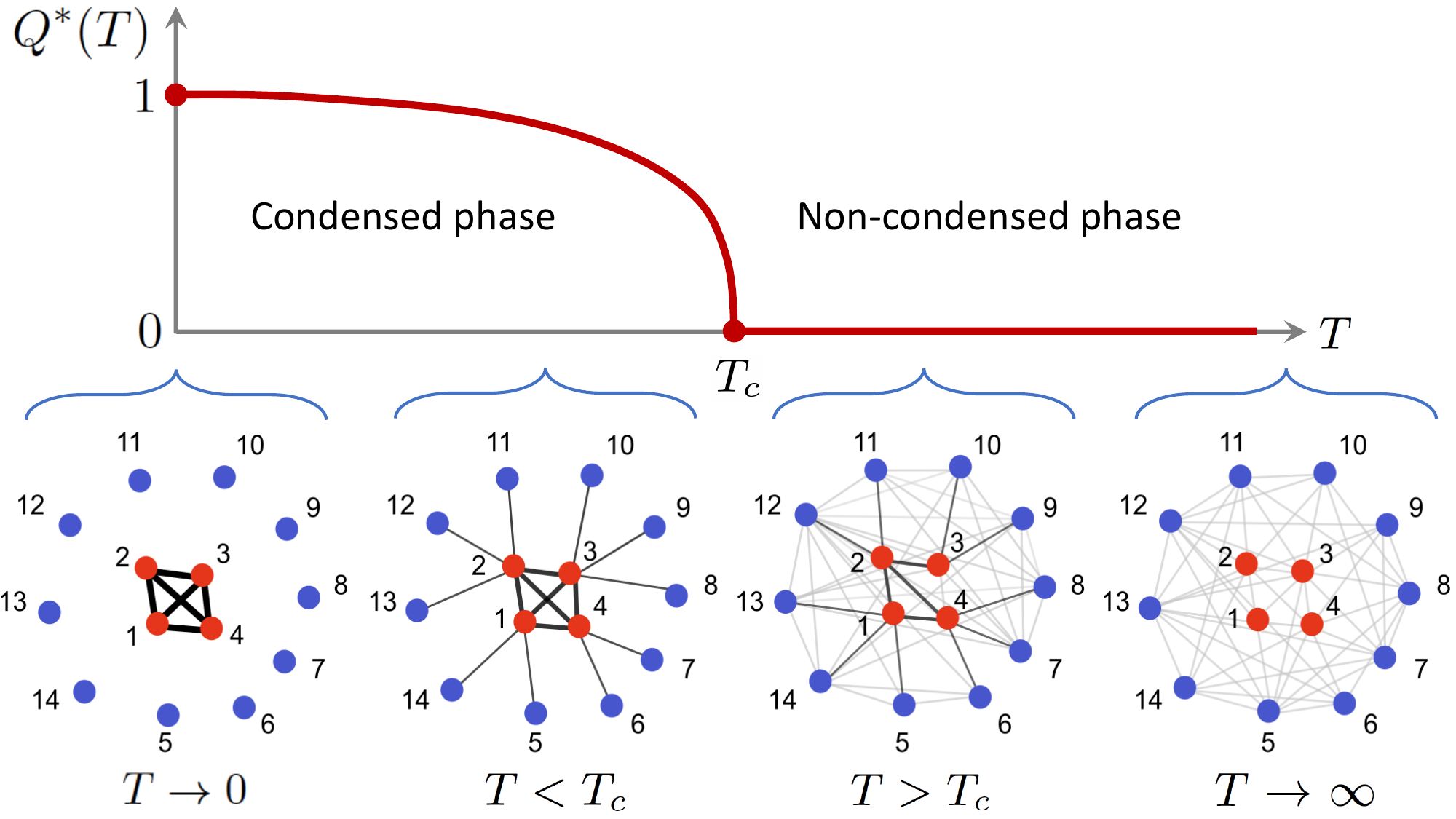}
  \caption{Illustration of possible realizations of a network as a function of temperature, from higher (right) to lower (left) values of $T$. As a schematic example, a network with $n=14$ nodes, of which $n_+$ are core nodes and $n_-=10$ are peripheral nodes, is considered.
The order parameter $Q^*(T)$ is zero for temperatures above the critical temperature $T_c$, while it is positive for temperatures below $T_c$, increasing towards 1 at zero temperature.
At infinite temperature ($T\!\to\!\infty$), there is no distinction between core and periphery: all links have the same probability of existing and the same expected weight.
At lower but supercritical temperature ($T\!>\!T_c$), a quantitative (but not yet qualitative) distinction between core and periphery appears: core-core links have higher probability and expected weight than core-periphery links, which in turn have higher probability and expected weight than periphery-periphery links, however all these probabilities and expected link weights are of the same (finite) order $O(1)$. 
Below a certain critical temperature ($T\!<\!T_c$), the distinction between core and periphery becomes qualitative and more dramatic (the core forms a `condensate'): the expected link weights are of order $O(n^2)$ for core-core links (with the corresponding connection probabilities approaching one) and $O(1)$ for core-periphery and periphery-periphery links.
Finally, at zero temperature ($T\!\to\! 0$) the condensate decouples from the rest: peripheral nodes are completely disconnected and all links end up in the core, with an expected weight still of order $O(n^2)$.  
}
  \label{fig3}
\end{figure*}

In our model, BEC occurs below a critical temperature $T_c$ such that a finite fraction of the total weight $W^*$ condenses in the core, which remains of finite size (i.e. of a finite number $n_+$ of nodes) even when the size of the whole network diverges.
This corresponds to requiring that, as $n\to\infty$, $n_+$ remains finite as dictated by Eq.~\eqref{n+}, $W^*$ diverges, and $W_+^*(T)$ takes up a finite fraction of $W^*$.
Rigorously, we can define this fraction (for finite $n$) as
\begin{equation}
Q^*_n(T)\equiv\frac{W_+^*(T)}{W^*}
\label{eq:preorder}
\end{equation}
and use it to introduce the \emph{order parameter} as
\begin{equation}
Q^*(T)\equiv\lim_{n\to\infty}Q^*_n(T)=\lim_{n\to\infty}\frac{W_+^*(T)}{W^*}.
\label{eq:order}
\end{equation}
We can then define the BEC phase as a phase emerging below a certain critical temperature $T_c$ such that
\begin{equation}
Q^*(T\!<\!T_c)>0.
\label{eq:orderBEC0}
\end{equation}
By contrast, the non-BEC phase is such that
\begin{equation}
Q^*(T\!>\!T_c)=0.
\label{eq:orderNONBEC}
\end{equation}
A visual anticipation of the qualitative behaviour that our system will exhibit is provided in Fig.~\ref{fig3}. 

In conjunction with BEC, we will investigate ensemble (non)equivalence. Therefore, in each phase of the model we will consider the relative entropy between the microcanonical and canonical ensembles and the relative fluctuations of the constraints. 
The criterion for measure equivalence is based on the relative entropy in Eq.~\eqref{eq:Sorder}, and useful techniques for the calculation of the determinant of the covariance matrix $\mathbf{\Sigma}^*$ in each phase are provided in the Appendix. 
Clearly, from Eqs.~\eqref{covariance-matrix-dewd-diagonal} and~\eqref{eq_bibeta} we see that the diagonal entries $\Sigma_{ii}^*$ of $\mathbf{\Sigma}^*$ take two possible values:
\begin{equation}
\Sigma_{ii}^*(T)=\sigma^2_{\vec{\beta}^*(T)}(s_i)=\left\{\begin{array}{ll}\Sigma_{+}^*(T)&i=1,n_+\\
\Sigma_{-}^*(T)&i=n_++1,n\end{array}\right.
\end{equation}
where
\begin{eqnarray}
\Sigma_{\pm}^*(T)&=&\frac{(n_\pm-1)e^{-2\beta^*_\pm(T)}}{\left[1-e^{-2\beta^*_\pm(T)}\right]^2}+\frac{n_\mp e^{-\beta^*_+(T)-\beta^*_-(T)}}{\left[1-e^{-\beta^*_+(T)-\beta^*_-(T)}\right]^2}\nonumber\\
&=&(n_\pm-1){w^*_\pm}(T)\left[1+w^*_\pm(T)\right]\nonumber\\
&&+n_\mp{w^*_0}(T)\left[1+w^*_0(T)\right].\label{eq:squares}
\end{eqnarray}

Recalling Eq.~\eqref{partition-function dewd}, we remark that the canonical entropy $S^*_{\textrm{can}}(T)$ can be easily calculated from Eq.~\eqref{can-entropy-G} as the following sum of five terms:
\begin{eqnarray}
 S^*_{\textrm{can}}(T)&=&\vec{\beta}^*(T)\cdot\vec{s}^*(T)+\ln Z[\vec{\beta}^*(T)]\nonumber\\
&=&n_+\beta^*_+(T)s^*_+(T)\nonumber\\
&&+n_-\beta^*_-(T)s^*_-(T)\nonumber\\
&&+\frac{n_+(n_+-1)}{2}\ln \frac{1}{1-e^{-2\beta^*_+(T)}}\nonumber\\
&&+\frac{n_-(n_--1)}{2}\ln \frac{1}{1-e^{-2\beta^*_-(T)}}\nonumber\\
&&+{n_+n_-}\ln \frac{1}{1-e^{-\beta^*_+(T)-\beta^*_-(T)}},
\label{entropy+-}
\end{eqnarray}
while the microcanonical entropy $S^*_{\textrm{mic}}(T)$ is in general hard to compute, as it requires an explicit enumeration. However, its leading order can be obtained combining Eqs.~\eqref{Mic-shannon-entropy-1} and~\eqref{entropy+-}.

The relative fluctuations of the constraints take the form
\begin{equation}
r^*_i(T)=\left\{
\begin{array}{ll}r^*_+(T)&i=1,n_+\\
r^*_-(T)&i=n_++1,n
\end{array}\right.
\label{eq:r+-}
\end{equation}
where, from Eq.~\eqref{ratio-variance-streng1},
\begin{eqnarray}
r_\pm^*(T)&=&\frac{\sqrt{\Sigma_{\pm}^*(T)}}{s^*_\pm(T)}\label{eq:flutt}\\
&=&\sqrt{\frac{1}{s_\pm^*(T)}+\frac{(n_\pm-1)[w^*_\pm(T)]^2+n_\mp[w^*_0(T)]^2}{[s_\pm^*(T)]^2}}\nonumber\\
&=&\sqrt{\frac{1}{s_\pm^*(T)}+\frac{(n_\pm-1){[w^*_\pm(T)]}^2+n_\mp{[w^*_0(T)]}^2}{[(n_\pm-1){w^*_\pm(T)}+n_\mp{w^*_0(T)}]^2}}.\nonumber
\end{eqnarray}
Therefore in the thermodynamic limit the relative fluctuations of the constraints, as defined in Eq.~\eqref{eq:rho}, take only the two possible limiting values
\begin{equation}
\rho^*_i(T)=\left\{
\begin{array}{ll}\rho^*_+(T)&i=1,n_+\\
\rho^*_-(T)&i=n_++1,n
\end{array}\right.
\label{eq:rhoT}
\end{equation}
where
\begin{equation}
\rho^*_\pm(T)=\lim_{n\to\infty}r^*_\pm(T).
\label{eq:rho+-}
\end{equation}
Armed with the above general results, we can now study each phase in detail.

\subsection{Non-condensed phase\label{sec:noncondensed}}
Let us start from the non-BEC phase ($T\!>\!T_c$).
We first consider the finite-temperature case ($T_c\!<\!T\!<\!\infty$) and then the infinite-temperature limit ($T\!\to\!\infty$).
As all the interesting phenomenology (in terms of both BEC and EN) occurs in the thermodynamic limit $n\to\infty$, we look for the asymptotic behaviour of all quantities in that limit.

\subsubsection{Finite (supercritical) temperature: $T_c\!<\!T\!<\!\infty$\label{scialla}}
Since, by definition, when $T\!>\!T_c$ there is no concentration of `particles' of weight on any of the links, all the expected link weights must be separately finite, i.e. $w^*_+(T\!>\!T_c)$, $w^*_-(T\!>\!T_c)$ and $w^*_0(T\!>\!T_c)$ are all $O(1)$.
Consequently, from Eqs.~\eqref{pijpm} and~\eqref{pij0} we see that all the connection probabilities $p^*_+(T\!>\!T_c)$, $p^*_-(T\!>\!T_c)$ and $p^*_0(T\!>\!T_c)$ are strictly smaller than one, i.e. missing links can occur anywhere in the network.
Using this fact into Eqs.~\eqref{s_hub},~\eqref{s--},~\eqref{whub-1},~\eqref{W0} and~\eqref{W-}, and using Eqs.~\eqref{n+} and~\eqref{n-}, we immediately get the strength of nodes in the core, i.e.
\begin{eqnarray}
s^*_+(T\!>\!T_c)&=&(n_+-1)w^*_+(T\!>\!T_c)+n_-w^*_0(T\!>\!T_c)\nonumber\\
&\approx& n\,w^*_0(T\!>\!T_c)\nonumber\\
&=&O(n),\label{s*+>}
\end{eqnarray}
and in the periphery, i.e.
\begin{eqnarray}
s^*_-(T\!>\!T_c)&=&(n_--1)w^*_-(T\!>\!T_c) +n_+w^*_0(T\!>\!T_c)\nonumber\\
&\approx&n\,w^*_-(T\!>\!T_c)\nonumber\\
&=&O(n).
\label{s*->}
\end{eqnarray}
Similarly, for $W^*_\pm(T\!>\!T_c)$, $W^*_0(T\!>\!T_c)$ we get
\begin{eqnarray}
W^*_+(T\!>\!T_c)&=&n_+(n_+-1)w^*_+(T\!>\!T_c)/2=O(1),\,\,\label{eq:nBECW+}\\
W^*_-(T\!>\!T_c)&\approx&n^2w^*_-(T\!>\!T_c)/2=O(n^2),\label{eq:nBECW-}\\
W^*_0(T\!>\!T_c)&\approx&n_+\,n\,w^*_0(T\!>\!T_c)=O(n),\label{eq:nBECW0}
\end{eqnarray}
from which we see that in this phase the total weight $W^*$ is essentially all in the periphery, i.e. 
\begin{eqnarray}
W^*_-(T\!>\!T_c)&=&W^*-o(n^2)\approx W^*,\\
w^*_-(T\!>\!T_c)&=&w^*-o(1)\approx w^*.
\label{eq:ww}
\end{eqnarray}
We stress that the above result does not mean that the core is empty or that there are no connections between core and periphery. Rather, it indicates that the total weight of all core-core and core-periphery connections is asymptotically negligible with respect to the total weight located inside the periphery, simply because the number of periphery-periphery node pairs dominates the number of core-periphery and core-core pairs. In particular, the finite parameter $w^*_+(T\!>\!T_c)$ can take an arbitrarily large value, without `moving' the (finite and positive) value of the average link weight $w^*$. All positive values of $w^*_+(T\!>\!T_c)$ are therefore allowed. By contrast, $w^*_-(T\!>\!T_c)$ is forced to take (to leading order) only the value $w^*$.

Using Eqs.~\eqref{s*->},~\eqref{eq:nBECW-} and~\eqref{eq:nBECW0}, we write the order parameter as
\begin{eqnarray}
Q^*(T\!>\!T_c)&\equiv&\lim_{n\to\infty}\frac{W_+^*(T\!>\!T_c)}{W^*}\nonumber\\
&=&\lim_{n\to\infty}\frac{W^*-W_-^*(T\!>\!T_c)-W_0^*(T\!>\!T_c)}{W^*}\nonumber\\
&=&1-\lim_{n\to\infty}\frac{n^2\,w^*_-(T\!>\!T_c)}{2W^*}\nonumber\\
&=&1-\lim_{n\to\infty}\frac{n\,s^*_-(T\!>\!T_c)}{2W^*},\nonumber\\
&=&0,\label{eq:Q>}
\end{eqnarray}
confirming the definition of non-condensed phase in Eq.~\eqref{eq:orderNONBEC} and showing that, since both $s^*_-(T\!>\!T_c)$ and $W^*$ have by construction the same value in the canonical and microcanonical ensemble, the order parameter must be zero in both ensembles, for all values of $T\!>\!T_c$.
Therefore, whenever one ensemble is in the non-condensed phase, the other ensemble is the non-condensed phase as well.
Importantly, this allows us to refer to the conjugate canonical and microcanonical ensembles `in the non-condensed phase' consistently.

To solve the `inverse' problem, we use Eqs.~\eqref{eq_betas} and~\eqref{eq_beta0} and invert Eqs.~\eqref{s*+>} and~\eqref{s*->} to get
\begin{eqnarray}
\beta_-^*(T\!>\!T_c)&=&\frac{1}{2}\ln\left(1+\frac{1}{w^*_-(T\!>\!T_c)}\right)\nonumber\\
&\approx&\frac{1}{2}\ln\left(1+\frac{n}{s^*_-(T\!>\!T_c)}\right)
\label{eq:nBECbeta-}
\end{eqnarray}
and
\begin{eqnarray}
\beta_+^*(T\!>\!T_c)+\beta_-^*(T\!>\!T_c)&=&\ln\left(1+\frac{1}{w^*_0(T\!>\!T_c)}\right)\nonumber\\
&\approx&\ln\left(1+\frac{n}{s^*_+(T\!>\!T_c)}\right).\quad
\label{eq:nBECbeta0}
\end{eqnarray}
Then, subtracting Eq.~\eqref{eq:nBECbeta-} from Eq.~\eqref{eq:nBECbeta0}, we get
\begin{eqnarray}
\beta_+^*(T\!>\!T_c)&=&\frac{1}{2}\ln\left(1+\frac{1}{w^*_+(T\!>\!T_c)}\right)\nonumber\\
&\approx&\ln\frac{1+\frac{n}{s^*_+(T\!>\!T_c)}}{\sqrt{1+\frac{n}{s^*_-(T\!>\!T_c)}}}.
\label{eq:nBECbeta+}
\end{eqnarray}
Equations~\eqref{eq:nBECbeta-} and~\eqref{eq:nBECbeta+} express $\beta^*_\pm(T\!>\!T_c)$ as a function of the two (diverging) constraints $s^*_\pm(T\!>\!T_c)$, or equivalently as a function of the finite parameters $w^*_\pm(T\!>\! T_c)$, which have to be specified for all values of $T$. 

To solve the `direct' problem, we first note an important consequence of Eq.~\eqref{eq:ww}: in the large $n$ limit, $w^*_-(T\!>\!T_c)$ and $\beta^*_-(T\!>\!T_c)$ are independent of temperature. Indeed, using Eq.~\eqref{s*->} and~\eqref{eq:ww} into Eq.~\eqref{eq:nBECbeta-}, we get asympotically
\begin{equation}
\beta_-^*(T\!>\!T_c)\approx\frac{1}{2}\ln\left(1+\frac{1}{w^*}\right)
\label{eq:nBECbeta--}
\end{equation}
and, using Eq.~\eqref{eq:mimmo1},
\begin{equation}
\beta_+^*(T\!>\!T_c)\approx\frac{1}{2}\ln\left(1+\frac{1}{w^*}\right)-\frac{\epsilon^*}{2kT}.
\label{eq:nBECbeta++}
\end{equation}
When inserted into Eq.~\eqref{eq:mimmo0}, the above expressions allow us to directly obtain the chemical potential as
\begin{eqnarray}
\mu(T\!>\!T_c)&=&\epsilon^*-2kT\beta_-^*(T\!>\!T_c)\nonumber\\
&\approx&\epsilon^*-kT\ln\left(1+\frac{1}{w^*}\right)+o(T).\label{eq:mimmo>}
\end{eqnarray}
As anticipated, the above result provides the solution to the direct problem in terms of the two finite constants $\epsilon^*$ and $w^*$, and allows us to explore the model by varying $T$ throughout the non-condensed phase $T\!>\!T_c$. 

We now consider ensemble (non)equivalence.
Inserting Eqs.~\eqref{s*+>} and~\eqref{s*->} into Eq.~\eqref{eq:squares}, we obtain the variance of the strength of nodes in the core, i.e. 
\begin{eqnarray}
\Sigma_+^*(T\!>\!T_c)&\approx&n\,{w^*_0}(T\!>\!T_c)\left[1+w^*_0(T\!>\!T_c)\right]\nonumber\\
&\approx&{s^*_+(T\!>\!T_c)}\left[1+\frac{s^*_+(T\!>\!T_c)}{n}\right]\nonumber\\
&=&O(n),\label{eq:var+>}
\end{eqnarray}
and in the periphery, i.e.
\begin{eqnarray}
\Sigma_-^*(T\!>\!T_c)&\approx&n\,{w^*_-}(T\!>\!T_c)\left[1+w^*_-(T\!>\!T_c)\right]\nonumber\\
&\approx&{s^*_-(T\!>\!T_c)}\left[1+\frac{s^*_-(T\!>\!T_c)}{n}\right]\nonumber\\
&=&O(n).\label{eq:var->}
\end{eqnarray}
As we show in Appendix~\ref{app:noncondensed}, it is possible to 
show that the leading term of the determinant of the covariance matrix $\mathbf{\Sigma}^*(T\!>\!T_c)$ in this non-condensed phase is
\begin{equation}
\det[\mathbf{\Sigma}^*(T\!>\!T_c)]=
\prod_{i=1}^n{\Sigma}^*_{ii}(T\!>\!T_c)+O(n^{n-2}).\label{det-production_main}
\end{equation}
Using Eqs.~\eqref{eq:squares},~\eqref{eq:var+>} and~\eqref{eq:var->} we obtain 
\begin{eqnarray}
\prod_{i=1}^n{\Sigma}^*_{ii}(T\!>\!T_c)&=&[\Sigma_{+}^*(T\!>\!T_c)]^{n_+}[\Sigma_{-}^*(T\!>\!T_c)]^{n_-}\nonumber\\
&=&O(n^n).\label{semidet-BEC}
\end{eqnarray}
Inserting this result into Eq.~\eqref{det-production_main}, we find 
\begin{equation}\label{det-BEC>}
\det[\mathbf{\Sigma}^*(T\!>\!T_c)]=O(n^n)+O(n^{n-2})=O(n^n),
\end{equation}
showing that the determinant is dominated by the diagonal entries of $\mathbf{\Sigma}^*(T\!>\!T_c)$.
Taking the logarithm, we obtain
\begin{equation}
\ln\det \mathbf{\Sigma}^*(T\!>\!T_c)=O(n\ln n)
\label{eq:ultimate>}
\end{equation}
which, when compared with Eq.~\eqref{eq:ultimate}, shows that the system is under ensemble nonequivalence.
We note that the $O(n\ln n)$ scaling of $\ln\det \mathbf{\Sigma}^*(T\!>\!T_c)$ ensures that Eq.~\eqref{eq_KLleading} holds, so the leading order of the relative entropy can be calculated exactly as
\begin{equation}
D^*(T\!>\!T_c)\approx\frac{1}{2}\ln \det\mathbf{\Sigma}^*(T\!>\!T_c)\approx\frac{1}{2}n\ln n
\label{eq_KLleading>}
\end{equation}
where we have used Eqs.~\eqref{eq:var+>} and~\eqref{eq:var->} into Eq.~\eqref{det-production_main}.
The above result is in line with the scaling of the relative entropy found in the case of binary networks with a constraint on the node degrees in the dense regime~\cite{squartini2015breaking,garlaschelli2016ensemble,roccaverde2019breaking}.
Another similarity between the two models is the order of the canonical entropy:
\begin{equation}
S^*_{\textrm{can}}(T\!>\!T_c)=O(n^2),
\label{eq:Scann2>}
\end{equation}
which can be easily seen from Eq.~\eqref{entropy+-} using $\beta^*_{\pm}(T\!>\!T_c)=O(1)$ and $s^*_\pm(T\!>\!T_c)=O(n)$, as found in Eqs.~\eqref{s*+>},~\eqref{s*->},~\eqref{eq:nBECbeta-} and~\eqref{eq:nBECbeta+}. Then Eq.~\eqref{Mic-shannon-entropy-1} also implies
\begin{equation}
S^*_{\textrm{mic}}(T\!>\!T_c)=O(n^2).
\label{eq:Smicn2>}
\end{equation}
Note that, even if the relative entropy is subleading with respect to the canonical and microcanonical entropies, it is still superextensive in the number $n$ of units of the system, thereby breaking ensemble equivalence as in binary networks with fixed degrees. 
Therefore the result in Eq.~\eqref{eq:ultimate>} is another observation, for the first time in weighted networks, of the fact that ensemble equivalence can be broken by the presence of an extensive number of local constraints, even away from phase transitions.

Coming to the relative fluctuations of the constraints, we see from Eqs.~\eqref{eq:flutt} and~\eqref{eq:rho+-} that
\begin{eqnarray}
r_\pm^*(T\!>\!T_c)&=&\frac{\sqrt{\Sigma_\pm^*(T\!>\!T_c)}}{s^*_\pm(T\!>\!T_c)}\nonumber\\
&\approx&\sqrt{\frac{n+s^*_\pm(T\!>\!T_c)}{n\,s^*_\pm(T\!>\!T_c)}}\nonumber\\
&=&O\left(\frac{1}{\sqrt{n}}\right)\label{eq:flutt>}
\end{eqnarray}
and
\begin{equation}
\rho^*_\pm(T\!>\!T_c)=0.
\label{eq:rho+->}
\end{equation}
We therefore observe that in the non-condensed phase the decay of the relative fluctuations of each constraint is of the same order $O(1/\sqrt{n})$ as generally observed for the global constraint (total energy) in a system with short-range interactions away from phase transitions. 
However, while in the traditional situation the vanishing of the relative fluctuations implies that the relative entropy is subextensive and that the relative entropy density vanishes in the thermodynamic limit (as discussed in Sec.~\ref{sec:math}), here the relative entropy density does not vanish and the ensembles are not equivalent.
Therefore we find that, \emph{in systems with an extensive number of local constraints, the vanishing of even all the relative fluctuations does not ensure ensemble equivalence.}

\subsubsection{Infinite temperature: $T\!\to\!\infty$\label{sec:infinite}}
The extreme regime of the non-condensed phase is the infinite-temperature case, which can be explored by taking the limit $T\!\to\!\infty$ in the solution to the `direct' problem provided by Eq.~\eqref{eq:mimmo>}.
In such a limit, Eq.~\eqref{eq_epsilon} implies that $\beta_+^*(T\!>\!T_c)$ and $\beta_-^*(T\!>\!T_c)$ converge to the same value $\beta^*_{\infty}$ given by
\begin{equation}
\beta_+^*(T\!\to\!\infty)=\beta_-^*(T\!\to\!\infty)=\beta^*_{\infty}\equiv\frac{1}{2}\ln\left(1+\frac{1}{w^*}\right).
\label{eq:equalbeta}
\end{equation}
Then, through Eqs.~\eqref{w+},~\eqref{w-},~\eqref{w0},~\eqref{pijpm} and~\eqref{pij0}, we get
\begin{eqnarray}
w^*_+(T\!\to\!\infty)=w^*_-(T\!\to\!\infty)=w^*_0(T\!\to\!\infty)&=&w^*,\qquad
\label{eq:allequal}\\
p^*_+(T\!\to\!\infty)=p^*_-(T\!\to\!\infty)=p^*_0(T\!\to\!\infty)&=&p^*,
\end{eqnarray}
i.e. all node pairs have the same expected link weight $w^*$ and connection probability $p^*$ given by
\begin{equation}
p^*=\frac{w^*}{1+w^*}.
\end{equation}
This is the characteristic situation in the infinite-temperature limit of Bose-Einstein statistics, where each particle is equally likely distributed across all energy levels. Here, this situation translates in the graph becoming completely homogeneous: the distinction between core and periphery disappears as the finite difference between energy levels becomes entirely dominated by the diverging temperature.
The expected strength of every node has the same value $s^*\equiv (n-1)w^*$:
\begin{equation}
s_+^*(T\!\to\!\infty)=s^*_-(T\!\to\!\infty)=s^*=(n-1)w^*,
\label{eq:allequals}
\end{equation}
i.e. the strength sequence becomes a constant vector.

Clearly, the above result does not change the value of the order parameter in Eq.~\eqref{eq:Q>}:
\begin{equation}
Q^*(T\!\to\!\infty)=0.
\end{equation}
Similarly, the final results in eqs.~\eqref{eq:ultimate>} and~\eqref{eq:rho+->} about the simultaneous breakdown of ensemble equivalence and the vanishing of all relative fluctuations carry over to the infinite-temperature limit, so in principle we do not have to further discuss this case. 
However, the fact that the strength sequence becomes a constant vector allows us to calculate many of the properties of the model exactly, so this example is a very transparent and instructive one. It is therefore worth considering it in some more detail, also because some of the following results will be useful in the (much less trivial) zero-temperature limit as well.

In particular, Eqs.~\eqref{eq:equalbeta} and~\eqref{eq:allequals} imply that Eqs.~\eqref{covariance-matrix-dewd-diagonal} and~\eqref{non-diagonal-matrix-dewd} can be rewritten as
\begin{equation}\label{dewd-cov-dagonal-parameter}
{\Sigma}^*_{ii}(T\!\to\!\infty)=\frac{(n-1)e^{-{2\beta^*_{\infty}}}}{(1-e^{-{2\beta^*_{\infty}}})^2}
=w^*(1+w^*)(n-1)
\end{equation}
for all $i=1,n$ and 
\begin{equation}\label{Non-diagonal-1}
{\Sigma}^*_{ij}(T\!\to\!\infty)=\frac{e^{-{2\beta^*_{\infty}}}}{(1-e^{-{2\beta^*_{\infty}}})^2}
=w^*(1+w^*)
\end{equation}
for all $i\ne j$ respectively.
In Appendix~\ref{app:infinite} we show that the above expressions can be used to calculate the determinant of $\mathbf{\Sigma}^*(T\!\to\!\infty)$ exactly as
\begin{equation}\label{determina-sigma-mn-decompose-1_main}
\det \mathbf{\Sigma}^*(T\!\to\!\infty)=2(n-1)(n-2)^{n-1}\left[w^*(1+w^*)\right]^n,
\end{equation}
from which we confirm, without having made the approximation in eq.~\eqref{det-production_main}, that
\begin{equation}
\ln\det \mathbf{\Sigma}^*(T\!\to\!\infty)=O(n\ln n)\label{stogranca}
\end{equation}
and that
\begin{equation}
D^*(T\!\to\!\infty)\approx\frac{1}{2}\ln \det\mathbf{\Sigma}^*(T\!\to\!\infty)\approx\frac{1}{2}n\ln n.
\label{eq_KLleadinginfty}
\end{equation}
Clearly, Eqs.~\eqref{eq:Scann2>} and~\eqref{eq:Smicn2>} hold in this limit as well:
\begin{equation}
S^*_{\textrm{can}}(T\!\to\!\infty)=O(n^2),\, S^*_{\textrm{mic}}(T\!\to\!\infty)=O(n^2).
\label{eq:Sn2infty}
\end{equation}

Finally, from eqs.~\eqref{eq:allequals} and~\eqref{dewd-cov-dagonal-parameter} we see that eq.~\eqref{eq:flutt} leads in this case to a unique value for the coefficient of variation of all the strengths:
\begin{equation}
r_i^*(T\!\to\!\infty)=\frac{\sqrt{\Sigma_{ii}^*(T\!\to\!\infty)}}{s_i^*(T\!\to\!\infty)}=\sqrt{\frac{1+w^*}{(n-1)w^*}}\quad\forall i,
\end{equation}
so that
\begin{equation}
\rho^*_i(T\!\to\!\infty)=0\quad\forall i,
\end{equation}
in accordance with eq.~\eqref{eq:rho+->}.

\subsection{Condensed phase\label{sec:condensed}}
We now consider the BEC phase $(T\!<\!T_c)$. We first derive general results and then discuss the finite-temperature case and the zero-temperature limit separately.

By the definition in Eq.~\eqref{eq:orderBEC0}, the condensed phase must be such that a positive fraction $Q^*(T\!<\!T_c)>0$ of the total weight lies in the core, i.e. (to leading order)
\begin{equation}
W^*_+(T\!<\!T_c) \approx Q_n^*(T\!<\!T_c)W^*\approx Q^*(T\!<\!T_c)W^*
\label{eq:emosocazzi}
\end{equation}
which necessarily means 
\begin{equation}
w^*_+(T\!<\!T_c)=O(n^2),\quad W^*_+(T\!<\!T_c) =O(n^2)
\label{eq_w+}
\end{equation}
and $p^*_+(T\!<\!T_c)\approx 1$, i.e. the core does not have missing links (the presence of core-core links is no longer a random event, while the weight of such links is still a random variable).
As expected, BEC corresponds to the divergence of $w^*_+(T\!<\!T_c)$, and we now see that the speed of this divergence is of order $n^2$ in our model.
For convenience, we may define
\begin{equation}
\psi^*_+(T\!<\!T_c)=\lim_{n\to\infty}\frac{w^*_+(T\!<\!T_c)}{n^2}
\label{eq:psi}
\end{equation}
which is finite and positive, so that
\begin{equation}
w^*_+(T\!<\!T_c)\approx \psi^*_+(T\!<\!T_c) n^2.
\label{eqwpsi}
\end{equation}
Combining Eqs.~\eqref{w0} and~\eqref{eq_w+} we see that
\begin{eqnarray}
w^*_0(T\!<\!T_c)\approx\frac{1}{\sqrt{1+1/w^*_-(T\!<\!T_c)}-1}\label{attila},
\end{eqnarray}
which inserted into Eq.~\eqref{eq:components} shows that, to leading order,
\begin{equation}
w^*\approx n_+(n_+-1)\psi^*_+(T\!<\!T_c)+w^*_-(T\!<\!T_c),
\label{panzanella}
\end{equation}
an expression that relates the finite parameters of the model with each other in the condensed phase.
Therefore we see from Eq.~\eqref{eq:emosocazzi} that
\begin{equation}
w^*_-(T\!<\!T_c)\approx [1-Q_n^*(T\!<\!T_c)]w^*
\label{eq_w-}
\end{equation}
and
\begin{equation}
W^*_-(T\!<\!T_c)\approx [1-Q_n^*(T\!<\!T_c)]W^*.
\label{eq_W-}
\end{equation}
Inserting Eq.~\eqref{eq_w-} into Eq.~\eqref{attila} yields
\begin{eqnarray}
w^*_0(T\!<\!T_c)\approx\frac{1}{\sqrt{1+\frac{1}{ [1-Q_n^*(T\!<\!T_c)]w^*}}-1}\label{eq_w0}
\end{eqnarray}
and
\begin{eqnarray}
W^*_0(T\!<\!T_c)\approx\frac{n_+n_-}{\sqrt{1+\frac{1}{ [1-Q_n^*(T\!<\!T_c)]w^*}}-1}\label{eq_W0}.
\end{eqnarray}
The above expressions show that neither $w^*_-(T\!<\!T_c)$ nor $w^*_0(T\!<\!T_c)$ diverge, indicating that BEC occurs only in the ground state and that $p^*_-(T\!<\!T_c)<1$ and $p^*_0(T\!<\!T_c)<1$, i.e. there can be missing links in the periphery and between core and periphery.
Moreover, we see that $W^*_0(T\!<\!T_c)$ is subleading with respect to both $W^*_+(T\!<\!T_c)$ and $W^*_-(T\!<\!T_c)$: although individual core-periphery links have an expected weight $w^*_0(T\!<\!T_c)$ larger than the expected weight $w^*_-(T\!<\!T_c)$ of individual periphery-periphery links, the number $n_+n_-$ of core-periphery links is of smaller order with respect to the number $n_-(n_--1)/2$ of periphery-periphery links, and as a result the total weight of all core-periphery links is of smaller order as well. 

To obtain $s^*_\pm(T\!<\!T_c)$ to leading order, we use Eqs.~\eqref{s_hub} and~\eqref{s--}:
\begin{eqnarray}
s^*_+(T\!<\!T_c)&=&(n_+-1)w^*_+(T\!<\!T_c)+n_-w^*_0(T\!<\!T_c)\nonumber\\
&\approx&(n_+-1)\psi^*_+(T\!<\!T_c)n^2,
\label{s*+<}
\end{eqnarray}
\begin{eqnarray}
s^*_-(T\!<\!T_c)&=&(n_--1)w^*_-(T\!<\!T_c) +n_+w^*_0(T\!<\!T_c)\nonumber\\
&\approx&n\,w^*[1-Q_n^*(T\!<\!T_c)].
\label{s*-<}
\end{eqnarray}
Now, combining the above expressions, we see that the order parameter defined in Eq.~\eqref{eq:order} can be written as
\begin{eqnarray}
Q^*(T\!<\!T_c)&=&\lim_{n\to\infty}\frac{W_+^*(T\!<\!T_c)}{W_+^*(T\!<\!T_c)+W_-^*(T\!<\!T_c)}\nonumber\\
&=&\lim_{n\to\infty}\frac{1}{1+\frac{n_-(n_--1)w^*_-(T\!<\!T_c)}{n_+(n_+-1)w^*_+(T\!<\!T_c)}}\nonumber\\
&=&\lim_{n\to\infty}\frac{1}{1+\frac{n_-s^*_-(T\!<\!T_c)}{n_+s^*_+(T\!<\!T_c)}}\nonumber\\
&=&\frac{1}{1+\frac{w^*_-(T\!<\!T_c)}{n_+(n_+-1)\psi^*_+(T\!<\!T_c)}}\nonumber\\
&=&\frac{n_+(n_+-1)\psi^*_+(T\!<\!T_c)}{w^*}>0.
\label{eq:orderBEC}
\end{eqnarray}
Besides quantifying the order parameter, the above calculation shows that, since the value of $Q^*(T\!<\!T_c)$ only depends on the values of $s^*_+(T\!<\!T_c)$ and $s^*_-(T\!<\!T_c)$ (which by construction are the same in the canonical and microcanonical ensembles), whenever one ensemble is in the BEC phase, the other ensemble is the BEC phase as well, for all temperatures $T\!<\!T_c$.
As for the non-condensed, this allows us to refer to the conjugate canonical and microcanonical ensembles being `in the same phase' consistently. 
Inverting Eq.~\eqref{eq:orderBEC}, we can also express the parameter $\psi^*_+(T\!<\!T_c)$ in terms of the order parameter as follows:
\begin{equation}
\psi^*_+(T\!<\!T_c)=\frac{w^*\,Q^*(T\!<\!T_c)}{n_+(n_+-1)}.\label{arosicone}
\end{equation}

The `inverse' problem is solved by inverting Eqs.~\eqref{s*+<} and~\eqref{s*-<} and using them into Eq.~\eqref{eq_betas} to get
\begin{eqnarray}
\beta_+^*(T\!<\!T_c)
&\approx&\frac{1}{2}\ln\left(1+\frac{1}{n^2 \psi^*_+(T\!<\!T_c)}\right)\nonumber\\
&\approx&\frac{1}{2n^2 \psi^*_+(T\!<\!T_c)}\nonumber\\
&\approx&\frac{n_+-1}{2s^*_+(T\!<\!T_c)}\nonumber\\
&\approx&\frac{n_+(n_+-1)}{2n^2 w^*\,Q_n^*(T\!<\!T_c)},\label{alimo}
\end{eqnarray}
\begin{eqnarray}
\beta_-^*(T\!<\!T_c)&=&\frac{1}{2}\ln\left(1+\frac{1}{w^*_-(T\!<\!T_c)}\right)\nonumber\\
&\approx&\frac{1}{2}\ln\left(1+\frac{n}{s^*_-(T\!<\!T_c)}\right)\nonumber\\
&\approx&\frac{1}{2}\ln\left(1+\frac{1}{w^*\left[1-Q_n^*(T\!<\!T_c)\right]}\right).\label{taccitu}
\end{eqnarray}
The above equations solve the inverse problem by expressing $\beta^*_\pm(T\!<\!T_c)$ as a function of the constraints $s^*_\pm(T\!<\!T_c)$, which can in turn be expressed either in terms of the finite parameters $\psi^*_+(T\!<\!T_c)$ and $w^*_-(T\!<\!T_c)$ or in terms of $Q_n^*(T\!<\!T_c)$ and the temperature-independent parameter $w^*$. 

Again, the `direct' problem requires finding the chemical potential $\mu^*(T)$ as a function of $\epsilon^*$, $w^*$ and $T$.
From Eq.~\eqref{eq:mimmo0} we get
\begin{eqnarray}
\mu^*(T\!<\!T_c)&=&-2kT\beta^*_+(T\!<\!T_c)\nonumber\\
&\approx&-\frac{kT}{n^2\psi^*_+(T\!<\!T_c)}\nonumber\\
&\approx&-kT\frac{n_+(n_+-1)}{n^2 w^*\,Q_n^*(T\!<\!T_c)}.
\label{eq:mimmo<}
\end{eqnarray}

We now consider the variance of the constraints. Inserting Eqs.~\eqref{eq_w+},~\eqref{eq_w-} and~\eqref{eq_w0} into Eqs.~\eqref{eq:squares}, we immediately see that
\begin{eqnarray}
\Sigma_{+}^*(T\!<\!T_c)&\approx&(n_+-1){[\psi^*_+(T\!<\!T_c) ]}^2 n^4,\label{tumadre}\\
\Sigma_{-}^*(T\!<\!T_c)&\approx&w^*_-(T\!<\!T_c)[1+w^*_-(T\!<\!T_c)]n.\label{tupadre}
\end{eqnarray}

\subsubsection{Finite (subcritical) temperature: $0\!<\!T\!<\!T_c$\label{scialla<}}
In this regime we have 
\begin{equation}
0<Q^*(0\!<\!T\!<\!T_c)<1,
\end{equation}
which, as clear from Eqs.~\eqref{eq_w-} and~\eqref{eq_w0}, implies 
\begin{eqnarray}
w^*_-(0\!<\!T\!<\!T_c)&\approx& [1-Q^*(0\!<T\!<\!T_c)]w^*,\label{eq:anvedi1}\\
w^*_0(0\!<\!T\!<\!T_c)&\approx&\frac{1}{\sqrt{1+\frac{1}{ [1-Q^*(0\!<T\!<\!T_c)]w^*}}-1},\label{eq:anvedi2}
\end{eqnarray}
where both quantities are $O(1)$.
Using these results, it is possible to show (see Appendix~\ref{app:condensed}) that the leading term of the determinant of the covariance matrix between the constraints is 
\begin{equation}
\det[\mathbf{\Sigma}^*(0\!<\!T\!<\!T_c)]=O(n^{n+3n_+}),\label{jack}
\end{equation}
implying
\begin{equation}
\ln\det[\mathbf{\Sigma}^*(0\!<\!T\!<\!T_c)]=O(n\ln n),\label{stoca}
\end{equation}
which is the same scaling found in Eq.~\eqref{eq:ultimate>} for the non-condensed phase.
The criterion for measure equivalence in Eq.~\eqref{eq:ultimate} is again violated, showing that ensemble equivalence does not hold in the condensed case as well. 
The leading term of the relative entropy can still be calculated exactly from Eq.~\eqref{eq_KLleading} and is the same as the one found in Eq.~\eqref{eq_KLleading>} for the non-condensed phase:
\begin{equation}
D^*(0\!<\!T\!<\!T_c)\approx\frac{1}{2}\ln \det\mathbf{\Sigma}^*(0\!<\!T\!<\!T_c)\approx\frac{1}{2}n\ln n.
\label{eq_KLleading<}
\end{equation}
Similarly, the canonical entropy is still of order $O(n^2)$, as can be seen by inserting Eqs.~\eqref{s*+<},~\eqref{s*-<},~\eqref{alimo} and~\eqref{taccitu} into Eq.~\eqref{entropy+-}.
We therefore retrieve
\begin{eqnarray}
S^*_{\textrm{can}}(0\!<\!T\!<\!T_c)&=&O(n^2),\label{eq:Sn2<1}\\
S^*_{\textrm{mic}}(0\!<\!T\!<\!T_c)&=&O(n^2).\label{eq:Sn2<2}
\end{eqnarray}

Coming to the relative fluctuations, from Eqs.~\eqref{eq:flutt},~\eqref{eq:rho+-},~\eqref{tumadre} and~\eqref{tupadre} we obtain
\begin{eqnarray}
\rho_+^*(0\!<\!T\!<\!T_c)&=&\frac{1}{\sqrt{n_+-1}},\label{eq:rho+>}\\
\rho_-^*(0\!<\!T\!<\!T_c)&=&0.\label{eq:rho->}
\end{eqnarray}
The above result can be interpreted as follows. The term $\sum_{i\neq j}(\langle w_{ij}\rangle_{\vec{\beta}^*})^2/(\sum_{i\neq j}\langle w_{ij}\rangle_{\vec{\beta}^*})^2$ in Eq.~\eqref{ratio-variance-streng1} is an inverse participation ratio, taking values in the range $[(n-1)^{-1},1]$ and quantifying the inverse of the effective number of link weights contributing to the strength of node $i$~\cite{squartini2015unbiased}. 
Here, for a node in the core, there is a finite number $n_+-1$ of dominant link weights, each equal to $w^*_+=O(n^2)$, while the remaining $n_-$ weights are of smaller order. Taking the thermodynamic limit, these $n_+-1$ dominant weights lead to the value for $\rho_+^*$ in Eq.~\eqref{eq:rho+>}.
By contrast, for a node in the periphery, all the expected link weights are of the same order, so the inverse participation ratio, and consequently the value of $\rho_-^*$ in Eq.~\eqref{eq:rho->}, vanishes.
It should be noted that, even if the expected strength of the core nodes is much bigger than that of the periphery nodes, the relative fluctuations of the core nodes do not vanish, while those of the peripheral nodes do.

The fact that BEC occurs necessarily among the core nodes confirms that the units of the system are the nodes, and not the node pairs: the `ground state pairs' are necessarily all and only the pairs of `ground state nodes'. Indeed, one cannot decide arbitrarily which node pairs form the degenerate ground state where condensation occurs. This would have been possible only if node pairs were the fundamental units, by assigning the same degenerate ground state energy value to any set of node pairs (including pairs not necessarily involving the same set of nodes). For instance, it would have been possible to include the pairs $(i,j)$ and $(i,k)$, without necessarily including the pair $(j,k)$, in the degenerate ground state (which is instead unavoidable in our system).

\subsubsection{Zero temperature: $T\!\to\! 0$}
We finally consider the zero-temperature limit as the extreme case of the condensed phase. Importantly, we have to be careful how we approach the two limits $T\!\to\! 0$ and $n\to \infty$. Indeed, we are going to show that taking the limit $T\!\to\! 0$ while $n$ is kept fixed leads to results that cannot be subsequently carried over to the thermodynamic limit by taking the limit $n\to\infty$. Since we are interested precisely in the thermodynamic limit, we have to take a different route. To show the difference, we consider the zero-temperature limit first in the case of finite $n$ and then in the case of growing $n$.

If $n$ is finite, the zero-temperature limit simply represents the situation where the only populated state is the degenerate ground state corresponding to the links in the core, i.e. 
\begin{equation}
Q_n^*(T\!\to\! 0)=1
\end{equation}
by construction. All other links are not present.
As usual, if the ground state is not degenerate, then the microcanonical entropy is zero, while if the ground state is degenerate, then the microcanonical entropy approaches a value called \emph{residual entropy} which, for a system of fixed size, is a constant that depends only on the degeneracy (these statements usually go under the names of \emph{Third Law of Thermodynamics} or, somehow improperly, \emph{Nernst Theorem})~\cite{kittel}.
In our setting the ground state is non-degenerate only if $n_+=2$, in which case the link between the two core nodes is the one with minimum energy. 
In the general case $n_+>2$, the ground state is degenerate and both the microcanonical and canonical entropies are strictly positive. 
In any case, for finite $n$ the zero-temperature limit is characterized by the fact that, in both the canonical and microcanonical ensembles, all nodes in the periphery are \emph{deterministically} isolated, i.e. necessarily isolated in all realizations of the network. The periphery becomes completely disconnected, both internally and from the core. 
Note that this is one of the degenerate situations (mentioned in Sec.~\ref{sec:math}) where, even if the constraints are in principle all mutually independent, for certain degenerate parameter value(s) some of them become `hard' in both ensembles, thereby not contributing any difference between the two ensembles. Note that if we simply take the ideal limit $n\to\infty$ starting from this zero-temperature state, we would be considering the degenerate situation where an infinite number of isolated `peripheral' nodes are added to the fully connected core. These nodes are unavoidably disconnected in both ensembles, so their contribution to the system is purely formal. The only variability (hence the only possible source of nonequivalence) comes from the core, which keeps having a finite number $n_+$ of nodes: as an extreme signature of BEC, the condensate behaves as an effectively lower-dimensional object.

In order to access the thermodynamic limit, we therefore have to consider from the beginning the case where $n$ can grow indefinitely. 
We are going to show that the main difference arises from the fact that the temperature can only correspond to graphical strength sequences, which on the other hand depend on $n$. Therefore one should expect a certain $n$-dependent temperature $T_n$. At that point, a temperature value $T_n>0$ that is small but finite when $n$ is finite (allowing for certain populated excited states besides the ground state) may actually approach zero as $n$ diverges, i.e. $\lim_{n\to\infty}T_n=0$. The corresponding excited states will effectively become part of the accessible configurations in the zero-temperature limit and contribute an extra residual entropy. 

To study this scenario, we start from the consideration that if the two limits $T\!\to\! 0$ and $n\to \infty$ were taken simultaneously for quantities that depend on both $Q^*_n$ and $n$, e.g. terms such as $n(1-Q^*_n)$, we would encounter indeterminate expressions. We therefore need to understand how, as $T$ goes to zero, $Q^*_n$ goes to one as a function of $n$, for $n$ large.
We recall that our starting point is always the value $\vec{s}^*(T)$ of the constraints. The temperature $T$ is a parameter that allows us to vary $\vec{s}^*(T)$, while keeping it graphic, i.e. realizable in at least one configuration of the network. For large $n$, we therefore have to identify the possible states of the network, hence the values of $\vec{s}^*(T)$, closest to zero temperature, i.e. when $T\!\simeq\!0$ (we will use the symbol `$\simeq$' to denote this near-zero-temperature behaviour of any quantity, thereby keeping the notation distinct from the symbol `$\approx$' that will still denote the leading order of any quantity for large $n$). This is easily done by realizing that, if we start from some $\vec{s}^*(T\!>\!0)$ and decrease the temperature towards zero, the lowest excited state accessible to the network (before all links condense in the core) is one where only the smallest possible number $\Delta W^*$ of the $W^*$ particles of weight remain out of the core, while keeping the strength sequence $\vec{s}^*(T\!\simeq\! 0)$ in the form given by Eq.~\eqref{eq_bis}. This state is necessarily such that $s_-^*(T\!\simeq\!0)=1$ (which is the minimum non-zero value of the strength, recalling that the strength is a non-negative integer by construction) and can be realized in multiple ways: either by connecting the $n_-$ peripheral nodes in pairs, thus creating $n_-/2$ periphery-periphery links of unit weight and energy $\epsilon^*$ (in which case $\Delta W^*=n_-/2$), or by connecting each peripheral node to a core node, thus creating $n_-$ core-periphery links of unit weight and energy $\epsilon^*/2$ (in which case $\Delta W^*=n_-$), or finally by combining both types of situations. Recalling the discussion in Sec.~\ref{sec:corper}, in all cases the $\Delta W^*$ links outside of the core have collectively the same energy $\epsilon^*n_-/2$ while the links in the core have zero energy; indeed, all such configurations are equiprobable. If we also consider the next excited states with $s^*_-(T\!\simeq\!0)=2, 3, \dots$, in general we will have $\Delta W^*(T\!\simeq\! 0)=n_-\ell/2$ where $\ell$ is a small (in a sense that will be clear in a moment) integer. 

The above considerations imply that the possible values of $Q^*_n$ close to zero temperature are of the form
\begin{eqnarray}
Q^*_n(T\!\simeq\! 0)&=& \frac{W^*-\Delta W^*(T\!\simeq\! 0)}{W^*}\nonumber\\
&=&1-\frac{n_-\ell/2}{W^*}\nonumber\\
&\approx&1-\frac{\ell}{n w^*}.
\label{eq:gioiapura}
\end{eqnarray}
Basically, the above expression makes it explicit that, since the strength is a discrete quantity, technically the temperature can only take discrete values in order to keep the strength sequence graphic, so the role of $T$ is taken up by $\ell$ (which is an integer) and a low temperature corresponds to a `small', i.e. finite or at most $o(n)$, value of $\ell$. 
Indeed, in the thermodynamic limit we recover
\begin{equation}
Q^*(T\!\simeq\! 0)=\lim_{n\to\infty} Q^*_n(T\!\simeq\! 0)=1,
\end{equation}
confirming that $\ell=o(n)$ leads to the correct zero-temperature limit. At the same time, Eq.~\eqref{eq:gioiapura} shows that, to recover any finite-temperature value $Q^*(T\!>\!0)<1$, we would need $\ell$ to grow linearly in $n$ in the thermodynamic limit (note that $\ell$ cannot grow faster than $n$, because $\Delta W^*$ cannot grow faster than $n^2$, which is the order of $W^*$).
The $\ell=o(n)$ regime considered here is therefore genuinely different from the positive-temperature cases discussed so far. 

Having characterized the zero-temperature limit in this way, we can calculate
\begin{equation}
\lim_{n\to\infty} n [1-Q^*_n(T\!\simeq\! 0)]=\frac{\ell}{w^*},
\end{equation}
from which we can obtain various asymptotic expressions.
Indeed, from Eqs.~\eqref{eq_w-} and~\eqref{eq_W-}
we obtain
\begin{equation}
w^*_-(T\!\simeq\! 0)\approx [1-Q^*_n(T\!\simeq\!0)]w^*\approx\frac{\ell}{n}
\label{eq_w-n}
\end{equation}
and
\begin{equation}
W^*_-(T\!\simeq\!0)\approx\frac{\ell n}{2}.
\label{eq_W-n}
\end{equation}
Similarly, expanding Eqs.~\eqref{eq_w0} and~\eqref{eq_W0} for $Q_n^*$ close to 1, we get
\begin{eqnarray}
w^*_0(T\!\simeq\! 0)\approx\sqrt{[1-Q^*_n(T\!\simeq\!0)]w^*}\approx \sqrt{\frac{\ell}{n}}\label{eq_w0n}
\end{eqnarray}
and
\begin{eqnarray}
W^*_0(T\!\simeq\! 0)\approx  n_+\sqrt{\ell n}\label{eq_W0n}.
\end{eqnarray}
Since both $W^*_-(T\!\simeq\! 0)$ and $W^*_0(T\!\simeq\! 0)$ are subleading with respect to $W^*_+(T\!\simeq\! 0)$, we have $W^*_+(T\!\simeq\! 0)\approx W^*$ which can be rewritten as $w_+^*(T\!\simeq\!0)n_+(n_+-1)/2\approx w^*n(n-1)/2$. This implies
\begin{equation}
w^*_+(T\!\simeq\!0)\approx\frac{w^*n^2}{n_+(n_+-1)}
\label{eq_w+n}
\end{equation}
and, from Eq.~\eqref{eq:psi},
\begin{equation}
\psi^*_+(T\!\simeq\!0)=\frac{w^*}{n_+(n_+-1)},
\label{eq:psi0}
\end{equation}
consistently with Eq.~\eqref{arosicone}.

We now note that, inserting Eqs.~\eqref{eq_w-n} and~\eqref{eq_w+n} into Eq.~\eqref{eq:mimmo2}, we obtain the anticipated dependence of $T_n$ (for $T_n\simeq 0$) on $n$:
\begin{equation}
e^{-\epsilon^*/kT_n}=\frac{w^*_-(T_n\!\simeq\! 0)[1+w^*_+(T_n\!\simeq\! 0)]}{w^*_+(T_n\!\simeq\! 0)[1+w^*_-(T_n\!\simeq\! 0)]}\approx\frac{\ell}{n}.
\end{equation}
Inverting, we find how the temperature approaches zero as $n$ grows:
\begin{equation}
T_n\approx \frac{\epsilon^*}{k\ln (n/\ell)}\approx \frac{\epsilon^*}{k\ln n}.
\label{eq:Ta0}
\end{equation}
The above result, which is independent of $\ell$, connects the thermodynamic limit $n\to\infty$ with the zero-temperature limit $T\to 0$ in our setting and confirms that it would be inappropriate to first identify the ground state are the core links by taking the limit $T\to 0$ and subsequently let $n$ grow. On the contrary, the zero-temperature state turns out to be the entire set of configurations obtained displacing a certain number of units of weight out of the core and such that $\ell=o(n)$. Inserting Eqs.~\eqref{eq_w-n},~\eqref{eq_w0n} and~\eqref{eq_w+n} into Eqs.~\eqref{pijpm} and~\eqref{pij0} we can characterize these configurations through the connection probabilities
\begin{eqnarray}
p^*_+(T\!\simeq\! 0)&\approx& 1-\frac{n_+(n_+-1)}{w^*n^2},\label{pij+0}\\
p^*_-(T\!\simeq\! 0)&\approx& \frac{\ell}{n},\label{pij-0}\\
p^*_0(T\!\simeq\! 0)&\approx& \sqrt{\frac{\ell}{n}},\label{pij00}
\end{eqnarray}
which in the thermodynamic limit behave as expected for the ground state, i.e. $p^*_+(T\!\to\! 0)=1$, $p^*_-(T\!\to\! 0)=0$ and $p^*_0(T\!\to\! 0)=0$.

Using Eqs.~\eqref{eq_w-n} and~\eqref{eq_w0n}, we can calculate the strengths from Eqs.~\eqref{s*+<} and~\eqref{s*-<} as
\begin{eqnarray}
s^*_+(T\!\simeq\! 0)&\approx&(n_+-1)\psi^*_+(T\!\simeq\! 0)n^2\approx\frac{w^*}{n_+}n^2\label{s+simeq}\\
s^*_-(T\!\simeq\! 0)&\approx& n w_-^*(T\!\simeq\! 0)+n_+ w^*_0(T\!\simeq\! 0)\approx \ell.\label{s-simeq}
\end{eqnarray}
The `inverse' problem is solved by Eqs.~\eqref{alimo} and~\eqref{taccitu}, which now become
\begin{eqnarray}
\beta_+^*(T\!\simeq\!0)&\approx&\frac{n_+(n_+-1)}{2n^2 w^*},\label{alimo0}\\
\beta_-^*(T\!\simeq\!0)&\approx&\frac{1}{2}\ln \frac{n}{\ell}\approx\frac{1}{2}\ln n.\label{taccitu0}
\end{eqnarray}
By contrast, the solution to the `direct' problem is given through the chemical potential, obtained inserting Eqs.~\eqref{eq:Ta0} and~\eqref{alimo0} into Eq.~\eqref{eq:mimmo<}:
\begin{eqnarray}
\mu^*(T\!\simeq\!0)&=&-2kT\beta^*_+(T\!\simeq \!0)\nonumber\\
&\approx&-\frac{\epsilon^* n_+(n_+-1)}{w^* n^2\ln n}.
\label{eq:mimmo<0}
\end{eqnarray}

The variances of the constraints can be calculated inserting Eqs.~\eqref{eq_w-n} and~\eqref{eq:psi0} into Eqs.~\eqref{tumadre} and~\eqref{tupadre}. This yields
\begin{eqnarray}
\Sigma^*_+(T\!\simeq\! 0)&\approx&\frac{(w^*)^2}{n^2_+(n_+-1)}n^4,\label{sigma+simeq}\\
\Sigma^*_-(T\!\simeq\! 0)&\approx& \ell.\label{sigma-simeq}
\end{eqnarray}
Using the above relationships, it is possible to show (see Appendix~\ref{app:zero}) that the leading order of the determinant of the covariance matrix is
\begin{equation}
\det[\mathbf{\Sigma}^*(T\!\simeq\!0)]=O(n^{4n_+}\ell^{n-n_+})\label{jack2}.
\end{equation}
This leads to
\begin{equation}
\ln\det[\mathbf{\Sigma}^*(T\!\simeq\!0)]=O(n\ln\ell),\label{picchio}
\end{equation}
which is of smaller order compared to the scaling $O(n\ln n)$ found in Eqs.~\eqref{eq:ultimate>},~\eqref{stogranca} and~\eqref{stoca} for all positive temperatures, but still signalling the breakdown of ensemble equivalence.
Equation~\eqref{picchio} implies that the requirements ensuring the validity of Eq.~\eqref{eq_KLleading} are not met, therefore in this case the leading term of the relative entropy cannot be calculated exactly. However, the leading order is still given by Eq.~\eqref{eq:Sorder}
\begin{equation}
D^*(T\!\simeq\!0)=O\big(\ln\det \mathbf{\Sigma}^*(T\!\simeq\!0)\big)=O(n\ln\ell).
\end{equation}

The canonical entropy $S^*_{\textrm{can}}(T\!\simeq\!0)$ can be calculated by using Eqs.~\eqref{alimo0} and~\eqref{taccitu0} to rewrite the sum of the five terms appearing into Eq.~\eqref{entropy+-} as the following sum:
\begin{eqnarray}
 S^*_{\textrm{can}}(T\!\simeq\!0)&=&\frac{n_+(n_+-1)}{2}+\frac{\ell}{2}n\ln n+n_+(n_+-1)\ln n\nonumber\\
&&+\frac{\ell}{2}n+n_+\sqrt{n\ell}=O(\ell n\ln n),
\end{eqnarray}
which, unless $\ell=O(n/\ln n)$ or bigger, is different from the scaling $O(n^2)$ obtained for finite temperatures in Eqs.~\eqref{eq:Scann2>} and~\eqref{eq:Sn2<1}.
This slower increase of the canonical entropy with $n$ confirms that in the zero-temperature limit the system behaves as a Bose-Einstein condensate, a phenomenon that determines a strong reduction in the dimensionality of the space of allowed configurations. 
Note that since the relative entropy is of order given by Eq.~\eqref{picchio}, which is smaller than the order of $S^*_{\textrm{can}}(T\!\simeq\!0)$, the leading term of the microcanonical entropy must be
\begin{equation}
 S^*_{\textrm{mic}}(T\!\simeq\!0)\approx \frac{\ell}{2}n\ln n=O(\ell n\ln n).
\end{equation}

Combining Eqs.~\eqref{s+simeq}-\eqref{sigma-simeq}, we finally obtain the relative canonical fluctuations
\begin{eqnarray}
\rho^*_+(T\!\simeq\! 0)&=&\frac{\sqrt{n_+}}{n_+-1},\label{eq:rho+>0}\\
\rho^*_-(T\!\simeq\! 0)&=& \frac{1}{\sqrt{\ell}},\label{eq:rho->0}
\end{eqnarray}
which differ from the results in Eqs.~\eqref{eq:rho+>} and~\eqref{eq:rho->} obtained in the (subcritical) finite-temperature case. In particular, now both $\rho^*_-(T\!\simeq\! 0)$ and $\rho^*_-(T\!\simeq\! 0)$ (if $\ell$ is finite) are non-zero. Note that, while Eq.~\eqref{eq:rho->} can be formally retrieved from Eq.~\eqref{eq:rho->0} by letting $\ell$ grow linearly in $n$, Eq.~\eqref{eq:rho+>} cannot be retrieved from Eq.~\eqref{eq:rho+>0}, because the assumption $\ell=o(n)$ has already been exploited in the derivation.

\subsection{Critical temperature: $T\!=\!T_c$}
Having characterized the model in all regimes, we can now discuss more easily what happens right at the critical temperature $T\!=\!T_c$. Clearly, a very interesting question is whether the phase transition is of first or second order. A first-order phase transition is obtained when the order parameter $Q^*(T)$ jumps discontinuously from zero to a strictly positive value as $T$ is lowered through $T_c$. In such a case, the left and right limits of $Q^*(T)$ at $T\!=\!T_c$ are different:
\begin{equation}
Q^*(T\!\to\! T_c^-)>Q^*(T\!\to\! T_c^+)=0.
\label{eq:firstorder}
\end{equation}
By contrast, the phase transition is of second order if the order parameter increases continuously from zero to positive values as $T$ is lowered through $T_c$:
\begin{equation}
Q^*(T\!\to\! T_c^-)=Q^*(T\!\to\! T_c^+)=0.
\label{eq:secondorder}
\end{equation}

In principle, in our setting we can engineer the order of the phase transition as we like: as clear from Eq.~\eqref{eq:orderBEC}, the value of the order parameter for values slight below the critical temperature is governed by the value of $\psi^*_+(T\lesssim T_c)$ defined in Eq.~\eqref{eq:psi}. So, if we choose $\psi^*_+(T\!\to\! T_c^-)>0$ the transition will be first-order, while if we choose $\psi^*_+(T\!\to\! T_c^-)=0$ the transition will be second-order.
While both choices are possible, the case $\psi^*_+(T\!\to\! T_c^-)>0$ is somewhat unnatural, since it would `forbid' all those strength sequences $\vec{s}^*(T\!<\!T_c)$ that, while being both graphic and perfectly consistent with the definition of `condensed' given in Sec.~\ref{scialla<}, are such that asymptotically $w^*_+(T\!<\!T_c)<n^2\psi^*_+(T\!\to\! T_c^-)$ or equivalently, by virtue of Eq.~\eqref{panzanella}, such that $w^*_-(T\!<\!T_c)< w^*- n_+(n_+-1)\psi^*_+(T\!\to\! T_c^-)$. Note that the latter inequality implies that $w^*_-(T)$ would experience a finite jump from $w^*- n_+(n_+-1)\psi^*_+(T\!\to\! T_c^-)<w^*$ to $w^*$ as $T$ is raised from a value just below $T_c$ to a value just above $T_c$: as discussed in Sec.~\ref{scialla}, $w^*$ is (to leading order) the only allowed value for $w^*_-(T)$ above the critical temperature.

Therefore we find more appropriate to choose $\psi^*_+(T)$ such that $\psi^*_+(T\!\to\! T_c^-)=0$. In this way, all values of $w^*_-(T\!<\!T_c)$ in the range $[0,w^*]$ are allowed and there is no discontinuity for $w^*_-(T)$ at $T_c$: Eq.~\eqref{panzanella} implies that its left limit is $w^*_-(T\!\to\! T_c^-)\approx w^*$, which coincides with its right limit $w_-^*(T\!\to\! T_c^+)\approx w^*$ implied by Eq.~\eqref{eq:ww}.
With this choice, we can locate the critical temperature $T_c$ by equating the right and left limits of Eq.~\eqref{eq:mimmo2}: since $\lim_{T\!\to\! T_c^-}e^{-\epsilon^*/kT}=\lim_{T\to T_c^+}e^{-\epsilon^*/kT}=e^{-\epsilon^*/kT_c}$, the right and left limits of 
\begin{equation}
\frac{w^*_-(T)[1+w^*_+(T)]}{w^*_+(T)[1+w^*_-(T)]}
\end{equation}
must coincide. This implies 
\begin{equation}
\frac{1+w^*_+(T\!\to\! T_c^+)}{w^*_+(T\!\to\! T_c^+)}=
\frac{1+w^*_+(T\!\to\! T_c^-)}{w^*_+(T\!\to\! T_c^-)}=1
\label{taspettoforiscola}
\end{equation}
since in the thermodynamic limit $w^*_+(T\!\to\! T_c^-)=\infty$.
The above expression in turn implies $w^*_+(T\!\to\! T_c^+)=\infty$.
Note that this is consistent with the fact that, as discussed in Sec.~\ref{scialla}, $w^*_+(T\!>\!T_c)$ is finite but can be arbitrarily large, while not altering (to leading order) the average weight $w^*$. So, $w^*_+(T\!>\!T_c)$ can grow indefinitely as $T$ decreases towards $T_c$, and take an infinite limit as $T\!\to\! T_c^-$, consistently with the fact that, for even lower temperatures, $w^*_+(T)$ diverges with speed $n^2$ as dictated by Eq.~\eqref{eq_w+}.
Inserting Eq.~\eqref{taspettoforiscola} into Eq.~\eqref{eq:mimmo2}, we get
\begin{equation}
e^{-\epsilon^*/kT_c}=\frac{w^*}{1+w^*}
\label{eq:mimmo3}
\end{equation}
as $n\to\infty$. We therefore obtain
\begin{equation}
T_c=\frac{\epsilon^*}{k\ln\left(1+\frac{1}{w^*}\right)},
\label{eq:Tc}
\end{equation}
finally showing how the critical temperature depends on the expected link weight $w^*$, on the energy difference $\epsilon^*$ between periphery-periphery and core-core links, and on the constant $k$ converting the units of the `cost of links' (energy) to those of the temperature.
These results explain much of the information anticipated previously in Fig.~\ref{fig3} and summarized therein.

Adopting the view that the order parameter is continuous through the critical value $T_c$, we notice that the `direct' solutions in Eqs.~\eqref{eq:mimmo>} and~\eqref{eq:mimmo<}, as well as the behaviour at the critical point $T\!=\!T_c$, can be combined into the general `phenomenological' expression
\begin{equation}
\mu^*(T)\approx\epsilon^*-kT\ln\left(1+\frac{1}{w^*\left[1-Q_n^*(T)\right]}\right)+o(T)\quad
\label{eq:mimmoforever}
\end{equation}
which is valid for all values of the temperature. Indeed, when $T\ge T_c$ the order parameter is zero and the above expression reduces to Eq.~\eqref{eq:mimmo>}, while when $T\!<\!T_c$ the order parameter takes the positive value in Eq.~\eqref{eq:orderBEC} and the above expression reduces to Eq.~\eqref{eq:mimmo<}. 
The extreme limits $T\to\infty$ and $T\to 0$ can be retrieved from Eq.~\eqref{eq:mimmoforever} as well.

\section{Conclusions\label{sec_conclusions}}
We have investigated the breakdown of equivalence between canonical and microcanonical ensembles of weighted networks with local constraints on the strength of each node (\emph{weighted configuration model}~\cite{squartini2017maximum}). While ensemble nonequivalence in the corresponding \emph{binary configuration model} (i.e. binary networks with given node degrees) had already been studied in detail~\cite{squartini2015breaking,garlaschelli2016ensemble,garlaschelli2018covariance,roccaverde2019breaking,dionigi2020spectral},  a similar analysis for weighted networks had not been carried out so far.
As a unique and novel ingredient in the case considered here, weighted networks can undergo BEC, a phase transition that is impossible to observe in the unweighted case.
BEC emerges when a finite fraction of the total weight of all links condenses in a finite number of links. 
We constructed the simplest model exhibiting such behaviour: a network with a finite core, an infinite periphery, and a temperature-dependent strength sequence.
This setting allows us to combine for the first time, in a single model, two completely different mechanisms that can potentially destroy the equivalence of ensembles: a phase transition (a condition exhibited in the earliest observations of ensemble nonequivalence~\cite{ellis2000large,blume1971ising,barre2001inequivalence,ellis2004thermodynamic,lynden1999negative,chavanis2003gravitational,d2000negative,barre2007ensemble,radin2013phase,ellis2002nonequivalent,kastner2010nonequivalence,campa2009statistical}) and an extensive number of local constraints (an ingredient found in more recent investigations on network ensembles~\cite{squartini2015breaking,garlaschelli2016ensemble,roccaverde2019breaking,squartini2017reconnecting}).

We have considered two criteria for ensemble equivalence: the traditional and intuitive one based on the vanishing of the relative canonical fluctuations of the constraints in the thermodynamic limit~\cite{gibbs1902elementary} and the more recent and rigorous one based on the vanishing of the relative entropy density between microcanonical and canonical probability distributions (\emph{measure equivalence}) ~\cite{touchette2015equivalence}.
While in the standard situation (i.e. under only one or a finite number of global constraints) the vanishing of the relative fluctuations implies measure equivalence, the relationship between the two criteria had not been investigated in presence of an extensive number of local constraints yet.
Technically, while the relative fluctuations can be calculated exactly (as they are purely canonical quantities), the relative entropy requires in principle unfeasible microcanonical calculations but it can still be calculated asymptotically via a recently proposed saddle-point technique showing that its leading term is the logarithm of the determinant of the matrix of canonical covariances between the constraints.

We found that, for all positive temperatures, the relative entropy is $O(n\ln n)$ while the canonical and microcanonical  entropies are $O(n^2)$. These behaviours mimick the corresponding ones found for the binary configuration model in the dense regime~\cite{squartini2015breaking,garlaschelli2016ensemble,roccaverde2019breaking}. This result shows that, for all $T\!>\!0$ (including $T\!\to\!\infty$), the relative entropy is subleading with respect to the canonical and microcanonical entropies, but is still superextensive in the number of nodes $n$, which in all network models represents the number of units (physical size) of the system. 
In the zero-temperature limit, we found slower scalings for the canonical, microcanonical and relative entropies. This is due to the fact that, in both canonical and microcanonical ensembles, the peripheral nodes are asymptotically disconnected from all other nodes in each possible realization of the network. In this zero-temperature limit, the condensate effectively behaves as a lower-dimensional system, as commonly observed in the physics of BEC. Its entropy is the residual entropy resulting from the degeneracy of the ground state. The scaling of the relative entropy still indicates ensemble nonequivalence.
We note that in the binary configuration model (which obeys Fermi-Dirac rather than Bose-Einstein statistics) the zero-temperature phase is one where the canonical and microcanonical ensembles are instead \emph{identical}, because in both ensembles the pairs of nodes below a certain `Fermi energy' (whose value coincides with the chemical potential) are surely connected, while those above it are surely disconnected~\cite{garlaschelli2013low}.
Therefore we can conclude that, irrespective of BEC, at all temperatures ensemble equivalence is broken by the presence of an extensive number of local constraints, as in the binary configuration model (for which, however, BEC cannot occur). So the condensation phase transition (occuring at some critical temperature $T_c\!>\!0$) appears to have no effect on ensemble equivalence.

On the other hand, the calculation of the canonical relative fluctuations of the constraints shows that they are sensitive to the phase transition, while they cannot be used to characterize ensemble (non)equivalence as traditionally expected. Indeed, we found that in the non-condensed phase ($T\!>\!T_c$) the relative fluctuations of the strength of all the $n$ nodes vanish in the thermodynamic limit. By contrast, in the condensed phase ($T\!<\!T_c$) the relative fluctuations of the strength of the $n_+$ nodes in the core do not vanish, while those for the $n_-$ nodes in the periphery still do (except in the zero-temperature limit, for which not even the relative fluctuations for the peripheral nodes vanish).
Therefore, as the temperature is lowered below the critical temperature, there is a sudden change in the relative fluctuations but no change in the scaling of the relative entropy.
Conversely, as the temperature is further lowered to zero, there is a sudden change in the scaling of the relative entropy, while the relative fluctuations for the core nodes remain non-zero, albeit with  a different value. 
These results show that, at least in the dense case studied here, the relative entropy and the relative fluctuations capture different aspects of the phenomenology of the proposed model, the former being sensitive to the presence of local constraints and the latter being sensitive to the phase transition. In any case, in presence of an extensive number of local constraints the vanishing of (even all) the canonical relative fluctuations does not guarantee measure equivalence and is therefore no longer a valid criterion for ensemble equivalence as intuitively expected.

We stress that, while the network model presented here is deliberately simple from the structural point of view (a core-periphery network with local, but homogeneous, constraints), it could certainly serve as a reference for more complicated models (e.g. a core-periphery network with local and heterogeneous constraints). Indeed, ensemble nonequivalence will still be manifest in such a generalized model for all positive temperatures, because research on binary networks with given degrees has shown that nonequivalence is due to the locality of the constraints, and not to their specific value~\cite{squartini2015breaking,garlaschelli2016ensemble,roccaverde2019breaking,squartini2017reconnecting}. Additionally, since a more heterogeneous choice of the constraints can only increase the number of states with different energy in the network, we expect that BEC will still emerge below some critical temperature. In general, we expect a qualitatively similar behaviour to the one found here, with only quantitative differences.

The concept of ensemble equivalence is central for the foundations of statistical physics, irrespective of the particular system being considered. The findings documented here shed new light on the breakdown of EE, on the (possibly misleading) criteria used to detect it, and on the (so far undocumented) interplay between different mechanisms producing it. 
We hope they can inspire future research on these subjects.
 
\section*{Acknowledgments}
This work is supported by the Chinese Scholarship Council (No. 201606990001), the Dutch Econophysics Foundation (Stichting Econophysics, Leiden, the Netherlands) and the Netherlands Organization for Scientific Research (NWO/OCW).

\appendix
\section{Determinant of the covariance matrix}
$\mathbf{\Sigma}^*(T)$ is the $n\times n$ canonical covariance matrix between the strengths of all nodes, with entries
\begin{eqnarray}
{\Sigma}^*_{ij}(T)&=&\left\{\begin{array}{ll}
\textrm{Var}_{\vec{\beta}^*(T)}(s_i)&i=j\\
\textrm{Var}_{\vec{\beta}^*(T)}(w_{ij})&i\ne j
\end{array}\right.\nonumber\\
&=&\left\{\begin{array}{ll}
\Sigma^*_{ii}(T)&i=j\\
\langle w_{ij}\rangle_{\vec{\beta}^*(T)}\big[1+\langle w_{ij}\rangle_{\vec{\beta}^*(T)}\big]&i\ne j
\end{array}\right..
\label{eq:blocksigma}
\end{eqnarray}
Combining Eqs.~\eqref{covariance-matrix-dewd-diagonal} and~\eqref{non-diagonal-matrix-dewd} with the results discussed in Sec.~\ref{sec:corper} for our core-periphery model, it is easy to see that, for all values of temperature, $\mathbf{\Sigma}^*(T)$ is a combination of four blocks
\begin{equation}
\mathbf{\Sigma}^*(T)=\begin{bmatrix}
		\mathbf{A}(T) & \mathbf{B}(T)\\ 
		\mathbf{C}(T) & \mathbf{D}(T)
	  \end{bmatrix},
\label{eq:blocks}
\end{equation}
where $\mathbf{A}(T)$ is the $n_+\times n_+$ submatrix of covariances between the strengths of nodes in the core, with entries
\begin{equation}
A_{ij}(T)=\left\{\begin{array}{ll}
\Sigma^*_+(T)&i=j\\
w^*_+(T)[1+w^*_+(T)]&i\ne j
\end{array}\right.,
\label{eq:blockA}
\end{equation}
$\mathbf{B}(T)$ is the $n_+\times n_-$ submatrix of covariances between the strengths of nodes across core and periphery, with entries
\begin{equation}
B_{ij}(T)=w^*_0(T)[1+w^*_0(T)]\quad\forall i,j,
\label{eq:blockB}
\end{equation}
$\mathbf{C}(T)$ is a $n_-\times n_+$ matrix equal to the transpose of $\mathbf{B}(T)$, and
$\mathbf{D}(T)$ is the $n_-\times n_-$ submatrix of covariances between the strengths of nodes in the periphery, with entries
\begin{equation}
D_{ij}(T)=\left\{\begin{array}{ll}
\Sigma^*_-(T)&i=j\\
w^*_-(T)[1+w^*_-(T)]&i\ne j
\end{array}\right..
\label{eq:blockD}
\end{equation}
Depending on the range of temperature values of interest, different techniques become useful in order to calculate the determinant of $\mathbf{\Sigma}^*(T)$.
We therefore consider each regime separately below.

\subsection{Non-condensed phase\label{app:noncondensed}}
In the regime of finite supercritical temperature $(T\!>\!T_c)$ discussed in Sec.~\ref{scialla}, it is possible to show that the asymptotic behaviour of $\det \mathbf{\Sigma}^*(T\!>\!T_c)$ can be decomposed as the product of the diagonal elements of $\mathbf{\Sigma}^*(T\!>\!T_c)$, plus a correction. 
Our rationale for this decomposition comes from the fact that, as noted in Sec.~\ref{scialla},
$w^*_+(T\!>\!T_c)$, $w^*_-(T\!>\!T_c)$ and $w^*_0(T\!>\!T_c)$ are all $O(1)$, i.e. the expected link weights are all of the same, finite order. 
Consequently, the block structure depicted in Eq.~\eqref{eq:blocks} does not identify any particular difference in the order of magnitude of the entries of $\mathbf{\Sigma}^*(T\!>\!T_c)$. Rather, an important property of $\mathbf{\Sigma^*}(T\!>\!T_c)$ in this regime is that its diagonal entries are, on average, $n$ times bigger than its off-diagonal ones. Indeed, the off-diagonal entries are $O(1)$, while the diagonal ones are $O(n)$. Then the asymptotic behaviour of $\det\mathbf{\Sigma^*}(T\!>\!T_c)$ must be essentially dictated by the product of the diagonal entries of $\mathbf{\Sigma^*}(T\!>\!T_c)$. 

To make this intuition more rigorous, we
recall that if a $k\times k$ matrix $\mathbf{L}$ can be decomposed as
\begin{equation}
\mathbf{L}=\mathbf{M}+\varepsilon\mathbf{N},\label{decomposto}
\end{equation}
where $\mathbf{M}$ is a diagonal matrix with entries of finite order and $\varepsilon\mathbf{N}$ is a perturbation, then Jacobi's formula applies as follows:
\begin{eqnarray}
\det\mathbf{L}&=&
  \det(\mathbf{M}+\varepsilon\mathbf{N})\\
&=&\det\mathbf{M}+\varepsilon(\det\mathbf{M})\textrm{tr}(\mathbf{M}^{-1}\mathbf{N})+O(\varepsilon^2).\nonumber
\label{eq:jacobi}
\end{eqnarray}
Moreover, if the diagonal elements of $\mathbf{N}$ are equal to 0, then the product $\mathbf{M}^{-1}\mathbf{N}$ is a $k\times k$ zero matrix and therefore
\begin{equation}
  \textrm{tr}(\mathbf{M}^{-1}\mathbf{N})=0.
\end{equation}
Equation~\eqref{eq:jacobi} then becomes
\begin{eqnarray}
\det\mathbf{L}&=&\det\mathbf{M}+O(\varepsilon^2)\label{determin-abstract}\\
&=&\prod_{i=1}^k L_{ii}+O(\varepsilon^2).\nonumber
\end{eqnarray}

Turning to the matrix $\mathbf{\Sigma^*}(T\!>\!T_c)$, we note that the above hypotheses apply by setting
\begin{equation}
k\equiv n,\qquad \varepsilon\equiv\frac{1}{n},\qquad \mathbf{L}\equiv\frac{\mathbf{\Sigma}^*(T\!>\!T_c)}{n}
\end{equation}
and defining the entries of $\mathbf{M}$ and $\mathbf{N}$ as 
\begin{eqnarray}
{M}_{ij}&=&\delta_{ij}{\Sigma}^*_{ij}(T\!>\!T_c)/n,\\
{N}_{ij}&=&(1-\delta_{ij}){\Sigma}^*_{ij}(T\!>\!T_c),
\end{eqnarray}
where $\delta_{ij}$ is the Kronecker delta symbol.
Equation~\eqref{determin-abstract} then becomes
\begin{equation}
  \det\left(\frac{\mathbf{\Sigma}^*(T\!>\!T_c)}{n}\right)=\frac{1}{n^n}\prod_{i=1}^n{\Sigma}^*_{ii}(T\!>\!T_c)+O(n^{-2})\nonumber
\end{equation}
and, finally, 
\begin{eqnarray}
\det\mathbf{\Sigma}^*(T\!>\!T_c)&=&n^n \det\left(\frac{\mathbf{\Sigma}^*(T\!>\!T_c)}{n}\right)\label{det-production}\\
&=&
\prod_{i=1}^n{\Sigma}^*_{ii}(T\!>\!T_c)+O(n^{n-2}),\nonumber
\end{eqnarray}
proving Eq.~\eqref{det-production_main} used in the main text.

\subsection{Infinite-temperature limit\label{app:infinite}}
In the infinite-temperature limit discussed in Sec.~\eqref{sec:infinite}, the determinant can be calculated exactly as follows. From Eqs.~\eqref{dewd-cov-dagonal-parameter} and~\eqref{Non-diagonal-1} we see that, if we introduce a $k\times k$ matrix $\mathbf{Z}_k$ defined as
\begin{equation}\label{M_n-1}
\mathbf{Z}_k=\begin{pmatrix}
  k-1& 1&\cdots&1&1\\
  1& k-1&1&\cdots&1\\
  \vdots&&\ddots&&\vdots\\
 1& \cdots&1&k-1&1\\
  1& 1&\cdots&1&k-1\\
\end{pmatrix},
\end{equation}
then we can rewrite the covariance matrix as
\begin{equation}\label{covariance-decompose-1}
\mathbf{\Sigma}^*(T\!\to\!\infty)=w^*(1+w^*)\mathbf{Z}_n.
\end{equation}
Clearly, the calculation of $ \det\mathbf{\Sigma}^*(T\!\to\!\infty)$ reduces to the calculation of $\det\mathbf{Z}_n$:
\begin{equation}\label{determinant-decomposed-1}
  \det \mathbf{\Sigma}^*(T\!\to\!\infty)=\left[w^*(1+w^*)\right]^n\det\mathbf{Z}_n.
\end{equation}
To compute $\det\mathbf{Z}_k$ for arbitrary $k$, we note that 
\begin{equation}\label{Mn-divided-1}
  \mathbf{Z}_k=(k-2)\mathbf{I}_k+\mathbf{u}_k^T\mathbf{u}_k=(k-2)\left(\mathbf{I}_k+\frac{\mathbf{u}_k^T\mathbf{u}_k}{k-2}\right),
\end{equation}
where $\mathbf{I}_k$ is the $k\times k$ identity matrix and 
\begin{equation}
\mathbf{u}_k=(1,\cdots,1)
\label{eq:row}
\end{equation}
is the $k$-dimensional row vector with all unit entries. Then, using Sylvester's identity $\det(\mathbf{I}_k+\mathbf{X}\mathbf{Y})=\det(\mathbf{I}_l+\mathbf{Y}\mathbf{X})$ (where $\mathbf{X}$ is a $k\times l$ matrix, $\mathbf{Y}$ is an $l\times k$ matrix, and $\mathbf{I}_k$ and $\mathbf{I}_l$ are $k\times k$ and $l\times l$ identity matrices respectively) with $l=1$, $\mathbf{X}=\mathbf{u}_k^T$, $\mathbf{Y}=\mathbf{u}_k$ and $\mathbf{I}_l=1$, we get
\begin{eqnarray}
\det \mathbf{Z}_k&=&(k-2)^k\det\left( \mathbf{I}_k+\frac{ \mathbf{u}_k^T}{\sqrt{k-2}}\frac{ \mathbf{u}_k}{\sqrt{k-2}}\right)\nonumber\\
&=&(k-2)^k\det\left( 1+\frac{ \mathbf{u}_k}{\sqrt{k-2}}\frac{ \mathbf{u}_k^T}{\sqrt{k-2}}\right)\nonumber\\
&=&(k-2)^k\left(1+\frac{k}{k-2}\right)\nonumber\\
&=&2(k-1)(k-2)^{k-1}.\label{determinant-1}
\end{eqnarray}
Combining Eqs.~\eqref{determinant-decomposed-1} and~\eqref{determinant-1}, and setting $k=n$, we obtain exactly
Eq.~\eqref{determina-sigma-mn-decompose-1_main} used in the main text.

\subsection{Condensed phase\label{app:condensed}}
In the regime of subcritical temperature $(T<T_c)$ discussed in Sec.~\ref{sec:condensed}, 
the block structure indicated in Eq.~\eqref{eq:blocks} becomes particularly relevant, as it captures the important differences in the order of magnitude of both diagonal and off-diagonal entries of $\mathbf{\Sigma}^*(T\!<\!T_c)$ calculated using Eqs.~\eqref{eqwpsi},~\eqref{eq_w-},~\eqref{eq_w0},~\eqref{tumadre} and~\eqref{tupadre}.
We first express each block conveniently and then proceed to the calculation of the determinant.
Inserting Eqs.~\eqref{eqwpsi} and~\eqref{tumadre} into Eq.~\eqref{eq:blockA}, we obtain
\begin{eqnarray}
\mathbf{A}(T\!<\!T_c)\approx {[\psi^*_+(T\!<\!T_c) ]}^2 n^4\,\mathbf{Z}_{n_+},
\label{eq:AA}
\end{eqnarray}
where $\mathbf{Z}_{k}$ is still the matrix defined in Eq.~\eqref{M_n-1}.
Next, we note from Eq.~\eqref{eq:blockB} that 
\begin{equation}
\mathbf{B}(T\!<\!T_c)=w^*_0(T\!<\!T_c)[1+w^*_0(T\!<\!T_c)]\mathbf{u}^T_{n_+}\mathbf{u}_{n_-},
\label{eq:Brow}
\end{equation}
where $\mathbf{u}_{k}$ is still given by Eq.~\eqref{eq:row}. Similarly,
\begin{equation}
\mathbf{C}(T\!<\!T_c)=w^*_0(T\!<\!T_c)[1+w^*_0(T\!<\!T_c)]\mathbf{u}^T_{n_-}\mathbf{u}_{n_+}.
\label{eq:Crow}
\end{equation}
Finally, inserting Eq.~\eqref{tupadre} into Eq.~\eqref{eq:blockD}, we obtain
\begin{eqnarray}
\mathbf{D}(T\!<\!T_c)\approx w^*_-(T\!<\!T_c)[1+w^*_-(T\!<\!T_c)]\,\mathbf{Z}_{n_-}.
\label{eq:DD}
\end{eqnarray}

Now, since $\mathbf{A}(T\!<\!T_c)$ is invertible, it is useful to exploit the block structure of $\mathbf{\Sigma}^*(T\!<\!T_c)$ by expressing its determinant as 
\begin{equation}
\det \mathbf{\Sigma}^*(T\!<\!T_c)=\det {\mathbf{A}}(T\!<\!T_c)\,\det\overline{\mathbf{A}}(T\!<\!T_c),\label{eq:detshur}
\end{equation}
where
\begin{eqnarray}
\overline{\mathbf{A}}(T\!<\!T_c)&\equiv&\mathbf{D}(T\!<\!T_c)\label{eq:shur}\\
&&-\mathbf{C}(T\!<\!T_c)\mathbf{A}^{-1}(T\!<\!T_c)\mathbf{B}(T\!<\!T_c)\nonumber
\end{eqnarray}
is the so-called Shur complement of $\mathbf{A}(T\!<\!T_c)$.
To calculate $\det {\mathbf{A}}(T\!<\!T_c)$, we use Eq.~\eqref{eq:AA} and immediately obtain
\begin{eqnarray}
  \det \mathbf{A}(T\!<\!T_c)&\approx&{[\psi^*_+(T\!<\!T_c) ]}^{2n_+} n^{4 n_+}\det\mathbf{Z}_{n_+}\nonumber\\
&=&O(n^{4 n_+})\label{determinant-decomposed-1new}
\end{eqnarray}
where, using Eq.~\eqref{determinant-1},
\begin{eqnarray}
\det \mathbf{Z}_{n_+}=2(n_+-1)(n_+-2)^{n_+-1}.\label{determinant-1new}
\end{eqnarray}
To calculate $\det \overline{\mathbf{A}}(T\!<\!T_c)$, we first use Eq.~\eqref{eq:AA} and obtain
\begin{eqnarray}
\mathbf{A}^{-1}(T\!<\!T_c)\approx {[\psi^*_+(T\!<\!T_c) ]}^{-2} n^{-4}\,\mathbf{Z}^{-1}_{n_+},
\label{eq:AA-1}
\end{eqnarray}
where, using Eq.~\eqref{M_n-1}, $\mathbf{Z}^{-1}_{k}$ is easily calculated by direct inversion of $\mathbf{Z}_{k}$ as
\begin{eqnarray}
\mathbf{Z}^{-1}_{k}&=&c_k\begin{pmatrix}
 2k-3& -1&\cdots&-1&-1\\
  -1& 2k-3&-1&\cdots&-1\\
  \vdots&&\ddots&&\vdots\\
 -1& \cdots&-1&2k-3&-1\\
  -1& -1&\cdots&-1&2k-3\\
\end{pmatrix}\nonumber\\
&=&\frac{\mathbf{I}_k}{k-2}-c_k\mathbf{u}_k^T\mathbf{u}_k\nonumber\\
&=&\frac{1}{k-2}\left[\mathbf{I}_k-\frac{\mathbf{u}_k^T\mathbf{u}_k}{2(k-1)}\right]\label{inverse}
\end{eqnarray}
with
\begin{equation}
c_k=\frac{1}{2(k-1)(k-2)}.
\end{equation}
Inserting Eqs.~\eqref{eq:Brow},\eqref{eq:Crow},~\eqref{eq:DD} and~\eqref{eq:AA-1} into Eq.~\eqref{eq:shur}, and noticing that 
\begin{eqnarray}
\mathbf{u}^T_{n_-}\mathbf{u}_{n_+} \mathbf{Z}^{-1}_{n_+} \mathbf{u}^T_{n_+}\mathbf{u}_{n_-}&=&c_{n_+} n_+(n_+-2)\mathbf{u}^T_{n_-}\mathbf{u}_{n_-}\nonumber\\
&=&\frac{n_+}{2(n_+-1)}\mathbf{u}^T_{n_-}\mathbf{u}_{n_-},
\end{eqnarray}
we can obtain the Shur complement of $\mathbf{A}(T\!<\!T_c)$ as
\begin{eqnarray}
\overline{\mathbf{A}}(T\!<\!T_c)
&\approx&w^*_-(T\!<\!T_c)[1+w^*_-(T\!<\!T_c)]\,\mathbf{Z}_{n_-}\label{eq:cazzo}\\
&&-\frac{n_+[w^*_0(T\!<\!T_c)]^2[1+w^*_0(T\!<\!T_c)]^2}{2(n_+-1){[\psi^*_+(T\!<\!T_c) ]}^2 n^4}\mathbf{u}^T_{n_-}\mathbf{u}_{n_-}\nonumber
\end{eqnarray}
from which we can calculate $\det \overline{\mathbf{A}}(T\!<\!T_c)$.
We have to distinguish the cases $0\!<\!T\!<\!T_c$ and $T\!\simeq\!0$, as they are characterized by different scalings of $w^*_-(T\!<\!T_c)$ and $w^*_0(T\!<\!T_c)$. In the rest of this section we consider the case of finite temperature, while the zero-temperature limit is considered in the next section.

When $0\!<\!T\!<\!T_c$, we recall from Eqs.\eqref{eq:anvedi1} and~\eqref{eq:anvedi2} that both $w^*_-(0\!<\!T\!<\!T_c)$ and $w^*_0(0\!<\!T\!<\!T_c)$ are $O(1)$. From Eq.~\eqref{eq:cazzo} we therefore see that all the off-diagonal entries of $\overline{\mathbf{A}}(0\!<\!T\!<\!T_c)$ are $O(1)$, while all the diagonal ones are $O(n_-)$.
This implies that we can use the decomposition in Eq.~\eqref{decomposto} where
\begin{equation}
k\equiv n_-,\qquad \varepsilon\equiv\frac{1}{n_-},\qquad \mathbf{L}\equiv\frac{\overline{\mathbf{A}}(0\!<\!T\!<\!T_c)}{n_-}.
\end{equation}
Equation~\eqref{determin-abstract} then implies
\begin{eqnarray}
\det\overline{\mathbf{A}}(0\!<\!T\!<\!T_c)&=&
\prod_{i=1}^{n_-}\overline{A}_{ii}(0\!<\!T\!<\!T_c)+O\big(n_-^{n_--2}\big)\nonumber\\
&=&O(n^{n_-}).\label{eq:Abarra}
\end{eqnarray}
Combining Eqs.~\eqref{determinant-decomposed-1new} and~\eqref{eq:Abarra} into Eq.~\eqref{eq:detshur}, we finally obtain the full determinant of $\mathbf{\Sigma}^*(0\!<\!T\!<\!T_c)$. We are interested only in its scaling with $n$, which is
\begin{equation}
\det \mathbf{\Sigma}^*(0\!<\!T\!<\!T_c)=O(n^{4 n_++n_-})=O(n^{n+3 n_+}),
\end{equation}
proving Eq.~\eqref{jack} used in the main text.

\subsection{Zero-temperature limit\label{app:zero}}
In the zero-temperature limit, all calculations of the previous section remain valid until and including Eq.~\eqref{eq:cazzo}. The scaling of the entries of $\overline{\mathbf{A}}(T\!\simeq\! 0)$ will however be different.
Indeed, we recall from Eqs.\eqref{eq_w-n} and~\eqref{eq_w0n} that $w^*_-(T\!\simeq\! 0)\approx{\ell}/{n}$ and $w^*_0(T\!\simeq\! 0)\approx\sqrt{{\ell}/{n}}$. Inserted into Eq.~\eqref{eq:cazzo}, these expressions imply that all the diagonal entries of $\overline{\mathbf{A}}(T\!\simeq\! 0)$ are asymptotically equal to $\ell$, while all the off-diagonal ones are $O(\ell/n)$.
We can therefore use the decomposition in Eq.~\eqref{decomposto} where
\begin{equation}
k\equiv n_-,\qquad \varepsilon\equiv\frac{1}{n},\qquad \mathbf{L}\equiv{\overline{\mathbf{A}}(T\!\simeq\!0)}.
\end{equation}
Equation~\eqref{determin-abstract} then implies
\begin{eqnarray}
\det\overline{\mathbf{A}}(T\!\simeq\!0)&=&
\prod_{i=1}^{n_-}\overline{A}_{ii}(T\!\simeq\!0)+O\big(n^{-2}\big)\nonumber\\
&=&\ell^{n_-}+O(n^{-2}).\label{eq:Abarra2}
\end{eqnarray}
Combined with Eq.~\eqref{determinant-decomposed-1new} into Eq.~\eqref{eq:detshur}, the above result leads to the full determinant of $\mathbf{\Sigma}^*(T\!<\!T_c)$, whose scaling with $n$ is
\begin{equation}
\det \mathbf{\Sigma}^*(T\!\simeq\!0)=O(n^{4 n_+}\ell^n),
\end{equation}
proving Eq.~\eqref{jack2} used in the main text.


\begin{thebibliography}{41}%
\makeatletter
\providecommand \@ifxundefined [1]{%
 \@ifx{#1\undefined}
}%
\providecommand \@ifnum [1]{%
 \ifnum #1\expandafter \@firstoftwo
 \else \expandafter \@secondoftwo
 \fi
}%
\providecommand \@ifx [1]{%
 \ifx #1\expandafter \@firstoftwo
 \else \expandafter \@secondoftwo
 \fi
}%
\providecommand \natexlab [1]{#1}%
\providecommand \enquote  [1]{``#1''}%
\providecommand \bibnamefont  [1]{#1}%
\providecommand \bibfnamefont [1]{#1}%
\providecommand \citenamefont [1]{#1}%
\providecommand \href@noop [0]{\@secondoftwo}%
\providecommand \href [0]{\begingroup \@sanitize@url \@href}%
\providecommand \@href[1]{\@@startlink{#1}\@@href}%
\providecommand \@@href[1]{\endgroup#1\@@endlink}%
\providecommand \@sanitize@url [0]{\catcode `\\12\catcode `\$12\catcode
  `\&12\catcode `\#12\catcode `\^12\catcode `\_12\catcode `\%12\relax}%
\providecommand \@@startlink[1]{}%
\providecommand \@@endlink[0]{}%
\providecommand \url  [0]{\begingroup\@sanitize@url \@url }%
\providecommand \@url [1]{\endgroup\@href {#1}{\urlprefix }}%
\providecommand \urlprefix  [0]{URL }%
\providecommand \Eprint [0]{\href }%
\providecommand \doibase [0]{http://dx.doi.org/}%
\providecommand \selectlanguage [0]{\@gobble}%
\providecommand \bibinfo  [0]{\@secondoftwo}%
\providecommand \bibfield  [0]{\@secondoftwo}%
\providecommand \translation [1]{[#1]}%
\providecommand \BibitemOpen [0]{}%
\providecommand \bibitemStop [0]{}%
\providecommand \bibitemNoStop [0]{.\EOS\space}%
\providecommand \EOS [0]{\spacefactor3000\relax}%
\providecommand \BibitemShut  [1]{\csname bibitem#1\endcsname}%
\let\auto@bib@innerbib\@empty
\bibitem [{\citenamefont {Gibbs}(1902)}]{gibbs1902elementary}%
  \BibitemOpen
  \bibfield  {author} {\bibinfo {author} {\bibfnamefont {J.}~\bibnamefont
  {Gibbs}},\ }\href@noop {} {\bibfield  {journal} {\bibinfo  {journal} {The
  Collected Works of JW Gibbs (Yale University, New Haven, CT, 1957)}\ }\textbf
  {\bibinfo {volume} {2}} (\bibinfo {year} {1902})}\BibitemShut {NoStop}%
\bibitem [{\citenamefont {Jaynes}(1957)}]{jaynes1957information}%
  \BibitemOpen
  \bibfield  {author} {\bibinfo {author} {\bibfnamefont {E.~T.}\ \bibnamefont
  {Jaynes}},\ }\href@noop {} {\bibfield  {journal} {\bibinfo  {journal}
  {Physical review}\ }\textbf {\bibinfo {volume} {106}},\ \bibinfo {pages}
  {620} (\bibinfo {year} {1957})}\BibitemShut {NoStop}%
\bibitem [{\citenamefont {Campa}\ \emph {et~al.}(2009)\citenamefont {Campa},
  \citenamefont {Dauxois},\ and\ \citenamefont {Ruffo}}]{campa2009statistical}%
  \BibitemOpen
  \bibfield  {author} {\bibinfo {author} {\bibfnamefont {A.}~\bibnamefont
  {Campa}}, \bibinfo {author} {\bibfnamefont {T.}~\bibnamefont {Dauxois}}, \
  and\ \bibinfo {author} {\bibfnamefont {S.}~\bibnamefont {Ruffo}},\
  }\href@noop {} {\bibfield  {journal} {\bibinfo  {journal} {Physics Reports}\
  }\textbf {\bibinfo {volume} {480}},\ \bibinfo {pages} {57} (\bibinfo {year}
  {2009})}\BibitemShut {NoStop}%
\bibitem [{\citenamefont {Touchette}(2015)}]{touchette2015equivalence}%
  \BibitemOpen
  \bibfield  {author} {\bibinfo {author} {\bibfnamefont {H.}~\bibnamefont
  {Touchette}},\ }\href@noop {} {\bibfield  {journal} {\bibinfo  {journal}
  {Journal of Statistical Physics}\ }\textbf {\bibinfo {volume} {159}},\
  \bibinfo {pages} {987} (\bibinfo {year} {2015})}\BibitemShut {NoStop}%
\bibitem [{\citenamefont {Squartini}\ \emph
  {et~al.}(2015{\natexlab{a}})\citenamefont {Squartini}, \citenamefont
  {de~Mol}, \citenamefont {den Hollander},\ and\ \citenamefont
  {Garlaschelli}}]{squartini2015breaking}%
  \BibitemOpen
  \bibfield  {author} {\bibinfo {author} {\bibfnamefont {T.}~\bibnamefont
  {Squartini}}, \bibinfo {author} {\bibfnamefont {J.}~\bibnamefont {de~Mol}},
  \bibinfo {author} {\bibfnamefont {F.}~\bibnamefont {den Hollander}}, \ and\
  \bibinfo {author} {\bibfnamefont {D.}~\bibnamefont {Garlaschelli}},\
  }\href@noop {} {\bibfield  {journal} {\bibinfo  {journal} {Physical review
  letters}\ }\textbf {\bibinfo {volume} {115}},\ \bibinfo {pages} {268701}
  (\bibinfo {year} {2015}{\natexlab{a}})}\BibitemShut {NoStop}%
\bibitem [{\citenamefont {Squartini}\ and\ \citenamefont
  {Garlaschelli}(2017{\natexlab{a}})}]{squartini2017reconnecting}%
  \BibitemOpen
  \bibfield  {author} {\bibinfo {author} {\bibfnamefont {T.}~\bibnamefont
  {Squartini}}\ and\ \bibinfo {author} {\bibfnamefont {D.}~\bibnamefont
  {Garlaschelli}},\ }\href@noop {} {\bibfield  {journal} {\bibinfo  {journal}
  {arXiv preprint arXiv:1710.11422}\ } (\bibinfo {year}
  {2017}{\natexlab{a}})}\BibitemShut {NoStop}%
\bibitem [{\citenamefont {Ellis}\ \emph {et~al.}(2000)\citenamefont {Ellis},
  \citenamefont {Haven},\ and\ \citenamefont {Turkington}}]{ellis2000large}%
  \BibitemOpen
  \bibfield  {author} {\bibinfo {author} {\bibfnamefont {R.~S.}\ \bibnamefont
  {Ellis}}, \bibinfo {author} {\bibfnamefont {K.}~\bibnamefont {Haven}}, \ and\
  \bibinfo {author} {\bibfnamefont {B.}~\bibnamefont {Turkington}},\
  }\href@noop {} {\bibfield  {journal} {\bibinfo  {journal} {Journal of
  Statistical Physics}\ }\textbf {\bibinfo {volume} {101}},\ \bibinfo {pages}
  {999} (\bibinfo {year} {2000})}\BibitemShut {NoStop}%
\bibitem [{\citenamefont {Blume}\ \emph {et~al.}(1971)\citenamefont {Blume},
  \citenamefont {Emery},\ and\ \citenamefont {Griffiths}}]{blume1971ising}%
  \BibitemOpen
  \bibfield  {author} {\bibinfo {author} {\bibfnamefont {M.}~\bibnamefont
  {Blume}}, \bibinfo {author} {\bibfnamefont {V.}~\bibnamefont {Emery}}, \ and\
  \bibinfo {author} {\bibfnamefont {R.~B.}\ \bibnamefont {Griffiths}},\
  }\href@noop {} {\bibfield  {journal} {\bibinfo  {journal} {Physical review
  A}\ }\textbf {\bibinfo {volume} {4}},\ \bibinfo {pages} {1071} (\bibinfo
  {year} {1971})}\BibitemShut {NoStop}%
\bibitem [{\citenamefont {Barr{\'e}}\ \emph {et~al.}(2001)\citenamefont
  {Barr{\'e}}, \citenamefont {Mukamel},\ and\ \citenamefont
  {Ruffo}}]{barre2001inequivalence}%
  \BibitemOpen
  \bibfield  {author} {\bibinfo {author} {\bibfnamefont {J.}~\bibnamefont
  {Barr{\'e}}}, \bibinfo {author} {\bibfnamefont {D.}~\bibnamefont {Mukamel}},
  \ and\ \bibinfo {author} {\bibfnamefont {S.}~\bibnamefont {Ruffo}},\
  }\href@noop {} {\bibfield  {journal} {\bibinfo  {journal} {Physical Review
  Letters}\ }\textbf {\bibinfo {volume} {87}},\ \bibinfo {pages} {030601}
  (\bibinfo {year} {2001})}\BibitemShut {NoStop}%
\bibitem [{\citenamefont {Ellis}\ \emph {et~al.}(2004)\citenamefont {Ellis},
  \citenamefont {Touchette},\ and\ \citenamefont
  {Turkington}}]{ellis2004thermodynamic}%
  \BibitemOpen
  \bibfield  {author} {\bibinfo {author} {\bibfnamefont {R.~S.}\ \bibnamefont
  {Ellis}}, \bibinfo {author} {\bibfnamefont {H.}~\bibnamefont {Touchette}}, \
  and\ \bibinfo {author} {\bibfnamefont {B.}~\bibnamefont {Turkington}},\
  }\href@noop {} {\bibfield  {journal} {\bibinfo  {journal} {Physica A:
  Statistical Mechanics and its Applications}\ }\textbf {\bibinfo {volume}
  {335}},\ \bibinfo {pages} {518} (\bibinfo {year} {2004})}\BibitemShut
  {NoStop}%
\bibitem [{\citenamefont {Lynden-Bell}(1999)}]{lynden1999negative}%
  \BibitemOpen
  \bibfield  {author} {\bibinfo {author} {\bibfnamefont {D.}~\bibnamefont
  {Lynden-Bell}},\ }\href@noop {} {\bibfield  {journal} {\bibinfo  {journal}
  {Physica A: Statistical Mechanics and its Applications}\ }\textbf {\bibinfo
  {volume} {263}},\ \bibinfo {pages} {293} (\bibinfo {year}
  {1999})}\BibitemShut {NoStop}%
\bibitem [{\citenamefont {Chavanis}(2003)}]{chavanis2003gravitational}%
  \BibitemOpen
  \bibfield  {author} {\bibinfo {author} {\bibfnamefont {P.-H.}\ \bibnamefont
  {Chavanis}},\ }\href@noop {} {\bibfield  {journal} {\bibinfo  {journal}
  {Astronomy \& Astrophysics}\ }\textbf {\bibinfo {volume} {401}},\ \bibinfo
  {pages} {15} (\bibinfo {year} {2003})}\BibitemShut {NoStop}%
\bibitem [{\citenamefont {d'Agostino}\ \emph {et~al.}(2000)\citenamefont
  {d'Agostino}, \citenamefont {Gulminelli}, \citenamefont {Chomaz},
  \citenamefont {Bruno}, \citenamefont {Cannata}, \citenamefont {Bougault},
  \citenamefont {Gramegna}, \citenamefont {Iori}, \citenamefont {Le~Neindre},
  \citenamefont {Margagliotti} \emph {et~al.}}]{d2000negative}%
  \BibitemOpen
  \bibfield  {author} {\bibinfo {author} {\bibfnamefont {M.}~\bibnamefont
  {d'Agostino}}, \bibinfo {author} {\bibfnamefont {F.}~\bibnamefont
  {Gulminelli}}, \bibinfo {author} {\bibfnamefont {P.}~\bibnamefont {Chomaz}},
  \bibinfo {author} {\bibfnamefont {M.}~\bibnamefont {Bruno}}, \bibinfo
  {author} {\bibfnamefont {F.}~\bibnamefont {Cannata}}, \bibinfo {author}
  {\bibfnamefont {R.}~\bibnamefont {Bougault}}, \bibinfo {author}
  {\bibfnamefont {F.}~\bibnamefont {Gramegna}}, \bibinfo {author}
  {\bibfnamefont {I.}~\bibnamefont {Iori}}, \bibinfo {author} {\bibfnamefont
  {N.}~\bibnamefont {Le~Neindre}}, \bibinfo {author} {\bibfnamefont
  {G.}~\bibnamefont {Margagliotti}},  \emph {et~al.},\ }\href@noop {}
  {\bibfield  {journal} {\bibinfo  {journal} {Physics Letters B}\ }\textbf
  {\bibinfo {volume} {473}},\ \bibinfo {pages} {219} (\bibinfo {year}
  {2000})}\BibitemShut {NoStop}%
\bibitem [{\citenamefont {Barr{\'e}}\ and\ \citenamefont
  {Gon{\c{c}}alves}(2007)}]{barre2007ensemble}%
  \BibitemOpen
  \bibfield  {author} {\bibinfo {author} {\bibfnamefont {J.}~\bibnamefont
  {Barr{\'e}}}\ and\ \bibinfo {author} {\bibfnamefont {B.}~\bibnamefont
  {Gon{\c{c}}alves}},\ }\href@noop {} {\bibfield  {journal} {\bibinfo
  {journal} {Physica A: Statistical Mechanics and its Applications}\ }\textbf
  {\bibinfo {volume} {386}},\ \bibinfo {pages} {212} (\bibinfo {year}
  {2007})}\BibitemShut {NoStop}%
\bibitem [{\citenamefont {Radin}\ and\ \citenamefont
  {Sadun}(2013)}]{radin2013phase}%
  \BibitemOpen
  \bibfield  {author} {\bibinfo {author} {\bibfnamefont {C.}~\bibnamefont
  {Radin}}\ and\ \bibinfo {author} {\bibfnamefont {L.}~\bibnamefont {Sadun}},\
  }\href@noop {} {\bibfield  {journal} {\bibinfo  {journal} {Journal of Physics
  A: Mathematical and Theoretical}\ }\textbf {\bibinfo {volume} {46}},\
  \bibinfo {pages} {305002} (\bibinfo {year} {2013})}\BibitemShut {NoStop}%
\bibitem [{\citenamefont {Ellis}\ \emph {et~al.}(2002)\citenamefont {Ellis},
  \citenamefont {Haven},\ and\ \citenamefont
  {Turkington}}]{ellis2002nonequivalent}%
  \BibitemOpen
  \bibfield  {author} {\bibinfo {author} {\bibfnamefont {R.~S.}\ \bibnamefont
  {Ellis}}, \bibinfo {author} {\bibfnamefont {K.}~\bibnamefont {Haven}}, \ and\
  \bibinfo {author} {\bibfnamefont {B.}~\bibnamefont {Turkington}},\
  }\href@noop {} {\bibfield  {journal} {\bibinfo  {journal} {Nonlinearity}\
  }\textbf {\bibinfo {volume} {15}},\ \bibinfo {pages} {239} (\bibinfo {year}
  {2002})}\BibitemShut {NoStop}%
\bibitem [{\citenamefont {Kastner}(2010)}]{kastner2010nonequivalence}%
  \BibitemOpen
  \bibfield  {author} {\bibinfo {author} {\bibfnamefont {M.}~\bibnamefont
  {Kastner}},\ }\href@noop {} {\bibfield  {journal} {\bibinfo  {journal}
  {Physical review letters}\ }\textbf {\bibinfo {volume} {104}},\ \bibinfo
  {pages} {240403} (\bibinfo {year} {2010})}\BibitemShut {NoStop}%
\bibitem [{\citenamefont {Squartini}\ and\ \citenamefont
  {Garlaschelli}(2017{\natexlab{b}})}]{squartini2017maximum}%
  \BibitemOpen
  \bibfield  {author} {\bibinfo {author} {\bibfnamefont {T.}~\bibnamefont
  {Squartini}}\ and\ \bibinfo {author} {\bibfnamefont {D.}~\bibnamefont
  {Garlaschelli}},\ }\href@noop {} {\emph {\bibinfo {title} {Maximum-Entropy
  Networks: Pattern Detection, Network Reconstruction and Graph
  Combinatorics}}}\ (\bibinfo  {publisher} {Springer},\ \bibinfo {year}
  {2017})\BibitemShut {NoStop}%
\bibitem [{\citenamefont {Garlaschelli}\ \emph {et~al.}(2016)\citenamefont
  {Garlaschelli}, \citenamefont {Den~Hollander},\ and\ \citenamefont
  {Roccaverde}}]{garlaschelli2016ensemble}%
  \BibitemOpen
  \bibfield  {author} {\bibinfo {author} {\bibfnamefont {D.}~\bibnamefont
  {Garlaschelli}}, \bibinfo {author} {\bibfnamefont {F.}~\bibnamefont
  {Den~Hollander}}, \ and\ \bibinfo {author} {\bibfnamefont {A.}~\bibnamefont
  {Roccaverde}},\ }\href@noop {} {\bibfield  {journal} {\bibinfo  {journal}
  {Journal of Physics A: Mathematical and Theoretical}\ }\textbf {\bibinfo
  {volume} {50}},\ \bibinfo {pages} {015001} (\bibinfo {year}
  {2016})}\BibitemShut {NoStop}%
\bibitem [{\citenamefont {Roccaverde}(2019)}]{roccaverde2019breaking}%
  \BibitemOpen
  \bibfield  {author} {\bibinfo {author} {\bibfnamefont {A.}~\bibnamefont
  {Roccaverde}},\ }\href@noop {} {\bibfield  {journal} {\bibinfo  {journal}
  {Indagationes Mathematicae}\ }\textbf {\bibinfo {volume} {30}},\ \bibinfo
  {pages} {7} (\bibinfo {year} {2019})}\BibitemShut {NoStop}%
\bibitem [{\citenamefont {Park}\ and\ \citenamefont
  {Newman}(2004)}]{park2004statistical}%
  \BibitemOpen
  \bibfield  {author} {\bibinfo {author} {\bibfnamefont {J.}~\bibnamefont
  {Park}}\ and\ \bibinfo {author} {\bibfnamefont {M.~E.}\ \bibnamefont
  {Newman}},\ }\href@noop {} {\bibfield  {journal} {\bibinfo  {journal}
  {Physical Review E}\ }\textbf {\bibinfo {volume} {70}},\ \bibinfo {pages}
  {066117} (\bibinfo {year} {2004})}\BibitemShut {NoStop}%
\bibitem [{\citenamefont {Garlaschelli}\ and\ \citenamefont
  {Loffredo}(2009)}]{garlaschelli2009generalized}%
  \BibitemOpen
  \bibfield  {author} {\bibinfo {author} {\bibfnamefont {D.}~\bibnamefont
  {Garlaschelli}}\ and\ \bibinfo {author} {\bibfnamefont {M.~I.}\ \bibnamefont
  {Loffredo}},\ }\href@noop {} {\bibfield  {journal} {\bibinfo  {journal}
  {Physical review letters}\ }\textbf {\bibinfo {volume} {102}},\ \bibinfo
  {pages} {038701} (\bibinfo {year} {2009})}\BibitemShut {NoStop}%
\bibitem [{\citenamefont {Bianconi}\ and\ \citenamefont
  {Barab{\'a}si}(2001)}]{bianconi2001bose}%
  \BibitemOpen
  \bibfield  {author} {\bibinfo {author} {\bibfnamefont {G.}~\bibnamefont
  {Bianconi}}\ and\ \bibinfo {author} {\bibfnamefont {A.-L.}\ \bibnamefont
  {Barab{\'a}si}},\ }\href@noop {} {\bibfield  {journal} {\bibinfo  {journal}
  {Physical review letters}\ }\textbf {\bibinfo {volume} {86}},\ \bibinfo
  {pages} {5632} (\bibinfo {year} {2001})}\BibitemShut {NoStop}%
\bibitem [{\citenamefont {Garlaschelli}\ \emph {et~al.}(2013)\citenamefont
  {Garlaschelli}, \citenamefont {Ahnert}, \citenamefont {Fink},\ and\
  \citenamefont {Caldarelli}}]{garlaschelli2013low}%
  \BibitemOpen
  \bibfield  {author} {\bibinfo {author} {\bibfnamefont {D.}~\bibnamefont
  {Garlaschelli}}, \bibinfo {author} {\bibfnamefont {S.~E.}\ \bibnamefont
  {Ahnert}}, \bibinfo {author} {\bibfnamefont {T.}~\bibnamefont {Fink}}, \ and\
  \bibinfo {author} {\bibfnamefont {G.}~\bibnamefont {Caldarelli}},\
  }\href@noop {} {\bibfield  {journal} {\bibinfo  {journal} {Entropy}\ }\textbf
  {\bibinfo {volume} {15}},\ \bibinfo {pages} {3148} (\bibinfo {year}
  {2013})}\BibitemShut {NoStop}%
\bibitem [{\citenamefont {Newman}\ \emph {et~al.}(2006)\citenamefont {Newman},
  \citenamefont {Barab{\'a}si},\ and\ \citenamefont
  {Watts}}]{newman2006structure}%
  \BibitemOpen
  \bibfield  {author} {\bibinfo {author} {\bibfnamefont {M.~E.}\ \bibnamefont
  {Newman}}, \bibinfo {author} {\bibfnamefont {A.-L.~E.}\ \bibnamefont
  {Barab{\'a}si}}, \ and\ \bibinfo {author} {\bibfnamefont {D.~J.}\
  \bibnamefont {Watts}},\ }\href@noop {} {\emph {\bibinfo {title} {The
  structure and dynamics of networks.}}}\ (\bibinfo  {publisher} {Princeton
  university press},\ \bibinfo {year} {2006})\BibitemShut {NoStop}%
\bibitem [{\citenamefont {Serrano}\ \emph {et~al.}(2006)\citenamefont
  {Serrano}, \citenamefont {Bogu{\~n}{\'a}},\ and\ \citenamefont
  {Pastor-Satorras}}]{serrano2006correlations}%
  \BibitemOpen
  \bibfield  {author} {\bibinfo {author} {\bibfnamefont {M.~{\'A}.}\
  \bibnamefont {Serrano}}, \bibinfo {author} {\bibfnamefont {M.}~\bibnamefont
  {Bogu{\~n}{\'a}}}, \ and\ \bibinfo {author} {\bibfnamefont {R.}~\bibnamefont
  {Pastor-Satorras}},\ }\href@noop {} {\bibfield  {journal} {\bibinfo
  {journal} {Physical Review E}\ }\textbf {\bibinfo {volume} {74}},\ \bibinfo
  {pages} {055101} (\bibinfo {year} {2006})}\BibitemShut {NoStop}%
\bibitem [{\citenamefont {Squartini}\ \emph
  {et~al.}(2015{\natexlab{b}})\citenamefont {Squartini}, \citenamefont
  {Mastrandrea},\ and\ \citenamefont {Garlaschelli}}]{squartini2015unbiased}%
  \BibitemOpen
  \bibfield  {author} {\bibinfo {author} {\bibfnamefont {T.}~\bibnamefont
  {Squartini}}, \bibinfo {author} {\bibfnamefont {R.}~\bibnamefont
  {Mastrandrea}}, \ and\ \bibinfo {author} {\bibfnamefont {D.}~\bibnamefont
  {Garlaschelli}},\ }\href@noop {} {\bibfield  {journal} {\bibinfo  {journal}
  {New Journal of Physics}\ }\textbf {\bibinfo {volume} {17}},\ \bibinfo
  {pages} {023052} (\bibinfo {year} {2015}{\natexlab{b}})}\BibitemShut
  {NoStop}%
\bibitem [{\citenamefont {Garlaschelli}\ and\ \citenamefont
  {Loffredo}(2008)}]{garlaschelli2008maximum}%
  \BibitemOpen
  \bibfield  {author} {\bibinfo {author} {\bibfnamefont {D.}~\bibnamefont
  {Garlaschelli}}\ and\ \bibinfo {author} {\bibfnamefont {M.~I.}\ \bibnamefont
  {Loffredo}},\ }\href@noop {} {\bibfield  {journal} {\bibinfo  {journal}
  {Physical Review E}\ }\textbf {\bibinfo {volume} {78}},\ \bibinfo {pages}
  {015101} (\bibinfo {year} {2008})}\BibitemShut {NoStop}%
\bibitem [{max()}]{maxsam}%
  \BibitemOpen
  \href@noop {} {\enquote {\bibinfo {title}
  {https://it.mathworks.com/matlabcentral/fileexchange/46912-max-sam-package-zip},}\
  }\BibitemShut {NoStop}%
\bibitem [{meh()}]{meh}%
  \BibitemOpen
  \href@noop {} {\enquote {\bibinfo {title} {https://meh.imtlucca.it},}\
  }\BibitemShut {NoStop}%
\bibitem [{\citenamefont {Kullback}\ and\ \citenamefont
  {Leibler}(1951)}]{kullback1951information}%
  \BibitemOpen
  \bibfield  {author} {\bibinfo {author} {\bibfnamefont {S.}~\bibnamefont
  {Kullback}}\ and\ \bibinfo {author} {\bibfnamefont {R.~A.}\ \bibnamefont
  {Leibler}},\ }\href@noop {} {\bibfield  {journal} {\bibinfo  {journal} {The
  annals of mathematical statistics}\ }\textbf {\bibinfo {volume} {22}},\
  \bibinfo {pages} {79} (\bibinfo {year} {1951})}\BibitemShut {NoStop}%
\bibitem [{\citenamefont {Einstein}(1924)}]{BEC}%
  \BibitemOpen
  \bibfield  {author} {\bibinfo {author} {\bibfnamefont {A.}~\bibnamefont
  {Einstein}},\ }\href@noop {} {\bibfield  {journal} {\bibinfo  {journal}
  {K{\"o}nigliche Preu{\ss}ische Akademie der Wissenschaften.}\ }\textbf
  {\bibinfo {volume} {261-267}} (\bibinfo {year} {1924})}\BibitemShut {NoStop}%
\bibitem [{\citenamefont {Navez}\ \emph {et~al.}(1997)\citenamefont {Navez},
  \citenamefont {Bitouk}, \citenamefont {Gajda}, \citenamefont {Idziaszek},\
  and\ \citenamefont {Rza{\.z}ewski}}]{navez1997fourth}%
  \BibitemOpen
  \bibfield  {author} {\bibinfo {author} {\bibfnamefont {P.}~\bibnamefont
  {Navez}}, \bibinfo {author} {\bibfnamefont {D.}~\bibnamefont {Bitouk}},
  \bibinfo {author} {\bibfnamefont {M.}~\bibnamefont {Gajda}}, \bibinfo
  {author} {\bibfnamefont {Z.}~\bibnamefont {Idziaszek}}, \ and\ \bibinfo
  {author} {\bibfnamefont {K.}~\bibnamefont {Rza{\.z}ewski}},\ }\href@noop {}
  {\bibfield  {journal} {\bibinfo  {journal} {Physical review letters}\
  }\textbf {\bibinfo {volume} {79}},\ \bibinfo {pages} {1789} (\bibinfo {year}
  {1997})}\BibitemShut {NoStop}%
\bibitem [{\citenamefont {Holthaus}\ \emph {et~al.}(1998)\citenamefont
  {Holthaus}, \citenamefont {Kalinowski},\ and\ \citenamefont
  {Kirsten}}]{holthaus1998condensate}%
  \BibitemOpen
  \bibfield  {author} {\bibinfo {author} {\bibfnamefont {M.}~\bibnamefont
  {Holthaus}}, \bibinfo {author} {\bibfnamefont {E.}~\bibnamefont
  {Kalinowski}}, \ and\ \bibinfo {author} {\bibfnamefont {K.}~\bibnamefont
  {Kirsten}},\ }\href@noop {} {\bibfield  {journal} {\bibinfo  {journal}
  {Annals of Physics}\ }\textbf {\bibinfo {volume} {270}},\ \bibinfo {pages}
  {198} (\bibinfo {year} {1998})}\BibitemShut {NoStop}%
\bibitem [{\citenamefont {Mullin}\ and\ \citenamefont
  {Fernandez}(2003)}]{mullin2003bose}%
  \BibitemOpen
  \bibfield  {author} {\bibinfo {author} {\bibfnamefont {W.}~\bibnamefont
  {Mullin}}\ and\ \bibinfo {author} {\bibfnamefont {J.}~\bibnamefont
  {Fernandez}},\ }\href@noop {} {\bibfield  {journal} {\bibinfo  {journal}
  {American Journal of Physics}\ }\textbf {\bibinfo {volume} {71}},\ \bibinfo
  {pages} {661} (\bibinfo {year} {2003})}\BibitemShut {NoStop}%
\bibitem [{\citenamefont {Chatterjee}\ and\ \citenamefont
  {Diaconis}(2014)}]{chatterjee2014fluctuations}%
  \BibitemOpen
  \bibfield  {author} {\bibinfo {author} {\bibfnamefont {S.}~\bibnamefont
  {Chatterjee}}\ and\ \bibinfo {author} {\bibfnamefont {P.}~\bibnamefont
  {Diaconis}},\ }\href@noop {} {\bibfield  {journal} {\bibinfo  {journal}
  {Journal of Physics A: Mathematical and Theoretical}\ }\textbf {\bibinfo
  {volume} {47}},\ \bibinfo {pages} {085201} (\bibinfo {year}
  {2014})}\BibitemShut {NoStop}%
\bibitem [{\citenamefont {Tarasov}\ \emph {et~al.}(2015)\citenamefont
  {Tarasov}, \citenamefont {Kocharovsky},\ and\ \citenamefont
  {Kocharovsky}}]{tarasov2015grand}%
  \BibitemOpen
  \bibfield  {author} {\bibinfo {author} {\bibfnamefont {S.}~\bibnamefont
  {Tarasov}}, \bibinfo {author} {\bibfnamefont {V.~V.}\ \bibnamefont
  {Kocharovsky}}, \ and\ \bibinfo {author} {\bibfnamefont {V.}~\bibnamefont
  {Kocharovsky}},\ }\href@noop {} {\bibfield  {journal} {\bibinfo  {journal}
  {Journal of Statistical Physics}\ }\textbf {\bibinfo {volume} {161}},\
  \bibinfo {pages} {942} (\bibinfo {year} {2015})}\BibitemShut {NoStop}%
\bibitem [{\citenamefont {Crisanti}\ \emph {et~al.}(2019)\citenamefont
  {Crisanti}, \citenamefont {Sarracino},\ and\ \citenamefont
  {Zannetti}}]{crisanti2019condensation}%
  \BibitemOpen
  \bibfield  {author} {\bibinfo {author} {\bibfnamefont {A.}~\bibnamefont
  {Crisanti}}, \bibinfo {author} {\bibfnamefont {A.}~\bibnamefont {Sarracino}},
  \ and\ \bibinfo {author} {\bibfnamefont {M.}~\bibnamefont {Zannetti}},\
  }\href@noop {} {\bibfield  {journal} {\bibinfo  {journal} {Physical Review
  Research}\ }\textbf {\bibinfo {volume} {1}},\ \bibinfo {pages} {023022}
  (\bibinfo {year} {2019})}\BibitemShut {NoStop}%
\bibitem [{\citenamefont {Kittel}\ and\ \citenamefont
  {Kroemer}(1980)}]{kittel}%
  \BibitemOpen
  \bibfield  {author} {\bibinfo {author} {\bibfnamefont {C.}~\bibnamefont
  {Kittel}}\ and\ \bibinfo {author} {\bibfnamefont {H.}~\bibnamefont
  {Kroemer}},\ }\href@noop {} {\emph {\bibinfo {title} {Thermal Physics}}}\
  (\bibinfo  {publisher} {W.H. Freeman, San Francisco},\ \bibinfo {year}
  {1980})\BibitemShut {NoStop}%
\bibitem [{\citenamefont {Garlaschelli}\ \emph {et~al.}(2018)\citenamefont
  {Garlaschelli}, \citenamefont {den Hollander},\ and\ \citenamefont
  {Roccaverde}}]{garlaschelli2018covariance}%
  \BibitemOpen
  \bibfield  {author} {\bibinfo {author} {\bibfnamefont {D.}~\bibnamefont
  {Garlaschelli}}, \bibinfo {author} {\bibfnamefont {F.}~\bibnamefont {den
  Hollander}}, \ and\ \bibinfo {author} {\bibfnamefont {A.}~\bibnamefont
  {Roccaverde}},\ }\href@noop {} {\bibfield  {journal} {\bibinfo  {journal}
  {Journal of Statistical Physics}\ }\textbf {\bibinfo {volume} {173}},\
  \bibinfo {pages} {644} (\bibinfo {year} {2018})}\BibitemShut {NoStop}%
\bibitem [{\citenamefont {Dionigi}\ \emph {et~al.}(2020)\citenamefont
  {Dionigi}, \citenamefont {Garlaschelli}, \citenamefont {den Hollander},\ and\
  \citenamefont {Mandjes}}]{dionigi2020spectral}%
  \BibitemOpen
  \bibfield  {author} {\bibinfo {author} {\bibfnamefont {P.}~\bibnamefont
  {Dionigi}}, \bibinfo {author} {\bibfnamefont {D.}~\bibnamefont
  {Garlaschelli}}, \bibinfo {author} {\bibfnamefont {F.}~\bibnamefont {den
  Hollander}}, \ and\ \bibinfo {author} {\bibfnamefont {M.}~\bibnamefont
  {Mandjes}},\ }\href@noop {} {\enquote {\bibinfo {title} {A spectral signature
  of breaking of ensemble equivalence for constrained random graphs},}\
  }\bibinfo {howpublished} {https://arxiv.org/abs/2009.05155} (\bibinfo {year}
  {2020})\BibitemShut {NoStop}%
\end{thebibliography}
\end{document}